\documentclass[nofootinbib,letterpaper]{revtex4}

\usepackage{graphicx}
\usepackage{dcolumn}
\usepackage{amsmath,amssymb,epsfig}
\usepackage{paralist}
\usepackage{wrapfig}
\usepackage{graphicx}
\usepackage{multirow}
\usepackage{color,soul}
\usepackage[normalem]{ulem}
\usepackage{mathtools}
\usepackage{tikz,pgfplots}

\allowdisplaybreaks

\renewcommand{\vec}[1]{\boldsymbol{\mathrm{#1}}}

\begin{document}

\title{Diffraction of electromagnetic waves by an extended gravitational  lens}

\author{Slava G. Turyshev$^{1}$, Viktor T. Toth$^2$
}

\affiliation{\vskip 3pt
$^1$Jet Propulsion Laboratory, California Institute of Technology,\\
4800 Oak Grove Drive, Pasadena, CA 91109-0899, USA
}%

\affiliation{\vskip 3pt
$^2$Ottawa, Ontario K1N 9H5, Canada
}%

\date{\today}

\begin{abstract}

We continue our study of the optical properties of the solar gravitational lens (SGL). Taking the next step beyond representing it as an idealized monopole, we now characterize the gravitational field of the Sun using an infinite series of multipole moments. We consider the propagation of electromagnetic (EM) waves in this gravitational field within the first post-Newtonian approximation of the general theory of relativity. The problem is formulated within the Mie diffraction theory. We solve Maxwell's equations for the EM wave propagating in the background of a static gravitational field of an extended gravitating body, while accounting for multipole contributions. Using a wave-theoretical approach and the eikonal approximation, we find an exact closed form solution for the Debye potentials and determine the EM field at an image plane in the strong interference region of the lens. The resulting EM field is characterized by a new diffraction integral. We study this solution and show how the presence of multipoles affects the optical properties of the lens, resulting in distinct diffraction patterns. We identify the gravitational deflection angle with the individual contributions due to each of the multipoles.  Treating the Sun as an extended, axisymmetric, rotating body, we show that each zonal harmonics causes light to diffract into an area whose boundary is a caustic of a particular shape. The appearance of the caustics modifies the point-spread function (PSF) of the lens, thus affecting its optical properties. The new wave-theoretical  solution allows the study gravitational lensing by a realistic lens that possesses an arbitrary number of gravitational multipoles. This {\em angular eikonal method} represents an improved treatment of realistic gravitational lensing. It may be used for a wave-optical description of many astrophysical lenses.

\end{abstract}


\maketitle

\section{Introduction}

Studied for over a century \cite{Einstein-1916,Einstein:1936}, gravitational lensing today is well understood \cite{Liebes:1964,Refsdal:1964,Schneider-Ehlers-Falco:1992}. It occurs when light travels in the vicinity of a gravitating body. In the post-Newtonian limit of the general theory of relativity, the gravitational field serves as a refracting medium \cite{Fock-book:1959,Landau-Lifshitz:1988} that deflects light rays towards the body.

\begin{figure}
\includegraphics[scale=0.25]{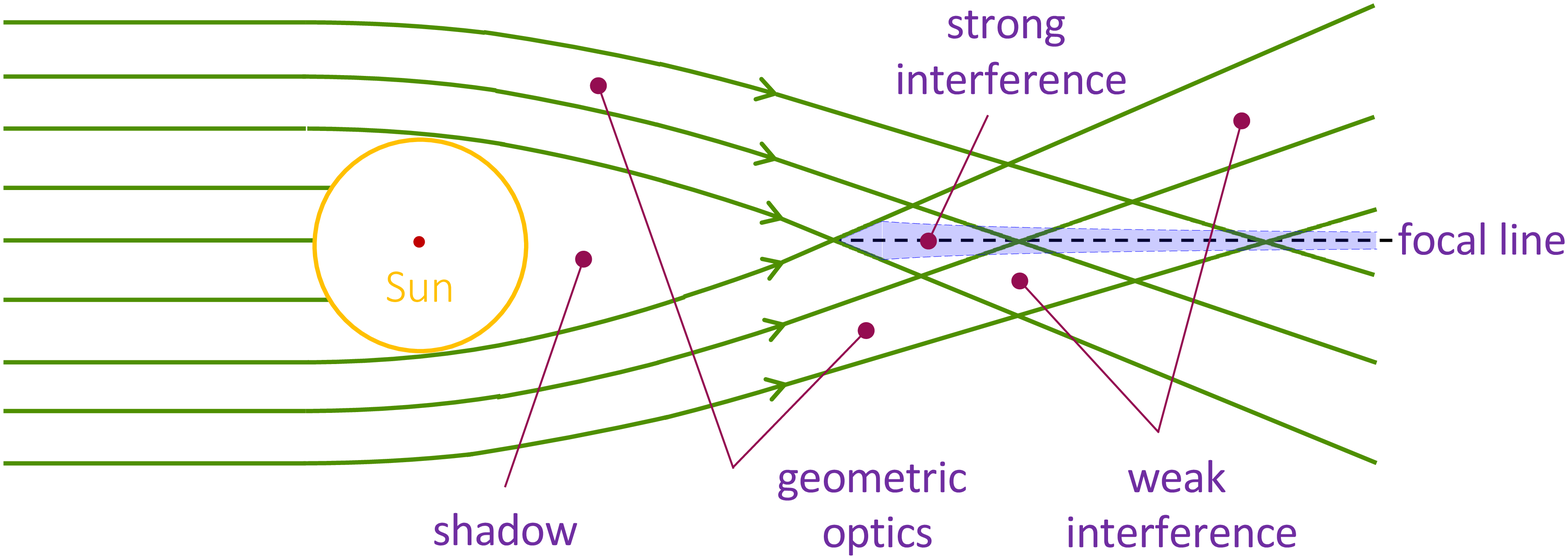}
\caption{\label{fig:regions}The different optical regions of the SGL
(from \cite{Turyshev-Toth:2019-fin-difract}) with the strong interference region formed beyond 547.8 AU.
}
\end{figure}

Following a methodical approach, we began our investigation by treating the solar gravitational field as a spherically symmetric field of a gravitational monopole, or point mass \cite{Turyshev:2017,Turyshev-Toth:2017}. After passing by such a monopole, light rays are focused in what we call the region of strong interference (Fig.~\ref{fig:regions}), with impressive optical properties including significant light amplification. However, even gravitational monopole lenses are subject to optical aberrations. As the deflection angle is inversely proportional to the impact parameter, light rays with larger impact parameters with respect to the lens are focused at larger distances from it. This causes spherical aberration, leading to blurred images and the requirement to employ appropriately designed deconvolution algorithms \cite{Turyshev-Toth:2020-extend}.

With this model, we were able to establish the basic properties of the solar gravitational lens (SGL) and understand image formation and image recovery.  We considered gravitational lensing by the Sun as the means to obtain high-resolution images of faint objects, such a  exoplanets. To enable practical applications of the SGL, we developed a wave-optical treatment of the diffraction of light in the presence of the solar gravity field. We studied the impact of the solar corona on light propagation in the vicinity of the Sun. We showed that diffraction in the solar atmosphere defocuses EM waves for wavelengths greater than 1 mm, but its impact is negligible at optical and IR wavelengths \cite{Turyshev-Toth:2019,Turyshev-Toth:2018-plasma}.  We extended our formulation to the case of extended sources at large but finite distances \cite{Turyshev-Toth:2019-fin-difract}. We studied image formation with the SGL \cite{Turyshev-Toth:2020-photom,Turyshev-Toth:2020-image} and addressed the realistic sensitivity of prospective imaging observations \cite{Turyshev-Toth:2020-extend}. In addition, we studied the image recovery process and have shown that the SGL may be used for multipixel imaging of exoplanets \cite{Toth-Turyshev:2020} that may be conducted in the context of a realistic space mission \cite{Turyshev-etal:2020-PhaseII}.

The next step is dictated by the realization that nothing is perfect in life, not even the Sun. Its rotation, the resulting oblateness, and its internal mass distribution result in a gravitational field that deviates from the idealized monopole. These deviations are small (in fact, the Sun is {\em almost} perfect), but given the distance and length scales involved, their impact cannot be neglected. The dimensionless magnitude of corrections due to deviations from the monopole is of ${\cal O}(10^{-7})$. This is not much until we consider that deflection of light by the SGL amounts to displacing a ray of light by at least as much as the solar radius by the time that ray approaches the focal region. An ${\cal O}(10^{-7})$ correction on this scale amounts to an additional deflection by tens if not hundreds of meters, which is quite significant when compared to the scale (typically, 1--10~km across) of the projected image size of a desired target.

Therefore, it is necessary to develop a formalism to modify the point-spread function (PSF) of the SGL, taking into account that, on the one hand, we deal with very large distances (measured in light years for distant imaging targets or in many hundreds of astronomical units (AU)) when it comes to the distance of the focal region from the Sun) and, on the other hand, distances measured on the scale of meters or less (such as the telescope aperture or the centimeter-scale Airy pattern that appears in the image plane, itself a result of observing a signal with a wavelength of ${\cal O}(1~\mu{\rm m})$). Consequently, even higher-order multipole moments (octupole, dodecapole, hexadecapole moments) of the Sun may have to be considered for accurate image modeling and reconstruction of some exoplanetary targets. These moments break the azimuthal symmetry of the PSF, introducing caustics instead of the regular Bessel $J_0$ pattern \cite{Turyshev-Toth:2017}. This is why we are turning the page on the chapter dealing with monopole gravitational lenses. With the present paper, we open a new, exciting area of investigation, aimed at developing a comprehensive description of realistic gravitational lenses possessing an arbitrary number of gravitational multipole moments.

This paper is organized as follows: In Section~\ref{sec:EM-waves} we discuss the solution of Maxwell's equations in the curved spacetime of the solar gravitational field, described at the first post-Newtonian approximation of the general theory of relativity. We develop a solution for the Debye potential using the eikonal approximation.
In Section~\ref{sec:Debye-sol}, we formulate a generic solution for EM waves in the field of a static, extended gravitational lens.
In Section~\ref{sec:EM-field-outside} we develop a general solution for the EM field, characterizing the scattering of EM waves on an extended lens.
In Section~\ref{sec:IF-region}, we study the EM field in the interference region. We develop a new integral formulation that describes light diffraction in the strong interference region.
In Section~\ref{sec:end} we discuss the results obtained and the next steps in our investigation.
To aid with the flow of material in this paper, we placed some important derivations in appendices. Appendix~\ref{app:Debye} discusses an approach to Maxwell's equations for EM waves propagating on the background of the static gravitational field of an extended lens. In Appendix~\ref{sec:eik-phase}, we discuss the eikonal phase for
\begin{inparaenum}[i)]
\item a generic axisymmetric gravitating body whose gravitational potential given by a set of gravitational multipoles, and
\item generic spatial-trace free (STF) tensors representing bodies with arbitrary gravitational fields.
\end{inparaenum}
Finally, in Appendix~\ref{app:pathintegral}, we explore the connection between our results and the path integral formalism.

\section{Electromagnetic waves in a static gravitational field}
\label{sec:EM-waves}

To describe the optical properties of the solar gravitational lens (SGL), we use a static harmonic metric\footnote{The notational conventions used in this paper are the same as in \cite{Landau-Lifshitz:1988,Turyshev-Toth:2013}: Latin indices ($i,j,k,...$) are spacetime indices that run from 0 to 3. Greek indices $\alpha,\beta,...$ are spatial indices that run from 1 to 3. In case of repeated indices in products, the Einstein summation rule applies: e.g., $a_mb^m=\sum_{m=0}^3a_mb^m$. Bold letters denote spatial (three-dimensional) vectors: e.g., ${\vec a} = (a_1, a_2, a_3), {\vec b} = (b_1, b_2, b_3)$. The dot ($\cdot$) and cross ($\times$) are used to indicate the Euclidean inner product and cross product of spatial vectors; following the convention of \cite{Fock-book:1959}, these are enclosed in round and square brackets, respectively. Latin indices are raised and lowered using the metric $g_{mn}$. The Minkowski (flat) spacetime metric is given by $\gamma_{mn} = {\rm diag} (1, -1, -1, -1)$, so that $\gamma_{\mu\nu}a^\mu b^\nu=-({\vec a}\cdot{\vec b})$. We use powers of the inverse of the speed of light, $c^{-1}$, and the gravitational constant, $G$, as bookkeeping devices for order terms: in the low-velocity ($v\ll c$), weak-field ($r_g/r=2GM/rc^2\ll 1$) approximation, a quantity of ${\cal O}(c^{-2})\simeq{\cal O}(G)$, for instance, has a magnitude comparable to $v^2/c^2$ or $GM/c^2r$. The notation ${\cal O}(a^k,b^\ell)$ is used to indicate that the preceding expression is free of terms containing powers of $a$ greater than or equal to $k$, and powers of $b$ greater than or equal to $\ell$. Other notations are explained in the paper.} in the first post-Newtonian approximation of the general theory of relativity. The line element for this metric may be given, in spherical coordinates $(r,\theta,\phi)$, as \cite{Fock-book:1959,Turyshev-Toth:2013}:
\begin{eqnarray}
ds^2&=&u^{-2}c^2dt^2-u^2\big(dr^2+r^2(d\theta^2+\sin^2\theta d\phi^2)\big),
\label{eq:metric-gen}
\end{eqnarray}
where, to the accuracy sufficient to describe light propagation in the solar system, the quantity $u$ can be given in terms of the Newtonian potential $U$ as
\begin{eqnarray}
u=1+c^{-2}U+{\cal O}(c^{-4}), \qquad \text{where} \qquad U({\vec x})=G\int\frac{\rho(x')d^3x'}{|{\vec x}-{\vec x}'|},
\label{eq:w-PN}
\end{eqnarray}
and $\rho({\vec x})$ is the mass density that is the source of the gravitational field.

The metric (\ref{eq:metric-gen})--(\ref{eq:w-PN}) allows us to consider effects on the propagation of light by the gravitational field of the Sun, due to an arbitrary static gravitational field. Furthermore, it was shown in \cite{Asada-Kasai:2020}  that to first order in the gravitational constant $G$, a rotating and a nonrotating lens cannot be distinguished. Thus, to the extent that it contributes to the quadrupole moment, solar rotation is automatically accounted for in our formalism.

The gravitational field of the Sun is weak: its potential is $GM/c^2r\lesssim 2\times 10^{-6}$ everywhere in the solar system. This allows us to carry out calculations to the first post-Newtonian order, while dropping higher-order terms.

We use the generally covariant form of Maxwell's equations for the electromagnetic (EM) field \cite{Landau-Lifshitz:1988,Turyshev-Toth:2017} and consider the propagation of an EM wave in the vacuum in the absence of charges and currents, i.e., $j^k=(\rho,{\vec j})=0$. As we showed in  \cite{Turyshev-Toth:2017,Turyshev-Toth:2019}, for the metric  (\ref{eq:metric-gen}) we obtain the following form for Maxwell's equations:
{}
\begin{eqnarray}
{\rm curl}\,{\vec D}&=&- u^2
\frac{\partial \,{\vec B}}{c\partial t}+{\cal O}(G^2),
\qquad ~{\rm div}\big(u^2\,{\vec D}\big)={\cal O}(G^2),
\label{eq:rotE_fl}\\[3pt]
{\rm curl}\,{\vec B}&=&u^2
\frac{\partial \,{\vec D}}{c\partial t}+{\cal O}(G^2),
\qquad \quad
{\rm div }\big(u^2\,{\vec B}\big)={\cal O}(G^2),
\label{eq:rotH_fl}
\end{eqnarray}
where the differential operators ${\rm curl}\, {\vec F}$  and ${\rm div} \,{\vec F}$ are with respect to the usual 3-space Euclidean flat metric (see \cite{Turyshev-Toth:2017} for technical details).

\subsection{Representation of the EM field in terms of Debye potentials}
\label{sec:debye}

To describe the problem of an EM wave propagating in the gravitational field of an extended lens that induces the static gravitational field with metric (\ref{eq:metric-gen}), we follow the Mie diffraction theory \cite{Mie:1908,Born-Wolf:1999} that allows us to determine the three-dimensional structure of the EM field diffracted on a spherical obstruction. This technique is done based representing the Maxwell equations (\ref{eq:rotE_fl})--(\ref{eq:rotH_fl}) in terms of the Debye potentials (see \cite{Turyshev-Toth:2017,Turyshev-Toth:2019}  and references therein).

Relying on the approach that we previously developed (see \cite{Turyshev-Toth:2017,Turyshev-Toth:2019}), in Appendix~\ref{app:Debye} we obtain the complete solution of these equations in terms of the electric and magnetic Debye potentials \cite{Born-Wolf:1999}, ${}^e\Pi$ and ${}^m\Pi$. We follow closely the derivation in \cite{Turyshev-Toth:2017} (see Appendix E therein) and also in \cite{Turyshev-Toth:2019} (see Appendix A therein).

We treat the lens as a compact gravitating body whose gravitational potential admits a representation in the form of an infinite series of zonal and tesseral harmonics (e.g., as given by (\ref{eq:pot_w_0})). The result is a system of equations for the components of a monochromatic EM field, characterized by the wavenumber $k=\omega/c$:
{}
\begin{eqnarray}
{\hat { D}}_r&=&
\frac{1}{u}\Big\{\frac{\partial^2 }{\partial r^2}
\Big[\frac{r\,{}^e{\hskip -1pt}\Pi}{u}\Big]+\Big(k^2 u^4-u\big(\frac{1}{u}\big)''\Big)\Big[\frac{r\,{}^e{\hskip -1pt}\Pi}{u}\Big]\Big\},
\label{eq:Dr-em0}\\[3pt]
{\hat {  D}}_\theta&=&
\frac{1}{ u^2r}\frac{\partial^2 \big(r\,{}^e{\hskip -1pt}\Pi\big)}{\partial r\partial \theta}+\frac{ik}{r\sin\theta}
\frac{\partial\big(r\,{}^m{\hskip -1pt}\Pi\big)}{\partial \phi},
\label{eq:Dt-em0}\\[3pt]
{\hat {  D}}_\phi&=&
\frac{1}{ u^2r\sin\theta}
\frac{\partial^2 \big(r\,{}^e{\hskip -1pt}\Pi\big)}{\partial r\partial \phi}-\frac{ik}{r}
\frac{\partial\big(r\,{}^m{\hskip -1pt}\Pi\big)}{\partial \theta},
\label{eq:Dp-em0}\\[3pt]
{\hat {  B}}_r&=&
\frac{1}{u}\Big\{\frac{\partial^2}{\partial r^2}\Big[\frac{r\,{}^m{\hskip -1pt}\Pi}{u}\Big]+\Big(k^2 u^4- u\big(\frac{1}{u}\big)''\Big)\Big[\frac{r\,{}^m{\hskip -1pt}\Pi}{u}\Big]\Big\},
\label{eq:Br-em0}\\[3pt]
{\hat {  B}}_\theta&=&
-\frac{ik}{r\sin\theta} \frac{\partial\big(r\,{}^e{\hskip -1pt}\Pi\big)}{\partial \phi}+\frac{1}{u^2r}
\frac{\partial^2 \big(r\,{}^m{\hskip -1pt}\Pi\big)}{\partial r\partial \theta},
\label{eq:Bt-em0}\\[3pt]
{\hat { B}}_\phi&=&
\frac{ik}{r}\frac{\partial\big(r\,{}^e{\hskip -1pt}\Pi\big)}{\partial \theta}+\frac{1}{u^2r\sin\theta}
\frac{\partial^2 \big(r\,{}^m{\hskip -1pt}\Pi\big)}{\partial r\partial \phi},
\label{eq:Bp-em0}
\end{eqnarray}
where  the electric and magnetic Debye potentials $\Pi({\vec r})=({}^e{\hskip -1pt}\Pi; {}^m{\hskip -1pt}\Pi )$ satisfy the following wave equation:
{}
\begin{eqnarray}
\big(\Delta+k^2u^4\big)\Big[\frac{\Pi}{u}\Big]= {\cal O}\Big(r_g^2,\frac{J_2}{r^3}\frac{\Pi}{u}\Big),
\label{eq:Pi-eq=}
\end{eqnarray}
with the quantity $u$ given by (\ref{eq:w-PN}). The Newtonian potential $U$ in (\ref{eq:w-PN}) at this point is unconstrained and can describe an arbitrary (weak, static) gravitational field. Here, $r_g=2GM/c^2$ is the Schwarzschild radius of the lens; $J_2$ characterizes the quadrupole component of the gravitational potential, $U$, of an extended gravitational lens.

Essentially the solution (\ref{eq:Dr-em0})--(\ref{eq:Bp-em0}) together with (\ref{eq:Pi-eq=}) was obtained under the thin lens or eikonal approximation where the primary emphasis was on the effect of the gravitational field on the phase of the EM wave rather than its amplitude. This approximation is well-justified as the source and the image plane are at very large distances from the lens. Our analysis showed that the effects of the higher order gravitational multipoles, starting from $J_2$, depend on the distance to the lens and, thus, may be neglected.

As a result, the entire solution to Maxwell's equations describing light propagation in the weak gravitational field with the post-Newtonian metric tensor (\ref{eq:metric-gen}) depends on the solution of the wave equation for the Debye potential (\ref{eq:Pi-eq=}). Using  the expression for $u$ from (\ref{eq:w-PN}), this equation is given as (see (\ref{eq:Pi-d+}))
\begin{eqnarray}
\Big(\Delta+k^2\big(1+\frac{4U}{c^2}\big)\Big)\Big[\frac{\Pi}{u}\Big]= {\cal O}\Big(r_g^2,\frac{J_2}{r^3}\frac{\Pi}{u}\Big).
\label{eq:Pi-eq+wew1*+}
\end{eqnarray}

Expressions (\ref{eq:Dr-em0})--(\ref{eq:Bp-em0}) together (\ref{eq:Pi-eq+wew1*+}) represent the solution of the Mie problem in terms of Debye potentials \cite{Mie:1908,Born-Wolf:1999}, in the presence of the gravitational field of an extended gravitating body, taken at the first post-Newtonian approximation of the general theory of relativity \cite{Turyshev:2017,Turyshev-Toth:2017} under the eikonal (or, essentially, the thin lens) approximation.

The set of equations (\ref{eq:Dr-em0})--(\ref{eq:Bp-em0}) with  (\ref{eq:Pi-eq+wew1*+}) determines the Debye potential for the entire problem. We see that the solution of (\ref{eq:Pi-eq+wew1*+}) now depends on the entire Newtonian potential, $U(\vec r)$, that may have arbitrary complexity. No exact solution of this time-independent Schr\"odinger equation exists. Thus, we need to develop an approximate solution that is suitable for our situation.  We found an approach to develop such a solution using the eikonal approximation.

\subsection{Separating variables in the equation for the Debye potential}
\label{sec:Deb-pot}

To consider the eikonal approximation, we present the Newtonian potential, $U$, as
{}
\begin{eqnarray}
U(\vec r)= \frac{GM}{r}+\delta U(\vec r),
\label{eq:U-dec}
\end{eqnarray}
where  the first term is the spherically symmetric monopole contribution and the second term, $\delta U(\vec r)$, represents the combined contribution of all the other terms in a suitable expansion of $U(\vec r)$.

If $\delta U(\vec r)$ is absent, (\ref{eq:Pi-d+}) reduces to the case of diffraction of the EM waves by a gravitational monopole (i.e., Schr\"odinger's equation with a Coulomb potential---see details in \cite{Turyshev-Toth:2017}):
{}
\begin{eqnarray}
\Big(\Delta +k^2\big(1+\frac{2r_g}{r}\big)\Big)\Big[\frac{\Pi_0}{u}\Big]=0.
\label{eq:Pi-d+0}
\end{eqnarray}

This equation describes light scattering that is dominated by a spherical relativistic potential due to a gravitational monopole  (which is equivalent to an attractive Coulomb potential, discussed in quantum mechanics \cite{Schiff:1968,Landau-Lifshitz:1989,Messiah:1968}).  In our case, this equation describes the incident wave that travels towards the lens from the source.

The solution to (\ref{eq:Pi-d+0}) is well known (see  \cite{Turyshev-Toth:2017} for details).  In this case, Eq.~(\ref{eq:Pi-eq+wew1*+}) is typically solved by separating variables \cite{Born-Wolf:1999}, which, in spherical polar coordinates, takes the form \cite{Turyshev-Toth:2017,Turyshev-Toth:2018-plasma}:
\begin{eqnarray}
\frac{\Pi_0}{u}=\frac{1}{r}R(r)\Theta(\theta)\Phi(\phi),
\label{eq:Pi*}
\end{eqnarray}
with integration constants and coefficients that are determined by boundary conditions. Direct substitution into (\ref{eq:Pi-d+0}) reveals that the functions $R, \Theta$ and $\Phi$ must satisfy the following ordinary differential equations:
{}
\begin{eqnarray}
\frac{d^2 R}{d r^2}+\Big(k^2(1+\frac{2r_g}{r}) -\frac{\alpha}{r^2}\Big)R&=&{\cal O}(r_g^2),
\label{eq:R-bar*}\\
\frac{1}{\sin\theta}\frac{d}{d \theta}\Big(\sin\theta \frac{d \Theta}{d \theta}\Big)+\big(\alpha-\frac{\beta}{\sin^2\theta}\big)\Theta&=&{\cal O}(r_g^2),
\label{eq:Th*}\\
\frac{d^2 \Phi}{d \phi^2}+\beta\Phi&=&{\cal O}(r_g^2).
\label{eq:Ph*}
\end{eqnarray}

As we discussed in \cite{Turyshev-Toth:2017}, the solution to (\ref{eq:Ph*}) is given as usual \cite{Born-Wolf:1999,Landau-Lifshitz:1989}:
{}
\begin{eqnarray}
\Phi_m(\phi)=e^{\pm im\phi}  \quad\rightarrow \quad \Phi_m(\phi)=a_m\cos (m\phi) +b_m\sin (m\phi),
\label{eq:Ph_m}
\end{eqnarray}
where $\beta=m^2$, $m$ is an integer and $a_m$ and $b_m$ are integration constants.

Equation (\ref{eq:Th*}) is well known for spherical harmonics. Single-valued solutions to this equation exist when $\alpha=l(l+1)$ with ($l>|m|,$ integer). With this condition, the solution to (\ref{eq:Th*}) becomes
{}
\begin{eqnarray}
\Theta_{lm}(\theta)&=&P^{(m)}_l(\cos\theta).
\label{eq:Th_lm}
\end{eqnarray}

We now focus on the equation for the radial function (\ref{eq:R-bar*}), where, because of (\ref{eq:Th*}), we have $\alpha=\ell(\ell+1)$.  As a result, (\ref{eq:R-bar*}) takes the form
{}
\begin{eqnarray}
\frac{d^2 R}{d r^2}+\Big(k^2(1+\frac{2r_g}{r})-\frac{\ell(\ell+1)}{r^2}\Big)R&=&{\cal O}(r^2_g).
\label{eq:R-bar-k*}
\end{eqnarray}
The solution to this equation is given in the form of a Coulomb function $F_\ell(kr_g,kr) $ \cite{Turyshev-Toth:2017}.

Collecting results for $\Phi_m(\phi)$, $\Theta_{lm}(\theta)$ and $R_\ell=F_\ell(kr_g,kr) $, we can assemble the ultimate solution to  (\ref{eq:Pi-d+0}), as was done in \cite{Turyshev-Toth:2017,Turyshev-Toth:2019}.  This solution is used to describe the electric and magnetic potentials of the incident wave, ${}^e{\hskip -1pt}\Pi_0$ and ${}^m{\hskip -1pt}\Pi_0$, which may be given in terms of a single potential $\Pi_0(r, \theta)$ (see  \cite{Turyshev-Toth:2017} for details):
{}
\begin{align}
  \left( \begin{aligned}
{}^e{\hskip -1pt}\Pi_0& \\
{}^m{\hskip -1pt}\Pi_0& \\
  \end{aligned} \right) =&  \left( \begin{aligned}
\cos\phi \\
\sin\phi  \\
  \end{aligned} \right) \,\Pi_0(r, \theta), &
\hskip 2pt {\rm where} \quad
\Pi_0 (r, \theta)=
\frac{E_0}{k^2}\frac{u}{r}\sum_{\ell=1}^\infty i^{\ell-1}\frac{2\ell+1}{\ell(\ell+1)}e^{i\sigma_\ell}
F_\ell(kr_g,kr) P^{(1)}_\ell(\cos\theta)+{\cal O}(r_g^2).
  \label{eq:Pi_ie*+*=}
\end{align}

Therefore, in the case when deviations from the monopole gravitational field represented by the term $\delta U(\vec r)$ in (\ref{eq:U-dec}) are absent, we can find a solution for the Debye potential (\ref{eq:Pi-d+0}) by separating variables with the ansatz (\ref{eq:Pi*}) that is used to deal with the Coulomb potential \cite{Turyshev:2017,Turyshev-Toth:2017}.  The structure of the resulting solution (\ref{eq:Pi_ie*+*=}) reflects the spherical symmetry that is preserved in this case.  The presence of the monopole is manifested by the potential term $2r_g/r$ in the equation for the radial function (\ref{eq:R-bar*}). Note that the other equations (\ref{eq:Th*}) and (\ref{eq:Ph*}) are not affected by gravity.

The situation changes drastically when the term $\delta U(\vec r)$ is present in (\ref{eq:U-dec}). In this case, (\ref{eq:Pi-eq+wew1*+}) becomes highly nonlinear and separation of variables (\ref{eq:Pi*}) does not work.  No exact solution of this equation is known. However, in some cases this equation may be solved using well-justified approximation methods. One such method, the eikonal approximation, is particularly useful for high-energy atomic scattering \cite{Friedrich-book-2006,Friedrich-book-2013,Burke-book-2011} and it is applicable in our case, which corresponds to the high-energy approximation in optical scattering \cite{Sharma-etal:1988,Sharma-Sommerford:1990,Sharma-Sommerford-book:2006}.

\subsection{Finding the Debye potential with the eikonal approximation}
\label{sec:eik-wfr}

We may now use the result (\ref{eq:Pi_ie*+*=}) as the basis to find solutions when $U$ is not restricted to a monopole gravitational field. We extend the discussion in the preceding subsection by considering the complete post-Newtonian potential $U$ of an extended body as the sum of two terms that includes the monopole field, $GM/r$, which is long-range  \cite{Turyshev-Toth:2017}, and deviations from the monopole, which constitute a short-range potential, $V_{\tt sr} (\vec r)=\delta U(\vec r)/c^2$. This yields the following form for the potential term in (\ref{eq:Pi-d+}):
{}
\begin{eqnarray}
\frac{4U}{c^2}&=& \frac{2r_g}{r}+ 4V_{\tt sr}.
\label{eq:shart-pot}
\end{eqnarray}
This decomposition allows us  to proceed with solving  (\ref{eq:Pi-d+}) that now takes the form
{}
\begin{eqnarray}
\Big(\Delta +k^2\big(1+\frac{2r_g}{r}+4V_{\tt sr}({ r})\big)\Big)
\Pi({\vec r}) = {\cal O}\Big(r_g^2,\frac{J_2}{r^3}\Pi\Big),
\label{eq:Pi-eq*0+*+1*}
\end{eqnarray}
where $V_{\tt sr}$ is from (\ref{eq:shart-pot}). In explicit form this short-range potential is given by either by (\ref{eq:V-sr-m2}) that is valid for any generic gravitational field, or expressed in terms of zonal harmonic coefficients $J_n$ using (\ref{eq:V-sr-m2A}), which is more suitable to describe the gravitational field of a rotating, axisymmetric mass, such as the Sun.

To solve (\ref{eq:Pi-eq*0+*+1*}), we will treat $V_{\tt sr} (\vec r)$ as a perturbation to the monopole term and will use the eikonal approximation \cite{Semon-Taylor:1977,Landau-Lifshitz:1988,Born-Wolf:1999,Sharma-Sommerford:1990,Grandy-book-2000}. To implement this approach, we consider a trial solution in the form
{}
\begin{eqnarray}
\Pi({\vec r})=\Pi_0({\vec r})\phi(\vec r),
\label{eq:Pi-eq*0+*+1}
\end{eqnarray}
where $\Pi_0({\vec r})$ is the ``free'' Debye potential  for the monopole gravitation given by (\ref{eq:Pi_ie*+*=}) \cite{Turyshev-Toth:2017,Turyshev-Toth:2019}. In other words, in the eikonal approximation the Debye potential $\Pi_0(\vec r)$, becomes ``distorted'' in the presence of the potential $V_{\tt sr}$ given in Eq.~(\ref{eq:V-sr-m2}), by $\phi$, a slowly varying function of $r$, such that
\begin{equation}
\left| \nabla^2 \phi \right|\ll k\left|\nabla \phi\right|.
\label{eq:eik2h+}
\end{equation}

When substituted into (\ref{eq:Pi-eq*0+*+1*}), the trial solution (\ref{eq:Pi-eq*0+*+1}) yields
{}
\begin{eqnarray}
\Big\{\Delta \Pi_0({\vec r})+k^2\big(1+\frac{2r_g}{r}\big)\Pi_0({\vec r})\Big\}\phi({\vec r})+\Pi_0({\vec r})\Delta \phi({\vec r})&+&\nonumber\\
+\,2\big({\vec \nabla}\Pi_0({\vec r})\cdot{\vec \nabla} \phi({\vec r})\big)
+4k^2V_{\tt sr}({\vec r})\Pi_0({\vec r})\phi(\vec r)&=&
{\cal O}\Big(r_g^2,\frac{J_2}{r^3}\Pi\Big).
\label{eq:eik4h0-nab}
\end{eqnarray}

As $\Pi_0({\vec r})$ is the solution of the homogeneous equation for the monopole gravitational field (\ref{eq:Pi_ie*+*=}) \cite{Turyshev-Toth:2017,Turyshev-Toth:2019}, the first term in (\ref{eq:eik4h0-nab}) is zero. Then, we neglect the second term, $\Pi_0({\vec r})\Delta \phi({\vec r})$, because of (\ref{eq:eik2h+}). As a result, from the last two terms we have
{}
\begin{eqnarray}
\big({\vec \nabla}\ln \Pi_0({\vec r})\cdot {\vec \nabla} \ln \phi(\vec r)\big)=-2k^2V_{\tt sr}({\vec r})+{\cal O}(r_g^2).
\label{eq:eik4h}
\end{eqnarray}

As we discussed above, we assume that contributions from deviations from the monopole are small and it is sufficient  to keep only terms to ${\cal O}\big(r_g^2,({J_2}/{r^3})\Pi\big)$. Thus, to formally solve (\ref{eq:eik4h}) we may present the solution for $\Pi_0({\vec r})$ at a large distance from the monopole, which yields the well-known solution for the incident wave in the presence of a gravitational monopole (see Eq.~(23) in \cite{Turyshev-Toth:2017}):
{}
\begin{eqnarray}
\Pi_0({\vec r})=e^{\pm ik\big(z-r_g\ln k(r-z)\big)} +{\cal O}(r_g^2 ).
\label{eq:eik4mu*}
\end{eqnarray}

To compute the gradient of $\Pi_0({\vec r})$, following \cite{Turyshev-Toth:2017}, we represent the unperturbed trajectory of a ray of light as
{}
\begin{eqnarray}
\vec{r}(t)&=&\vec{r}_{0}+\vec{k}c(t-t_0)+{\cal O}(r_g),
\label{eq:x-Newt0}
\end{eqnarray}
where $\vec k$ is the unit vector in the incident direction of the light ray's propagation path and $\vec r_0$ represents the starting point. Following \cite{Kopeikin:1997,Kopeikin-book-2011,Turyshev-Toth:2017}, we define ${\vec b}=[[{\vec k}\times{\vec r}_0]\times{\vec k}]$ to be the impact parameter of the unperturbed trajectory of the light ray. The vector ${\vec b}$ is directed from the origin of the coordinate system toward the point of the closest approach of the unperturbed path of light ray to that origin.

With (\ref{eq:x-Newt0}), we introduce the parameter $\tau=\tau(t)$ along the path of the light ray (see details in Appendix~B in \cite{Turyshev-Toth:2017}):
{}
\begin{eqnarray}
\tau &=&({\vec k}\cdot {\vec r})=({\vec k}\cdot {\vec r}_{0})+c(t-t_0),
\label{eq:x-Newt*=0}
\end{eqnarray}
which may be positive or negative. Note that $\tau=z\cos\alpha$ where $\alpha$ is the angle between ${\vec e}_z$ and ${\vec k}$. Furthermore, $\tau=z$ when the $z$-axis of the chosen Cartesian coordinate system is oriented along the incident direction of the light ray. We can see that the quantity $\tau$ evolves from a negative value (representing a source at a large distance from the lens, $\alpha \simeq\pi$), through $\tau=0$ (the shortest distance from the lens where $\alpha =\pi/2$), to positive values (with $\alpha \simeq 0$ at the image plane.)
The parameter $\tau$ allows us to rewrite (\ref{eq:x-Newt0}) as
{}
\begin{eqnarray}
{\vec r}(\tau)&=&{\vec b}+{\vec k} \tau+{\cal O}(r_g),
\qquad {\rm with} \qquad ||{\vec r}(\tau)|| \equiv r(\tau) =\sqrt{b^2+\tau^2}+{\cal O}(r_g).
\label{eq:b0}
\end{eqnarray}

Using  (\ref{eq:b0}), the gradient of $\Pi^{(1)}({\vec r})$ from (\ref{eq:eik4mu*}) may be computed as
{}
\begin{eqnarray}
{\vec \nabla}\ln \Pi_0({\vec r})=\pm ik\Big({\vec k}(1+\frac{r_g}{r})-\frac{r_g}{b^2}{\vec b}\big(1+\frac{\tau}{r}\big)\Big)+{\cal O}(r_g^2).
\label{eq:eik4h+}
\end{eqnarray}

As a result, (\ref{eq:eik4h}) takes the form
{}
\begin{eqnarray}
\pm ik\Big(\big({\vec k}(1+\frac{r_g}{r})-\frac{r_g}{b^2}{\vec b}\big(1+\frac{\tau}{r}\big)\big)\cdot {\vec \nabla} \ln \phi(\vec r)\Big)=-2V_{\tt sr}({\vec r})+{\cal O}(r_g^2).
\label{eq:eik4h**0}
\end{eqnarray}

As we want to identify the largest contribution from corrections to the monopole to light propagation, we keep only linear terms with respect to gravity. As a result, neglecting the $r_g$-dependent terms in (\ref{eq:eik4h**0}), we may present (\ref{eq:eik4h}) as
{}
\begin{eqnarray}
\pm ik({\vec k} \cdot {\vec \nabla}) \ln \phi=-2 k^2V_{\tt sr}+{\cal O}(r_g^2).
\label{eq:eik4h**}
\end{eqnarray}

We may now compute the eikonal phase due to the short-range potential $V_{\tt sr}$. Using the representation of the light ray's path as ${\vec r}=({\vec b},\tau)$ given by   (\ref{eq:b0}), we observe that (as was also shown in \cite{Turyshev-Toth:2017}) the gradient ${\vec \nabla}$ may be expressed in terms of the variables along the path as ${\vec \nabla}={\nabla}_b+{\vec k}\,d/d\tau +{\cal O}(r_g)$, where ${\nabla}_b$ is the gradient along the direction of the impact parameter ${\vec b}$ and $\tau$ being the parameter taken along the path. Thus, the differential operator on the left side of (\ref{eq:eik4h**})  is the derivative along the light ray's path, namely  $({\vec k} \cdot {\vec \nabla})=d/d\tau$.

As a result, for (\ref{eq:eik4h**}) we have
{}
\begin{equation}
\frac{d\ln\phi^\pm}{d\tau} =\pm ik \,2V_{\tt sr}+{\cal O}(r_g^2),
\label{eq:eik4h*}
\end{equation}
the solutions of which are
 \begin{eqnarray}
\phi^\pm({\vec b}, \tau)&=&\exp\Big(\pm ik \int^{\tau}_{\tau_0}  2V_{\tt sr}({\vec b},\tau') d\tau' \Big).
\label{eq:eik5h}
\end{eqnarray}
We therefore have the following two particular eikonal solutions of (\ref{eq:Pi-eq*0+*+1*}) for $\Pi(\vec r)$:
\begin{equation}
\Pi(\vec r)=\Pi_0(\vec r)\exp\Big(\pm i \xi_b({\vec b}, \tau) \Big)+{\cal O}\Big(r_g^2,\frac{J_2}{r^3}\Pi\Big),
\label{eq:eik6h}
\end{equation}
where we introduced the eikonal phase
\begin{equation}
\xi_b(\vec b,\tau) =\frac{k}{2}\int^{\tau}_{\tau_0}  2V_{\tt sr}({\vec b},\tau') d\tau' +{\cal O}(r_g^2).
\label{eq:eik7h}
\end{equation}

The solution given by (\ref{eq:eik6h})--(\ref{eq:eik7h}) was obtained under the eikonal condition (\ref{eq:eik2h+}) that allows us to consider the effect of the short-range potential due to gravitational multipoles (as shown in (\ref{eq:shart-pot})--(\ref{eq:Pi-eq*0+*+1*})) on the phase of the EM wave only, and not on its amplitude. This is similar to the thin lens approximation that is extensively used in the description of many problems on modern optics \cite{Born-Wolf:1999} and gravitational lensing \cite{Schneider-Ehlers-Falco:1992}. That fact is captured by (\ref{eq:eik7h}) where we assume that light moves in a straight line before it reaches the lens and then it changes direction at $\tau =({\vec k}\cdot {\vec r})=0$ and moves again on a straight line towards the observer. Thus, the phase shift   (\ref{eq:eik7h})  occurs only on the second part of the path.

Considering the structure of solutions (\ref{eq:Pi_ie*+*=}) and (\ref{eq:eik6h}), we note that the eikonal phase, $\xi_b(\vec b, \vec s)$ from (\ref{eq:eik7h}), depends on the vector of the impact parameter $\vec b$ and its orientation with respect to the solar rotational axis.
Thus, the presence of $\xi_b(\vec b, \vec s)$ in (\ref{eq:eik6h}) is understood in the context of solution (\ref{eq:Pi_ie*+*=}), where the sum over $\ell=kb$ also acts on the $b$-dependent eikonal phase. In general, this approach is similar to that of the Born approximation \cite{Sharma-Sommerford-book:2006,Friedrich-book-2006,Friedrich-book-2013,Messiah:1968} or path integrals in quantum mechanics \cite{Feynman:1948,Feynman-Hibbs:1991,Nakamura-Deguchi:1999,Yamamoto:2017}.  This point will become more evident in Section~\ref{sec:IF-region}.

In Appendix \ref{sec:potU}, we compute the eikonal phases for two possible forms of the gravitational potential, valid in the generic case. In the case of the spatial trace-free (STF) multipole moments from (\ref{eq:pot_stf}), the eikonal phase is given by (\ref{eq:eik-ph}).  However, in the case of the SGL, the gravitational potential of the Sun is that of an axisymmetric body best characterized using zonal harmonics (\ref{eq:pot_stis}) and may be expressed \cite{Roxburgh:2001,LePoncinLafitte:2007tx} in terms of the usual dimensionless multipole moments $J_\ell$:
{}
\begin{eqnarray}
U&=&\frac{GM}{r}\Big\{1-\sum_{\ell=2}^\infty J_\ell \Big(\frac{R}{r}\Big)^\ell P_\ell\Big(\frac{{\vec s}\cdot{\vec x}}{r}\Big)\Big\}+
{\cal O}(c^{-4}),
\label{eq:pot_stis2}
\end{eqnarray}
where ${\vec s}$ denotes the unit vector along the $x^3$-axis, $P_\ell$ are the Legendre polynomials and the quantities $M, J_2..., J_\ell$ correspond to the multipole moments. Note that, in the case of an axisymmetric and rotating body with ``north-south symmetry'' (i.e., a body that is symmetric under a reflection with respect to the plane of rotation), the expression (\ref{eq:pot_stis}) contains only the $\ell=2,4,6,8...$ even moments.

To determine the eikonal phase (\ref{eq:eik7h}), we use a heliocentric coordinate system with its $z$-axis aligned with the wavevector $\vec k$, so that $\vec k=(0,0,1)$. We introduce a unit vector in the direction of the impact parameter, $\vec b= b\vec n_\xi$, coordinates on the image plane, $\vec x$ that is located at the distance $z$ from the Sun, and the unit vector in the direction of the solar rotation axis, $\vec s$:
\begin{eqnarray}
{\vec b}&=&b(\cos\phi_\xi,\sin \phi_\xi,0),
\label{eq:notes-b}\\
{\vec x}&=&\rho(\cos\phi,\sin \phi,0),\\
{\vec s}&=&(\sin\beta_s\cos\phi_s,\sin\beta_s\sin\phi_s,\cos\beta_s).
\label{eq:notes}
\end{eqnarray}
In this coordinate system, the eikonal phase shift (\ref{eq:eik7h}) accumulated by an EM wave propagating in the gravitational field of an axisymmetric body takes an elegant form (see discussion in Appendix~\ref{sec:eik-phase} and the result (\ref{eq:eik-ph-axi*})):
 {}
\begin{eqnarray}
\xi_b(\vec b, \vec s)
&=& -kr_g\sum_{n=2}^\infty\frac{J_n}{n}\Big(\frac{R_\odot}{b}\Big)^n \sin^n\beta_s\cos[n(\phi_\xi-\phi_s)]
+{\cal O}(r_g^2).
\label{eq:eik-phase-axi}
\end{eqnarray}
This result provides the context for our investigation below as it shows the explicit dependencies of the eikonal phase on the orientation of the vector of the impact parameter with respect to the lens' rotational axis.

\section{Electromagnetic wave in the field of a static extended gravity lens}
\label{sec:Debye-sol}

Our next goal is to find a solution to the EM field in that region. We accomplish this objective using the approach developed for classical diffraction theory, by finding the set of equations that determine the EM field via Debye potentials and then matching these equations with the incident wave.


At this point, we already have all the key components  needed to develop the solution for the Debye potentials in the case of the long-range, spherically symmetric gravitational field produced by the solar monopole, and the static long-range gravitational field produced by deviations from the monopole, characterized using zonal harmonics. Following \cite{Turyshev-Toth:2017,Turyshev-Toth:2019},  a particular solution for the Debye potential, ${\Pi}$, is obtained by combining results for $\Phi(\phi)$ from (\ref{eq:Ph_m})  and $\Theta(\theta)$ from (\ref{eq:Th_lm}). The solution for the Debye potential takes the form
{}
\begin{eqnarray}
\frac{\Pi}{u}&=&\frac{1}{r} \sum_{\ell=0}^\infty\sum_{m=-\ell}^\ell
\mu_\ell R_\ell(r,\vec b)\big[ P^{(m)}_l(\cos\theta)\big]\big[a_m\cos (m\phi) +b_m\sin (m\phi)\big]+{\cal O}\Big(r_g^2,\frac{J_2}{r^3}\Pi\Big),
\label{eq:Pi-degn-sol-0}
\end{eqnarray}
where the yet to be constructed $R_\ell(r,\vec b)$ is the radial function and $\mu_\ell, a_m, b_m$ are arbitrary and as yet unknown constants.
Note that the structure of the solution (\ref{eq:Pi-degn-sol-0}) preserves the angular symmetries of the monopole case given by $\Phi(\phi)$ and $\Theta(\theta)$. The presence of the gravitational zonal harmonics is accounted for by the generalized radial function  $R_\ell(r,\vec b)$ that now depends on $\vec b$ via the eikonal phase, as shown in (\ref{eq:eik6h}).

In the vacuum, the solutions for the electric and magnetic potentials of the incident wave, ${}^e{\hskip -1pt}\Pi_0$ and ${}^m{\hskip -1pt}\Pi_0$, were found to be given in terms of a single potential $\Pi_0(r, \theta)$ that is given by (\ref{eq:Pi_ie*+*=}). In other words, the incident EM wave is not affected by the gravitational field from the zonal harmonics of the extended Sun. Its form is identical to that of the free EM wave propagating in monopole gravity, discussed in \cite{Turyshev-Toth:2017}.

Considering deviations from spherical symmery, we notice that,  for large $r$, the potential $ V_{\tt sr}(r)$ in (\ref{eq:Pi-eq*0+*+1*}) can be neglected in comparison to the Coulomb potential $U_{\tt c}(r )=2k^2r_g/r$ and this equation reduces to the Coulomb equation discussed in \cite{Turyshev-Toth:2017} with the solution given by  (\ref{eq:Pi_ie*+*=}). The solution of (\ref{eq:Pi-eq*0+*+1*}) that is regular at the origin can thus be written asymptotically as a linear combination of the regular and irregular Coulomb wave functions $F_e(kr_g, kr)$ and $G_\ell(kr_g, kr)$, respectively  \cite{Hull-Breit:1959,Friedrich-book-2006,Friedrich-book-2013,Burke-book-2011,Turyshev-Toth:2019}, which are solutions of (\ref{eq:Pi-eq*0+*+1*})
in the absence of the potential $V_{\tt sr}(r )$. Asymptotically, at large values of the argument $(kr)$, these functions behave as \cite{Turyshev-Toth:2017,Turyshev-Toth:2019}:
{}
\begin{eqnarray}
 F_\ell(kr_g, kr) &\sim& \sin\Big(k(r+r_g\ln2kr) +\frac{\ell(\ell+1)}{2kr}-\frac{\pi \ell}{2}+\sigma_\ell\Big),
 \label{eq:F-beh}\\
 G_\ell(kr_g, kr) &\sim& \cos\Big(k(r+r_g\ln2kr)+\frac{\ell(\ell+1)}{2kr}-\frac{\pi \ell}{2}+\sigma_\ell \Big).
\label{eq:G-beh}
\end{eqnarray}

In the case of centrally symmetric potentials, since the Coulomb potential falls off slower than the centrifugal potential (i.e., the $\ell(\ell+1)/r^2$ term in (\ref{eq:F-beh}) and (\ref{eq:G-beh})) at large distances, it dominates the asymptotic behavior of the effective potential in every partial wave. Hence, we can generally look for a solution satisfying the following boundary conditions \cite{Burke-book-2011}:
{}
\begin{eqnarray}
R_\ell (r) &\underset{r\rightarrow 0}{\sim}& nr^{\ell+1},
\label{eq:R-anz0}\\
R_\ell (r) &\underset{r\rightarrow \infty}{\sim}&  F_\ell(kr_g, kr) + \tan \delta_\ell \,G_\ell(kr_g, kr)
\underset{kr\rightarrow \infty}{\propto}
\sin\Big(k(r+r_g\ln2kr)+\frac{\ell(\ell+1)}{2kr}-\frac{\pi \ell}{2}+\sigma_\ell +\delta_\ell\Big),
\label{eq:R-anz}
\end{eqnarray}
where $n$ is a normalization factor and $F_\ell(kr_g, kr)$ and $G_\ell(kr_g, kr)$ are solutions of (\ref{eq:Pi-eq*0+*+1*}) in the absence of the potential $V_{\tt sr}(r)$, which, as we discussed above, are respectively regular and irregular at the origin. The real quantities $\delta_\ell(k)$ introduced by these equations are the phase shifts due to the short-range potential $V_{\tt sr}(r)$ (\ref{eq:V-sr-m2}) in the presence of the Coulomb potential $U_{\tt c}(r )=2k^2r_g/r$ in  (\ref{eq:Pi-eq*0+*+1*}). We note that $\delta_\ell(k)$ fully describes the non-Coulombic part of the scattering and vanishes when this short-range potential is absent.

In the case of generic gravitational fields, we can satisfy the conditions (\ref{eq:R-anz0})--(\ref{eq:R-anz}) by choosing the  function $R_\ell(r)$ as a linear combination of the two solutions (\ref{eq:eik6h}), where $\delta_\ell(k)$ is replaced by the eikonal phase, $\xi_b (\vec b)$. One way to do that is by relying on the two solutions to (\ref{eq:eik6h}), taken in the form of the incident and scattered waves \cite{Thomson-Nunes-book:2009}, which are correspondingly given by the functions $H^{-}_\ell(kr_g,kr)$ and  $H^{+}_\ell(kr_g,kr)$, and to show explicit dependence on the eikonal phase shift, $\xi_b(\vec b)$, which can be captured in the following form:
{}
\begin{eqnarray}
R_\ell(r,\vec b) = \frac{1}{2i}\Big(H^+_\ell(kr_g,kr)e^{i \xi_b (\vec b)} -H^-_\ell(kr_g,kr)e^{-i \xi_b(\vec b)}\Big),
\label{eq:R-L}
\end{eqnarray}
where the Coulomb--Hankel functions $H_\ell^{(\pm)}$ are related to the Coulomb functions by $H_\ell^{\pm}(kr_g,kr)=G_\ell(kr_g,kr)\pm iF_\ell(kr_g,kr)$ (for discussion, see Appendix~A of \cite{Turyshev-Toth:2017}) and their asymptotic behavior is given by (see Appendix F of \cite{Turyshev-Toth:2017}):
{}
\begin{equation}
H^\pm_\ell(kr_g,kr)\underset{kr\to\infty}{\sim} \exp\Big\{\pm i\Big(k(r+r_g\ln 2kr)+\frac{\ell(\ell+1)}{2kr}-\frac{\pi \ell}{2}+\sigma_\ell\Big)\Big\},
\label{eq:free_u+}
\end{equation}
where $\xi_b(\vec b)$ in (\ref{eq:R-L}) is the eikonal phase shift that is accumulated by the EM wave along its entire path. The expression for this quantity is given by (\ref{eq:eik7h}), which, for axisymmetric body, is computed by (\ref{eq:eik-phase-axi}).

The form of the radial function $R_\ell$ from (\ref{eq:R-L}) captures our expectation that, in the presence of a potential $V_{\tt sr}$ from (\ref{eq:V-sr-m2}),  the Coulomb--Hankel functions (which represent the radial free-particle wavefunction solutions of the homogeneous equation (\ref{eq:Pi-eq*0+*+1*})), become ``distorted'' by this short-range potential due to the gravitational mass multipoles. We can verify that $R_\ell$ in the form of (\ref{eq:R-L}) also satisfies the asymptotic boundary conditions (\ref{eq:R-anz0})--(\ref{eq:R-anz}). Indeed, as the gravitational potential for the inner region of the Sun vanishes, the eikonal phase $\xi_b$ is zero for $r<R_\odot$. Therefore, as $r\rightarrow 0$,  the radial function  (\ref{eq:R-L}) becomes $R_\ell(r,\vec b) \rightarrow F_\ell(kr_g,kr)$, where the function  $F_\ell(kr_g,kr)$ obeys the condition (\ref{eq:R-anz0}). Next, we consider another limit, when $r\rightarrow \infty$. Using the asymptotic behavior of $H^\pm_\ell$ from (\ref{eq:free_u+}), we see that, as $r\rightarrow \infty$, the radial function obeys the asymptotic condition (\ref{eq:R-anz}) taking the form where the phase shift $\delta_\ell$ is given by the eikonal phase $\xi_{b}$ introduced by (\ref{eq:eik7h}).

We may put the result (\ref{eq:R-L}) in the following equivalent form:
{}
\begin{eqnarray}
R_\ell (r,\vec b) &=&  \cos\xi_b(\vec b)F_\ell(kr_g, kr) + \sin\xi_b(\vec b)\,G_\ell(kr_g, kr),
\label{eq:R-L3}
\end{eqnarray}
which explicitly shows the phase shift, $\xi_b(\vec b)$, induced by the short-range extended gravity potential, clearly satisfying the boundary condition (\ref{eq:R-anz})  with the quantity $\xi(\vec b)$ from (\ref{eq:eik7h}) being the anticipated phase shift.

To match the potentials (\ref{eq:Pi-degn-sol-0}) of the incident and scattered waves outside, the latter must be expressed in a similar form but with arbitrary coefficients. Only the function $F_\ell(kr_g,kr)$ may be used in the expression for the potential inside the sphere  since $G_\ell(kr_g,kr)$ becomes infinite at the origin. On the other hand, the scattered wave must vanish at infinity. The Coulomb--Hankel functions $H^+_\ell(kr_g,kr)$ are characterized by precisely this property, which makes them suitable as representations of scattered waves. For large values of the argument $kr$, the result behaves as  $e^{ik(r+r_g\ln2kr)}$ and the Debye potential $\Pi\propto e^{ik(r+r_g\ln2kr)}/r$ for large $r$. Thus, at large distances from the sphere the scattered wave is spherical (with the $\ln$ term in the phase due to the modification by the Coulomb potential), with its center at the origin  $r=0$. Accordingly, we use it in the expression for the scattered wave.

Collecting results for the functions $\Phi(\phi)$ and $\Theta(\theta)$, respectively given by (\ref{eq:Ph_m}) and (\ref{eq:Th_lm}), and $R_\ell(r,\vec b)=H^+_\ell(kr_g, kr)e^{i\xi_{b}(\vec b)}$ from (\ref{eq:eik6h}), to ${\cal O}(r_g^2,({J_2}/{r^3})\Pi)$, we obtain the Debye potential for the scattered wave:
{}
\begin{eqnarray}
\Pi_{\tt s}&=&\frac{u}{r} \sum_{\ell=0}^\infty\sum_{m=-\ell}^\ell
a_\ell H^+_\ell(kr_g, kr)e^{i\xi_b(\vec b)}\big[ P^{(m)}_\ell(\cos\theta)\big]\big[a'_m\cos (m\phi) +b'_m\sin (m\phi)\big],
\label{eq:Pi-degn-sol-s}
\end{eqnarray}
where $a_\ell, a'_m, b'_m$ are arbitrary and as yet unknown constants.

Representing the potential via $F_\ell(kr_g,kr)$ is appropriate. The trial solution to (\ref{eq:Pi-eq*0+*+1*}) for the electric and magnetic Debye potentials relies on the radial function $R_\ell(r,\vec b)$ given by (\ref{eq:R-L3}) and has the form
{}
\begin{eqnarray}
\Pi_{\tt in}&=&\frac{u}{r} \sum_{\ell=0}^\infty\sum_{m=-\ell}^\ell
b_\ell \Big\{ \cos\xi_{ b}(\vec b)F_\ell(kr_g, kr) + \sin\xi_{b}(\vec b)\,G_\ell(kr_g, kr)\Big\}\big[ P^{(m)}_\ell(\cos\theta)\big]\big[a_m\cos (m\phi) +b_m\sin (m\phi)\big],
~~~~
\label{eq:Pi-degn-sol-in}
\end{eqnarray}
where $b_\ell, a_m, b_m$ are arbitrary and yet unknown constants.

The boundary (continuity) conditions (see discussion in \cite{Born-Wolf:1999,Turyshev-Toth:2017}), imposed on the quantities (\ref{eq:bound_cond}) at some radius $r=R_\odot^\star=R_\odot +r_g$, are written in full as
{}
\begin{eqnarray}
\frac{\partial }{\partial r}\Big[\frac{r\,{}^e{\hskip -1pt}\Pi_0}{u}+\frac{r{}^e{\hskip -1pt}\Pi_{\tt s}}{\sqrt{\epsilon}u}\Big]\Big|^{}_{r=R^\star_\odot}&=&\frac{\partial }{\partial r}\Big[\frac{r\,{}^e{\hskip -1pt}\Pi_{\tt in}}{\sqrt{\epsilon}u}\Big]\Big|^{}_{r=R^\star_\odot},
\label{eq:bound_cond-expand1+}\\[3pt]
\frac{\partial }{\partial r}\Big[\frac{r\,{}^m{\hskip -1pt}\Pi_0}{u}+\frac{r{}^m{\hskip -1pt}\Pi_{\tt s}}{\sqrt{\mu}u}\Big]\Big|^{}_{r=R^\star_\odot}&=&\frac{\partial }{\partial r}\Big[\frac{r\,{}^m{\hskip -1pt}\Pi_{\tt in}}{\sqrt{\mu}u}\Big]\Big|^{}_{r=R^\star_\odot},
\label{eq:bound_cond-expand2+}\\[3pt]
\Big[ \frac{r\,{}^e{\hskip -1pt}\Pi_0}{u}+\frac{r{}^e{\hskip -1pt}\Pi_{\tt s}}{\sqrt{\epsilon}u}\Big]\Big|^{}_{r=R^\star_\odot}&=&\Big[\frac{r\,{}^e{\hskip -1pt}\Pi_{\tt in}}{\sqrt{\epsilon}u}\Big]\Big|^{}_{r=R^\star_\odot},
\label{eq:bound_cond-expand3+}\\[3pt]
\Big[\frac{r\,{}^m{\hskip -1pt}\Pi_0}{u}+\frac{{}^m{\hskip -1pt}\Pi_{\tt s}}{\sqrt{\mu}u}\Big]\Big|^{}_{r=R^\star_\odot}&=&\Big[\frac{r\,{}^m{\hskip -1pt}\Pi_{\tt in}}{\sqrt{\mu}u}\Big]\Big|^{}_{r=R^\star_\odot}.
\label{eq:bound_cond-expand4+}
\end{eqnarray}

We now make use of the symmetry of the geometry of the problem \cite{Born-Wolf:1999} and by applying the boundary conditions (\ref{eq:bound_cond-expand1+})--(\ref{eq:bound_cond-expand4+}). We recall that we can use a single Debye potential $\Pi$ in (\ref{eq:Pi-degn-sol-s}) and (\ref{eq:Pi-degn-sol-in}) to represent electric and magnetic fields. We find that the constants $a_m$ and $b_m$ for the electric Debye potentials are $a_1=1$, $b_1=0$ and $a_m=b_m=0$ for $m\ge 2$. For the magnetic Debye potentials, we obtain $a_1=0$, $b_1=1$ and $a_m=b_m=0$ for $m\ge 2$. The values are identical for $a'_m$ and $b'_m$.

As a result, the solutions for the electric and magnetic potentials of the scattered wave, ${}^e{\hskip -1pt}\Pi_{\tt s}$ and ${}^m{\hskip -1pt}\Pi_{\tt s}$, may be given in terms of a single potential $\Pi_{\tt s}(r, \theta)$ (see  \cite{Turyshev-Toth:2017} for details), which, to ${\cal O}(r_g^2,({J_2}/{r^3})\Pi)$, is given by
\begin{align}
  \left( \begin{aligned}
{}^e{\hskip -1pt}\Pi_{\tt s}& \\
{}^m{\hskip -1pt}\Pi_{\tt s}& \\
  \end{aligned} \right) =&  \left( \begin{aligned}
\cos\phi\\
\sin\phi  \\
  \end{aligned} \right) \,\Pi_{\tt s}(r, \theta),& {\rm where} \hskip 30pt
 \Pi_{\tt s}(r, \theta)=\frac{u}{r} \sum_{\ell=1}^\infty
a_\ell H^+_\ell(kr_g, kr)e^{i\xi_{b}(\vec b)} P^{(1)}_\ell(\cos\theta). ~~~~~
  \label{eq:Pi_s+}
\end{align}

In a relevant scattering scenario, the EM wave and the Sun are well separated initially, so the Debye potential for the incident wave can be expected to have the same form as for the pure monopole case that includes only the Coulomb potential that is given by (\ref{eq:Pi_ie*+*=}). Therefore, the Debye potential for the inner region has the form:
\begin{align}
  \left( \begin{aligned}
{}^e{\hskip -1pt}\Pi_{\tt in}& \\
{}^m{\hskip -1pt}\Pi_{\tt in}& \\
  \end{aligned} \right) =&  \left( \begin{aligned}
\cos\phi\\
\sin\phi  \\
  \end{aligned} \right) \,\Pi_{\tt in}(r, \theta),
  \label{eq:Pi_in+}
\end{align}
with the potential $\Pi_{\tt in}$ given, to ${\cal O}(r_g^2,({J_2}/{r^3})\Pi)$, as
{}
\begin{eqnarray}
\Pi_{\tt in}(r, \theta)&=&\frac{u}{r} \sum_{\ell=1}^\infty
b_\ell \Big\{ \cos\xi_{b}(\vec b)F_\ell(kr_g, kr) + \sin\xi_{b}(\vec b)\,G_\ell(kr_g, kr)\Big\}P^{(1)}_\ell(\cos\theta).
\label{eq:Pi-in}
\end{eqnarray}

We thus expressed all the potentials in the series (\ref{eq:Pi-degn-sol-0}) and any unknown constants can now be determined easily. If we now substitute the expressions (\ref{eq:Pi_ie*+*=}), (\ref{eq:Pi_s+}) and (\ref{eq:Pi_in+})--(\ref{eq:Pi-in}) into the boundary conditions (\ref{eq:bound_cond-expand1+})--(\ref{eq:bound_cond-expand4+}), we obtain the following linear relationships between the coefficients $a_\ell$ and $b_\ell$:
{}
\begin{eqnarray}
\Big[\frac{E_0}{k^2}i^{\ell-1}\frac{2\ell+1}{\ell(\ell+1)}e^{i\sigma_\ell}
F'_\ell(kr_g,kr)+a_\ell \Big(H^+_\ell(kr_g,kr)e^{i\xi_{b}(\vec b)}\Big)'\Big]\Big|_{r=R^\star_\odot}&=&b_\ell R'_\ell(r,\vec b)\Big|_{r=R^\star_\odot},
\label{eq:bound-cond*1}\\
\Big[\frac{E_0}{k^2}i^{\ell-1}\frac{2\ell+1}{\ell(\ell+1)}e^{i\sigma_\ell}
F_\ell(kr_g,kr)+a_\ell H^+_\ell(kr_g,kr)e^{i\xi_{b}(\vec b)}\Big]\Big|_{r=R^\star_\odot}&=&b_\ell R_\ell(r,\vec b)\Big|_{r=R^\star_\odot},
\label{eq:bound-cond*2}
\end{eqnarray}
where $R_\ell(r)$ is from (\ref{eq:R-L3}) and $'=d/dr$. We now define, for convenience, $\alpha_\ell$ and $\beta_\ell$ as
{}
\begin{eqnarray}
a_\ell=\frac{E_0}{k^2}i^{\ell-1}\frac{2\ell+1}{\ell(\ell+1)}e^{i\sigma_\ell} \alpha_\ell \qquad {\rm and}\qquad b_\ell=\frac{E_0}{k^2}i^{\ell-1}\frac{2\ell+1}{\ell(\ell+1)}e^{i\sigma_\ell}\beta_\ell.
\label{eq:a-b}
\end{eqnarray}
From (\ref{eq:bound-cond*1})--(\ref{eq:bound-cond*2}), we have:
{}
\begin{eqnarray}
F'_\ell(R^\star_\odot)+\alpha_\ell {H^+_\ell}'(R^\star_\odot)e^{i\xi_{b}(\vec b)}&=&\beta_\ell R'_\ell(R^\star_\odot,\vec b),
\label{eq:bound-cond*1+}\\[3pt]
F_\ell(R^\star_\odot)+\alpha_\ell H^+_\ell(R^\star_\odot)e^{i\xi_{b}(\vec b)}&=&\beta_\ell R_\ell(R^\star_\odot,\vec b),
\label{eq:bound-cond*2+}
\end{eqnarray}
where $F_\ell(R^\star_\odot)=F_\ell(kr_g,kR^\star_\odot)$ and $H^+_\ell(R^\star_\odot)=H^+_\ell(kr_g,kR^\star_\odot)$ with similar definitions for the derivatives of these functions. Equations (\ref{eq:bound-cond*1+})--(\ref{eq:bound-cond*2+}) may now be solved to determine the two sets of coefficients $\alpha_\ell$ and $\beta_\ell$:
{}
\begin{eqnarray}
\alpha_\ell&=&e^{-i\xi_{b}(\vec b)}\frac{F_\ell(R^\star_\odot)R'_\ell(R^\star_\odot,\vec b)-F'_\ell(R^\star_\odot)R_\ell(R^\star_\odot,\vec b)}{R_\ell(R^\star_\odot,\vec b){H^+_\ell}'(R^\star_\odot)-R'_\ell(R^\star_\odot,\vec b)H^+_\ell(R^\star_\odot)},
\label{eq:a_l*}\\
\beta_\ell&=&\frac{F_\ell(R^\star_\odot){H^+}'_\ell(R^\star_\odot)-F'_\ell(R^\star_\odot)H^+_\ell(R^\star_\odot)}{R_\ell(R^\star_\odot,\vec b){H^+_\ell}'(R^\star_\odot)-R'_\ell(R^\star_\odot)H^+_\ell(R^\star_\odot,\vec b)}.
\label{eq:b_l*}
\end{eqnarray}

Taking into account the asymptotic behavior of all the functions involved: namely (\ref{eq:free_u+}) for $H^+_\ell$ and (\ref{eq:F-beh})--(\ref{eq:G-beh}) for $F_\ell$ and $G_\ell$, we have the following solution for the coefficients $\alpha_\ell$ and $\beta_\ell$:
 {}
\begin{eqnarray}
\alpha_\ell&=&\sin\xi_{b}(\vec b), \qquad \beta_\ell=e^{i\xi_{b}(\vec b)},
\label{eq:a_b_del}
\end{eqnarray}
with $\xi_b(\vec b)$ is the phase shift induced by the gravitational multipoles to the phase of the EM wave propagating through the solar system.

Therefore, using the value for $a_\ell$ from (\ref{eq:a-b}), together with $\alpha_\ell$ from (\ref{eq:a_b_del}), we determine that the solution for the scattered potential (\ref{eq:Pi_s+}) takes the form
{}
\begin{eqnarray}
\Pi_{\tt s}(r, \theta)&=& \frac{E_0}{k^2} \frac{u}{r}\sum_{\ell=1}^\infty
i^{\ell-1}\frac{2\ell+1}{\ell(\ell+1)}e^{i\sigma_\ell} \sin\xi_b(\vec b) H^+_\ell(kr_g, kr)e^{i\xi_{b}(\vec b)} P^{(1)}_\ell(\cos\theta),
\label{eq:Pi-s_a*=}
\end{eqnarray}
which we can present  as
{}
\begin{eqnarray}
\Pi_{\tt s}(r, \theta)&=&\frac{E_0}{2ik^2} \frac{u}{r}
\sum_{\ell=1}^\infty
i^{\ell-1}\frac{2\ell+1}{\ell(\ell+1)}e^{i\sigma_\ell} H^+_\ell(kr_g, kr) \Big(e^{2i\xi_b(\vec b)} -1\Big)P^{(1)}_\ell(\cos\theta).
\label{eq:Pi-s_a*}
\end{eqnarray}

In the region outside the Sun, $r>R^\star_\odot$, we may take the asymptotic form for the Coulomb--Hankel function and present (\ref{eq:Pi-s_a*}) as
{}
\begin{eqnarray}
\Pi_{\tt s}(r, \theta)&=&- \frac{E_0}{2k^2} \frac{u}{r}e^{ik(r+r_g\ln 2kr)}
\sum_{\ell=1}^\infty
\frac{2\ell+1}{\ell(\ell+1)}e^{i(2\sigma_\ell+\frac{\ell(\ell+1)}{2kr})} \Big(e^{2i\xi_b(\vec b)}-1\Big)P^{(1)}_\ell(\cos\theta).
\label{eq:Pi-s_ass}
\end{eqnarray}

As a result, using (\ref{eq:Pi_ie*+*=}) and (\ref{eq:Pi-s_a*}), we present the Debye potential in the region outside the Sun, $r>R_\odot$,  in the following form:
{}
\begin{eqnarray}
\Pi_{\tt out}(r, \theta)&=& \Pi_0(r,\theta)+\Pi_{\tt s}(r,\theta)=\nonumber\\
&=&
\frac{E_0}{k^2}\frac{u}{r}\sum_{\ell=1}^\infty i^{\ell-1}\frac{2\ell+1}{\ell(\ell+1)}e^{i\sigma_\ell}
\Big\{F_\ell(kr_g,kr) + \frac{1}{2i}\Big(e^{2i\xi_b(\vec b)}-1\Big) H^+_\ell(kr_g, kr)\Big\}P^{(1)}_\ell(\cos\theta).~~~~~~
\label{eq:Pi-s_a1*0}
\end{eqnarray}

Similarly,  substituting the value for $b_\ell$ from (\ref{eq:a-b}), together with $\beta_\ell$ from (\ref{eq:a_b_del}), we determine the solution for the inner Debye potential (\ref{eq:a_b_del}) in the form
{}
\begin{eqnarray}
\Pi_{\tt in}(r, \theta)&=&\frac{E_0}{k^2} \frac{u}{r} \sum_{\ell=1}^\infty
i^{\ell-1}\frac{2\ell+1}{\ell(\ell+1)}e^{i(\sigma_\ell +\xi_b(\vec b))} \Big\{ \cos\xi_b(\vec b)F_\ell(kr_g, kr) + \sin\xi_b(\vec b)\,G_\ell(kr_g, kr)\Big\}P^{(1)}_\ell(\cos\theta).
\label{eq:Pi-in+}
\end{eqnarray}

As solar gravity is rather weak, we may use the asymptotic expressions for $F_\ell, G_\ell$ and $H^\pm_\ell$ for $r\geq R_\odot$.  Therefore, the radial function  $R_\ell (r,\vec b) $ from (\ref{eq:R-L}) (or, equivalently, from (\ref{eq:R-L3})) may be given as
{}
\begin{align}
R_\ell(r,\vec b) &= \frac{1}{2i}\Big(H^+_\ell(kr_g,kr)e^{i \xi_{b} (\vec b)} -H^-_\ell(kr_g,kr)e^{-i \xi_{b}(\vec b)}\Big)= e^{-i \xi_b(\vec b)}\Big\{F_\ell(kr_g,kr) + \frac{1}{2i}\big(e^{2i \xi_b(\vec b)}-1\big)H^+_\ell(kr_g,kr)\Big\},~~~~~~~
\label{eq:R-L3+*}
\end{align}
where $\xi_{b}=\xi_{b}(\vec b)$ is the eikonal phase.

As  a result, outside the Sun, we may present (\ref{eq:Pi-in+}) in the following equivalent form:
{}
\begin{eqnarray}
\Pi_{\tt in}(r, \theta)&=&\frac{E_0}{k^2} \frac{u}{r} \sum_{\ell=1}^\infty
i^{\ell-1}\frac{2\ell+1}{\ell(\ell+1)}e^{i\sigma_\ell} \Big\{F_\ell(kr_g, kr)+ \frac{1}{2i}\Big(e^{2i\xi_b(\vec b)}-1\Big)H^+_\ell(kr_g, kr)\Big\}P^{(1)}_\ell(\cos\theta).
\label{eq:Pi-in+sl}
\end{eqnarray}

The solution for the Debye potential, $\Pi (r, \theta)$ from (\ref{eq:Pi-in+sl}), describing the propagation of the EM wave on the background of the static gravitational monopole and the short-range multipole gravitational field takes the form
{}
\begin{eqnarray}
\Pi_{\tt in}(r, \theta)&=&
\frac{E_0}{k^2}\frac{u}{r}\sum_{\ell=1}^\infty i^{\ell-1}\frac{2\ell+1}{\ell(\ell+1)}e^{i\sigma_\ell} F_\ell(kr_g,kr) P^{(1)}_\ell(\cos\theta)+\nonumber\\
&+&\frac{E_0}{2ik^2}\frac{u}{r}\sum_{\ell=1}^\infty i^{\ell-1}\frac{2\ell+1}{\ell(\ell+1)}e^{i\sigma_\ell }
\Big(e^{2i\xi_b(\vec b)}-1\Big)
H^{(+)}_\ell(kr_g,kr) P^{(1)}_\ell(\cos\theta) +{\cal O}\Big(r_g^2,\frac{J_2}{r^3}\Pi\Big).
  \label{eq:Pi_g+p}
\end{eqnarray}

The first term in (\ref{eq:Pi-in+sl}) is the Debye potential of an EM wave propagating  in a vacuum but modified by the gravity of extended Sun. The second term represents the effect of the solar gravitational multipoles on the propagation of the EM waves. Notice that, as the distance increases, this term approaches the form of the Debye potential $\Pi_{\tt s}$ for the scattered EM field given by (\ref{eq:Pi-s_ass}).

Thus, we have identified all the Debye potentials involved in the Mie problem \cite{Mie:1908}, namely the potential $\Pi_0$ given by (\ref{eq:Pi_ie*+*=}) representing the incident EM field, the potential $\Pi_{\tt s}$ from (\ref{eq:Pi-s_ass}) describing the scattered EM field, and the potential $\Pi_{\tt in}$ from (\ref{eq:Pi-in+sl}) total field.

\section{General solution for the EM field}
\label{sec:EM-field-outside}

To describe the scattering of light by the extended Sun, we use solutions for the Debye potential representing the scattered EM wave (\ref{eq:Pi-s_ass}),  and the EM wave  (\ref{eq:Pi_g+p}). The presence of the Sun itself is not yet captured. For this, we need to set additional boundary conditions that describe the interaction of the Sun with the incident radiation. Similarly to \cite{Turyshev-Toth:2017,Turyshev-Toth:2018-plasma}, we apply the fully absorbing boundary conditions that represent the physical size and the surface properties of the Sun \cite{Turyshev-Toth:2018-grav-shadow,Turyshev-Toth:2019}.

We begin with the area that lies outside the Sun where three regions are present, namely (i) the shadow region, (ii) the geometric optics region, and (iii) the interference region. Clearly, as far as imaging with the SGL is concerned,  the interference region is of the greatest importance. This is where the SGL focuses light coming from a distant object, forming an image.

\subsection{Fully absorbing boundary conditions}
\label{sec:bound-c}

Boundary conditions representing the opaque Sun were introduced in \cite{Herlt-Stephani:1976} and were used in \cite{Turyshev-Toth:2017,Turyshev-Toth:2018-plasma}. Here we use these conditions again. Specifically, to set the boundary conditions, we rely on the semiclassical analogy between the partial momentum, $\ell$, and the impact parameter, $b$, that is given as $\ell=kb$ \cite{Messiah:1968,Landau-Lifshitz:1989}.

To set the boundary conditions, we require that rays with impact parameters $b\le R_\odot^\star=R_\odot +r_g$ are completely absorbed by the Sun \cite{Turyshev-Toth:2017}. Thus, the fully absorbing boundary condition signifies that all the radiation intercepted by the body of the Sun is fully absorbed by it and no reflection or coherent reemission occurs. All intercepted radiation is transformed into some other forms of energy, notably heat. Thus, we require that no scattered waves exist with impact parameter $b\ll R_\odot^\star$ or, equivalently, for $\ell \leq kR_\odot^\star$. Such formulation relies on the concept of the semiclassical impact parameter $b$ and its relationship with the partial momentum, $\ell$, as $\ell=k b$. (A relevant discussion on this relation between $\ell$ and $b$ is on p.~29 of \cite{Grandy-book-2005} with reference to \cite{vandeHulst-book-1981}.)  In terms of the boundary conditions, this means that we need to subtract the scattered waves from the incident wave for $\ell \leq kR^\star_\odot$, as was discussed in \cite{Turyshev-Toth:2017}. Furthermore, as it was shown in \cite{Turyshev-Toth:2018-grav-shadow}, the fully absorbing boundary conditions introduce a fictitious EM field that precisely compensates the incident field in the area behind the Sun. This area has the shape of a rotational hyperboloid that starts directly at the solar surface behind the Sun and extends to the vertex of the hyperboloid at $z_0=R^2_\odot/2r_g\simeq$ 547.8~AU.

\subsection{The Debye potential for the region outside the Sun}

To implement the boundary conditions for the EM wave outside the Sun, we realize that the total EM field in this region is given as the sum of the incident and scattered waves, $\Pi=\Pi_0+\Pi_{\tt s}$, with these two potentials given by (\ref{eq:Pi_ie*+*=}) and (\ref{eq:Pi-s_ass}), correspondingly. Accordingly, we use (\ref{eq:Pi-s_a1*0}), which represents the Debye potential in the region of interest and is given as
{}
\begin{eqnarray}
\Pi(r, \theta)=\Pi_0(r,\theta)+\Pi_{\tt s}(r,\theta)&=& \frac{E_0}{k^2}\frac{u}{r}\sum_{\ell=1}^\infty i^{\ell-1}\frac{2\ell+1}{\ell(\ell+1)}e^{i\sigma_\ell}
\Big\{F_\ell(kr_g,kr) + \frac{1}{2i}\big(e^{2i\xi_b(\vec b)}-1\big) H^+_\ell(kr_g, kr)\Big\}P^{(1)}_\ell(\cos\theta).~~~~~~
\label{eq:Pi-s_a1*}
\end{eqnarray}

Next, relying on the representation of the regular Coulomb function $F_\ell$ via incoming, $H^{+}_\ell$, and outgoing, $H^{-}_\ell$, waves as $F_\ell=(H^{+}_\ell-H^{-}_\ell)/2i$ (discussed in \cite{Turyshev-Toth:2017} and also by the expression given after (\ref{eq:R-L})), we may express the Debye potential (\ref{eq:Pi-s_a1*}) as
{}
\begin{eqnarray}
\Pi(r, \theta)&=& \frac{E_0}{2ik^2}\frac{u}{r}\sum_{\ell=1}^\infty i^{\ell-1}\frac{2\ell+1}{\ell(\ell+1)}e^{i\sigma_\ell}
\Big\{e^{2i\xi_b(\vec b)}H^+_\ell(kr_g, kr)-H^-_\ell(kr_g, kr)\Big\}P^{(1)}_\ell(\cos\theta).
\label{eq:Pi-s_a*0}
\end{eqnarray}

This form of the combined Debye potential is convenient for implementing the fully absorbing  boundary conditions discussed in Sec.~\ref{sec:bound-c}. Specifically, subtracting from (\ref{eq:Pi-s_a*0})  the outgoing wave (i.e., $\propto H^{(+)}_\ell$) for the impact parameters $b\leq R_\odot^\star$ or equivalently for $\ell\in[1,kR_\odot^\star]$, we have
 {}
\begin{eqnarray}
\Pi(r, \theta)&=& \frac{E_0}{2ik^2}\frac{u}{r}\sum_{\ell=1}^\infty i^{\ell-1}\frac{2\ell+1}{\ell(\ell+1)}e^{i\sigma_\ell}
\Big\{e^{2i\xi_b(\vec b)}H^+_\ell(kr_g, kr)-H^-_\ell(kr_g, kr)\Big\}P^{(1)}_\ell(\cos\theta)-\nonumber\\
&&\hskip 10 pt -\, \frac{E_0}{2ik^2}\frac{u}{r}\sum_{\ell=1}^{kR_\odot^\star} i^{\ell-1}\frac{2\ell+1}{\ell(\ell+1)}e^{i\sigma_\ell}e^{2i\xi_b(\vec b)}H^+_\ell(kr_g, kr)P^{(1)}_\ell(\cos\theta),
\label{eq:Pi-s_a+0}
\end{eqnarray}
or, equivalently, coming back to the form (\ref{eq:Pi-s_a1*}),
 {}
\begin{eqnarray}
\Pi(r, \theta)&=&
\Pi_0(r,\theta)+ \frac{E_0}{2ik^2}\frac{u}{r}\sum_{\ell=1}^\infty i^{\ell-1}\frac{2\ell+1}{\ell(\ell+1)}e^{i\sigma_\ell}
\big(e^{2i\xi_b(\vec b)}-1\big) H^+_\ell(kr_g, kr)P^{(1)}_\ell(\cos\theta) -\nonumber\\
&&\hskip 36 pt -\,
\frac{E_0}{2ik^2}\frac{u}{r}\sum_{\ell=1}^{kR_\odot^\star} i^{\ell-1}\frac{2\ell+1}{\ell(\ell+1)}e^{i\sigma_\ell}e^{2i\xi_b(\vec b)}H^+_\ell(kr_g, kr)P^{(1)}_\ell(\cos\theta).
\label{eq:Pi-s_a+}
\end{eqnarray}

This is a rather complex expression. It requires the tools of numerical analysis to fully explore its behavior and the resulting EM field \cite{Kerker-book:1969,vandeHulst-book-1981,Grandy-book-2005}. However, in most practically important applications, we need to know the field in the forward direction. Furthermore, our main interest is to study the largest  impact of the extended gravity on light propagation, which corresponds to the smallest values of the impact parameter. In this situation, we may simplify the result (\ref{eq:Pi-s_a+}) by taking into account the asymptotic behavior of the function $H^{+}_\ell(kr_g,kr)$, considering the field at large heliocentric distances, such that $kr\gg\ell$, where $\ell$ is the order of the Coulomb function (see p.~631 of \cite{Morse-Feshbach:1953}). For  $kr\rightarrow\infty $ and also for $r\gg r_{\tt t}=\sqrt{\ell(\ell+1)}/k$ (see \cite{Turyshev-Toth:2017,Turyshev-Toth:2018-plasma}), such an expression is given in the form \cite{Turyshev-Toth:2019}):
{}
\begin{eqnarray}
\lim_{kr\rightarrow\infty} H^{\pm}_\ell(kr_g,kr)&\sim&
\exp\Big[\pm ik\big(r+r_g\ln2kr\big)+\frac{\ell(\ell+1)}{2kr}+\sigma_\ell-\frac{\pi \ell}{2}\Big)\Big] +{\cal O}\big((kr)^{-2}, r_g^2\big),
\label{eq:Fass*}
\end{eqnarray}
which includes the contribution from the centrifugal potential in the radial equation (\ref{eq:R-bar-k*}) (see e.g., Appendix~C of \cite{Turyshev-Toth:2019}, Appendix A in \cite{Turyshev-Toth:2018} or \cite{Kerker-book:1969}). In fact, expression (\ref{eq:Fass*}) extends the argument of (\ref{eq:free_u+}) to shorter distances, closer to the turning point of the potential (see the relevant discussion in Appendix~F of \cite{Turyshev-Toth:2017}). By including the extended centrifugal term in (\ref{eq:Fass*}) (i.e., shown by the terms with various powers of $\ell(\ell+1)/2kr$), we can now better describe  the bending of the trajectory of a light ray under the combined influence of extended gravity.

We use the approximate behavior of $H^{+}_\ell $ given by  (\ref{eq:Fass*}) and use it in (\ref{eq:Pi-s_a+}) to present the solution for the Debye potential in the following form:
{}
\begin{eqnarray}
\Pi (r, \theta)&=&
\Pi_0 (r, \theta)+\frac{ue^{ik(r+r_g\ln 2kr)}}{r}\Big\{\frac{E_0}{2k^2}\sum_{\ell=1}^{kR_\odot^\star} \frac{2\ell+1}{\ell(\ell+1)}e^{i\big(2\sigma_\ell+\frac{\ell(\ell+1)}{2k r}
\big)}P^{(1)}_\ell(\cos\theta)-\nonumber\\
&&\hskip80pt -\,\frac{E_0}{2k^2}\sum_{\ell=kR_\odot^\star}^{\infty} \frac{2\ell+1}{\ell(\ell+1)}e^{i\big(2\sigma_\ell+\frac{\ell(\ell+1)}{2k r}
\big)}\big(e^{i2\xi_b(\vec b)}-1\big)P^{(1)}_\ell(\cos\theta) \Big\}+{\cal O}\Big(r_g^2,\frac{J_2}{r^3}\Pi\Big)=\nonumber\\
&=&\Pi_0 (r, \theta)+\Pi_{\tt bc} (r, \theta)+\Pi _{\tt G} (r, \theta).
  \label{eq:Pi_g+p0}
\end{eqnarray}

The first term in (\ref{eq:Pi_g+p0}), $\Pi_0 (r, \theta)$, is the Debye potential that represents the incident EM wave propagating in the vacuum on the background of a post-Newtonian gravity field produced by a gravitational monopole. The solution for  $\Pi_0 (r, \theta)$ is known and is given by  (\ref{eq:Pi_ie*+*=}) in the form of infinite series with respect to partial momenta, $\ell$ (see \cite{Turyshev-Toth:2017,Turyshev-Toth:2019}).

The second term in (\ref{eq:Pi_g+p0}), $\Pi_{\tt bc} (r, \theta)$, is due to the physical obscuration introduced by the Sun and was derived by applying the fully absorbing boundary conditions. This term is responsible for the geometric shadow behind the Sun.

The third term in (\ref{eq:Pi_g+p0}), $\Pi _{\tt G} (r, \theta)$,  quantifies the contribution of the extended gravitational field to the scattering of the EM wave.

With the solution for the Debye potential given by (\ref{eq:Pi_g+p0}), and with the help of (\ref{eq:Dr-em0})--(\ref{eq:Bp-em0}) (also see \cite{Turyshev-Toth:2017}), we may now compute the EM field in the various regions involved. Given the smallness of the ratio $r_gJ_2R_\odot^2/{r^3}$, we may neglect the distance-dependent effects of the solar extended gravity on the amplitude of the EM wave.
Thus, the extended gravity contributes to the delay of the EM wave and is fully accounted for by the solution for the Debye potentials. Therefore, we can  use the following expressions to construct the EM field in the static,  gravity field produced by an extended gravity (see details in \cite{Turyshev-Toth:2017,Turyshev-Toth:2019}):
{}
\begin{align}
  \left( \begin{aligned}
{ \hat D}_r& \\
{ \hat B}_r& \\
  \end{aligned} \right) =&  \left( \begin{aligned}
\cos\phi \\
\sin\phi  \\
  \end{aligned} \right) \,e^{-i\omega t}\alpha(r, \theta,\phi), &
    \left( \begin{aligned}
{ \hat D}_\theta& \\
{ \hat B}_\theta& \\
  \end{aligned} \right) =&  \left( \begin{aligned}
\cos\phi \\
\sin\phi  \\
  \end{aligned} \right) \,e^{-i\omega t}\beta(r, \theta,\phi), &
    \left( \begin{aligned}
{ \hat D}_\phi& \\
{ \hat B}_\phi& \\
  \end{aligned} \right) =&  \left( \begin{aligned}
-\sin\phi \\
\cos\phi  \\
  \end{aligned} \right) \,e^{-i\omega t}\gamma(r, \theta,\phi),
  \label{eq:DB-sol00p*}
\end{align}
with the quantities $\alpha, \beta$ and $\gamma$ computed from the known Debye potential, $\Pi$, as
{}
\begin{eqnarray}
\alpha(r, \theta,\phi)&=&
\frac{1}{u}\Big\{\frac{\partial^2 }{\partial r^2}
\Big[\frac{r\,{\hskip -1pt}\Pi}{u}\Big]+k^2 u^4\Big[\frac{r\,{\hskip -1pt}\Pi}{u}\Big]\Big\}+{\cal O}\Big(\big(\frac{1}{u}\big)''\Big),
\label{eq:alpha*}\\
\beta(r, \theta,\phi)&=&\frac{1}{u^2r}
\frac{\partial^2 \big(r\,{\hskip -1pt}\Pi\big)}{\partial r\partial \theta}+\frac{ik\big(r\,{\hskip -1pt}\Pi\big)}{r\sin\theta},
\label{eq:beta*}\\[0pt]
\gamma(r, \theta,\phi)&=&\frac{1}{u^2r\sin\theta}
\frac{\partial \big(r\,{\hskip -1pt}\Pi\big)}{\partial r}+\frac{ik}{r}
\frac{\partial\big(r\,{\hskip -1pt}\Pi\big)}{\partial \theta}.
\label{eq:gamma*}
\end{eqnarray}

This completes the solution for the Debye potentials on the background of a spherically symmetric, static gravitational field of the Sun. We will use (\ref{eq:DB-sol00p*})--(\ref{eq:gamma*}) to compute the relevant EM fields.

\subsection{EM field in the shadow region}

In the shadow behind the Sun (i.e., for impact parameters $b\leq R^\star_\odot$) the EM field is represented by the Debye potential of the shadow, $\Pi_{\tt sh}$, which is given as
{}
\begin{eqnarray}
\Pi_{\tt sh} (r, \theta)&=&
\Pi_0 (r, \theta)+\frac{ue^{ik(r+r_g\ln 2kr)}}{r}\frac{E_0}{2k^2}\sum_{\ell=1}^{kR_\odot^\star} \frac{2\ell+1}{\ell(\ell+1)}e^{i\big(2\sigma_\ell+\frac{\ell(\ell+1)}{2kr}
\big)}P^{(1)}_\ell(\cos\theta)+{\cal O}\Big(r_g^2,\frac{J_2}{r^3}\Pi\Big),
  \label{eq:Pi_sh}
\end{eqnarray}
where $\Pi_0 (r, \theta)$ is well represented by (\ref{eq:Pi_ie*+*=}).
As discussed in \cite{Turyshev-Toth:2017,Turyshev-Toth:2018-grav-shadow}, the potential (\ref{eq:Pi_sh}) produces no EM field. In other words, there is no light in the shadow. Furthermore, as the solar boundary is rather diffuse, there is no expectation for a Poisson--Arago bright spot to form in this region.

\subsection{EM field outside the shadow}
\label{sec:EM-field}

In the region behind the Sun but outside the solar shadow (i.e., for light rays with impact parameters $b>R_\odot$) which includes both the geometric optics  and interference regions (in the immediate vicinity of the focal line), the EM field is derived from the Debye potential given by the remaining terms in (\ref{eq:Pi_g+p0}) to ${\cal O}(r_g^2,{J_2}/{r^3})$ as
{}
\begin{eqnarray}
\Pi (r, \theta)&=& \Pi_0 (r, \theta)-\frac{ue^{ik(r+r_g\ln 2kr)}}{r}\frac{E_0}{2k^2}\sum_{\ell=kR_\odot^\star}^{\infty} \frac{2\ell+1}{\ell(\ell+1)}e^{i\big(2\sigma_\ell+\frac{\ell(\ell+1)}{2kr}
\big)}\Big(e^{i2\xi_b(\vec b)}-1\Big)P^{(1)}_\ell(\cos\theta),
  \label{eq:Pi_g+!}
\end{eqnarray}
where for the geometric optics region the potential $\Pi_0 (r, \theta)$ is well represented by (\ref{eq:Pi_ie*+*=}).

Expression (\ref{eq:Pi_g+!}) is our main result for the regions outside  the shadow region, i.e., $b\geq R_\odot^\star$. It contains all the information needed to describe the total EM field originating from an incident Coulomb-modified plane wave that passed through the region of the extended solar gravity field, characterized by the distance dependence that diminishes as $r^{-3}$ or faster.

To evaluate the total solution for the Debye potential (\ref{eq:Pi_g+!}), we present it in the following compact form:
{}
\begin{eqnarray}
{\Pi (r, \theta)}&=&
{\Pi_0 (r, \theta)}+E_0f_{\tt G}(r,\theta,\phi)\frac{ue^{ik(r+r_g\ln 2kr)}}{r},  \label{eq:Pi-g+p}
\end{eqnarray}
where the extended gravity scattering amplitude $f_{\tt G}(r,\theta,\phi)$ is given by
{}
\begin{eqnarray}
  f_{\tt G}(r,\theta,\phi)&=&-\frac{1}{2k^2}\sum_{\ell=kR_\odot^\star}^\infty \frac{2\ell+1}{\ell(\ell+1)}e^{i\big(2\sigma_\ell+\frac{\ell(\ell+1)}{2kr}  \big)}\Big(e^{i2\xi_b(\vec b)}-1\Big)
P^{(1)}_\ell(\cos\theta) +{\cal O}\Big(r_g^2,\frac{J_2}{r^3}\Pi\Big).
  \label{eq:f-v*+}
\end{eqnarray}

We note that because of the contribution from the centrifugal potential in (\ref{eq:Fass*}), the scattering amplitude $  f_{\tt p}(r,\theta)$ is now also a function of the heliocentric distance \cite{Turyshev-Toth:2018-plasma}. This is not the case in typical problems describing nuclear and atomic scattering \cite{Messiah:1968,Landau-Lifshitz:1989,Burke:2011,Newton-book-2013}. However, as we observed in \cite{Turyshev-Toth:2017,Turyshev-Toth:2018,Turyshev-Toth:2018-plasma}, when we are interested in the trajectories of light rays, the presence of such dependence and especially the $\propto 1/r$ term in the phase of the scattering amplitude (\ref{eq:f-v*+}) allows us to properly describe the bending of the light rays in the presence of gravity together with the contribution from deviations from spherical symmetry.

As a result, the Debye potential takes the form
{}
\begin{align}
  \Pi_{\tt G} (r, \theta)=E_0f_{\tt G}(r,\theta,\phi)\frac{ue^{ik(r+r_g\ln 2kr)}}{r},
  \label{eq:Pi_ie*+8p*}
\end{align}
with the extended gravity scattering  amplitude  $f_{\tt G}(r,\theta)$ given by (\ref{eq:f-v*+}). We use these expressions to derive the components of the EM field produced by this wave. For this, we substitute (\ref{eq:Pi_ie*+8p*})--(\ref{eq:f-v*+})  in the expressions (\ref{eq:alpha*})--(\ref{eq:gamma*}) to derive the factors $\alpha(r,\theta,\phi), \beta(r,\theta,\phi)$ and $\gamma(r,\theta,\phi)$, which to ${\cal O}(r_g^2,J_2/r^3)$  are computed to be:
{}
\begin{eqnarray}
\alpha(r,\theta,\phi) &=& -E_0\frac{ue^{ik(r+r_g\ln 2kr)}}{k^2r^2}\sum_{\ell=kR^\star_\odot}^\infty(\ell+{\textstyle\frac{1}{2}})e^{i\big(2\sigma_\ell+\frac{\ell(\ell+1)}{2kr}
\big)}\Big(e^{i2\xi_b(\vec b)}-1\Big)P^{(1)}_\ell(\cos\theta)
\Big\{1-\frac{\ell(\ell+1)}{4k^2r^2}
\Big\},
  \label{eq:alpha*1*=}\\
\beta(r,\theta,\phi) &=& E_0\frac{ue^{ik(r+r_g\ln 2kr)}}{ikr}\times\nonumber\\
&&\hskip 10pt \times\,
\sum_{\ell=kR^\star_\odot}^\infty\frac{\ell+{\textstyle\frac{1}{2}}}{\ell(\ell+1)}e^{i\big(2\sigma_\ell+\frac{\ell(\ell+1)}{2kr}
\big)}\Big(e^{i2\xi_b(\vec b)}-1\Big)
\Big\{\frac{\partial P^{(1)}_\ell(\cos\theta)}
{\partial \theta}\Big(1-\frac{\ell(\ell+1)}{2u^2k^2r^2}
\Big)+\frac{P^{(1)}_\ell(\cos\theta)}{\sin\theta}
 \Big\},
  \label{eq:beta*1*=}\\
\gamma(r,\theta,\phi) &=& E_0\frac{ue^{ik(r+r_g\ln 2kr)}}{ikr}\times\nonumber\\
&&\hskip 10pt \times\,
\sum_{\ell=kR^\star_\odot}^\infty\frac{\ell+{\textstyle\frac{1}{2}}}{\ell(\ell+1)}e^{i\big(2\sigma_\ell+\frac{\ell(\ell+1)}{2kr}
\big)}\Big(e^{i2\xi_b(\vec b)}-1\Big)
\Big\{\frac{\partial P^{(1)}_\ell(\cos\theta)}
{\partial \theta}+\frac{P^{(1)}_\ell(\cos\theta)}{\sin\theta}\Big(1-\frac{\ell(\ell+1)}{2u^2k^2r^2}
\Big)
 \Big\},
  \label{eq:gamma*1*=}
\end{eqnarray}
where we neglected small terms that behave as $\propto i/(u^2kr)$; terms $\propto ikr_g/\ell^2$ were also omitted because of the large partial momenta involved, $\ell\geq kR^\star_\odot$. Terms in both of these groups are negligably small when compared to the leading terms in each of these expressions above (a similar conclusion was reached in \cite{Turyshev-Toth:2018-plasma,Turyshev-Toth:2019}.)

This is an important result as it allows us to describe the EM field in all the regions of interest for the SGL, namely the strong and weak  interference regions and the region of geometric optics.

\section{EM field in the interference region}
\label{sec:IF-region}

Results from the previous section allow us to study optical properties of the SGL in the case of extended gravitational lens. Our primary concern is the strong interference region: the area behind the Sun, reachable by light rays with impact parameters $b>R_\odot^\star$. The focal region of the SGL begins where $r>2b^2/2r_g$ and  $0\leq \theta\simeq \sqrt{2r_g/r}$.
The EM field is derived from the Debye potential (\ref{eq:Pi-g+p})--(\ref{eq:f-v*+}), given by the factors $\alpha$, $\beta$ and $\gamma$ from (\ref{eq:alpha*1*=})--(\ref{eq:gamma*1*=}). In the strong interference region, these expressions take the following form  \cite{Turyshev-Toth:2017,Turyshev-Toth:2019,Turyshev-Toth:2019-fin-difract}:
{}
\begin{eqnarray}
\alpha(r,\theta,\phi) &=& -E_0\frac{ue^{ik(r+r_g\ln 2kr)}}{k^2r^2}\sum_{\ell=kR^\star_\odot}^\infty(\ell+{\textstyle\frac{1}{2}})e^{i\big(2\sigma_\ell+\frac{\ell(\ell+1)}{2kr}
\big)}\Big(e^{i2\xi_b(\vec b)}-1\Big)P^{(1)}_\ell(\cos\theta)
\Big(1+{\cal O}\big(\frac{r_g}{r},r_g^2,J_2\big)\Big),
  \label{eq:alpha*1*}\\
\gamma(r,\theta,\phi) &=& \beta(r,\theta,\phi) = E_0\frac{ue^{ik(r+r_g\ln 2kr)}}{ikr}\times \nonumber\\
&&\hskip -10pt \times\,  \sum_{\ell=kR^\star_\odot}^\infty\frac{\ell+{\textstyle\frac{1}{2}}}{\ell(\ell+1)}e^{i\big(2\sigma_\ell+\frac{\ell(\ell+1)}{2kr}
\big)}\Big(e^{i2\xi_b(\vec b)}-1\Big)
\Big\{\frac{\partial P^{(1)}_\ell(\cos\theta)}
{\partial \theta} +\frac{P^{(1)}_\ell(\cos\theta)}{\sin\theta}
 \Big\}\Big(1+{\cal O}\big(\frac{r_g}{r},r_g^2,J_2\big)\Big).~~
  \label{eq:beta*1*}
\end{eqnarray}

We recognize that (\ref{eq:alpha*1*})--(\ref{eq:beta*1*}) represent the scattered EM field in the interference region.  As evident in the structure of the expression (\ref{eq:Pi-g+p}), the term  $(e^{i2\xi_b(\vec b)}-1)$ present in these expressions leads to formation of two waves -- that one that is $\propto e^{i2\xi_b(\vec b)}$ is the EM wave due to diffraction of light by the gravitational multipoles, while the $\propto -1$ results in the term cancels the  incident wave (see \cite{Turyshev-Toth:2017,Turyshev-Toth:2019}).  Thus, without loss of generality we may drop the term  $\propto -1$ in the term $(e^{i2\xi_b(\vec b)}-1)$. This will directly yield the solution with the corresponding scattering amplitude that can be used to characterize the EM field that was diffracted on the extended solar gravitational field.

At this point, we may replace the sums in (\ref{eq:alpha*1*})--(\ref{eq:beta*1*}) with an integral (accounting for the fact that $\ell\gg1$ and keeping the terms up to ${\cal O}(\theta)$) to be evaluated with the method of stationary phase (with $\beta(r,\theta,\phi) = \gamma(r,\theta,\phi)$):
{}
\begin{eqnarray}
\alpha(r,\theta,\phi) &=& -E_0\frac{ue^{ik(r+r_g\ln 2kr)}}{k^2r^2} \int_{\ell=kR^\star_\odot}^\infty \ell d\ell e^{i\big(2\sigma_\ell+\frac{\ell^2}{2kr}+2\xi_b(\vec b)\big)}P^{(1)}_\ell(\cos\theta)
\Big(1+{\cal O}\big(\theta, \frac{r_g}{r},r_g^2,J_2\big)\Big),
  \label{eq:alpha*2}\\
\gamma(r,\theta,\phi) &=&
E_0\frac{ue^{ik(r+r_g\ln 2kr)}}{ikr}\int_{\ell=kR^\star_\odot}^\infty
\frac{d\ell}{\ell} e^{i\big(2\sigma_\ell+\frac{\ell^2}{2kr}+
2\xi_b(\vec b)\big)}
\Big\{\frac{\partial P^{(1)}_\ell(\cos\theta)}
{\partial \theta} +\frac{P^{(1)}_\ell(\cos\theta)}{\sin\theta}
 \Big\}\Big(1+{\cal O}\big(\theta,\frac{r_g}{r},r_g^2,J_2\big)\Big).~~~~
  \label{eq:beta*2}
\end{eqnarray}

To evaluate these expressions in the interference region and for $0\leq \theta \simeq \sqrt{2r_g/r}$, we use the asymptotic representation for $P_\ell(\cos\theta)$ and $\ell\gg1$ from \cite{Bateman-Erdelyi:1953,Korn-Korn:1968,Kerker-book:1969,Abramovitz-Stegun:1965}
{}
\begin{eqnarray}
P_\ell(\cos\theta)&=& \sqrt{\frac{\theta}{\sin\theta}} J_0\big(\ell \theta\big)+{\cal O}(\theta^2).
\label{eq:Bess0}
\end{eqnarray}
For improved explicit two-term uniformly valid asymptotic  form of this expression, check \cite{Bakaleinikov:2020}.
We use the expression
{}
\begin{eqnarray}
P^{(1)}_\ell(\cos\theta)=-\frac{\partial P_\ell(\cos\theta)}{\partial\theta}=\ell J_1(\ell\theta)+
{\textstyle\frac{1}{6}}\theta J_0(\ell\theta) +{\cal O}(\theta^2),
\label{eq:Bess}
\end{eqnarray}
to derive
{}
\begin{eqnarray}
\frac{P^{(1)}_\ell(\cos\theta)}{\sin\theta}&=& {\textstyle\frac{1}{2}}\ell^2\Big(J_0(\ell \theta)+J_2(\ell \theta)\Big),
\label{eq:pi-l=}
\qquad
\frac{dP^{(1)}_\ell(\cos\theta)}{d\theta}= {\textstyle\frac{1}{2}}\ell^2\Big(J_0(\ell \theta)-J_2(\ell \theta)\Big).
\label{eq:tau-l=}
\end{eqnarray}
Using expressions (\ref{eq:Bess}) and (\ref{eq:pi-l=}) in (\ref{eq:alpha*2})--(\ref{eq:beta*2}), we have
{}
\begin{eqnarray}
\alpha(r,\theta,\phi) &=& -E_0\frac{ue^{ik(r+r_g\ln 2kr)}}{k^2r^2} \int_{\ell=kR^\star_\odot}^\infty \ell^2 d\ell \, J_1(\ell\theta) e^{i\big(2\sigma_\ell+\frac{\ell^2}{2kr}+2\xi_b(\vec b)\big)}
\Big(1+{\cal O}\big(\theta, \frac{r_g}{r},r_g^2,J_2\big)\Big),
  \label{eq:alpha*3=}\\
\gamma(r,\theta,\phi) &=&
E_0\frac{ue^{ik(r+r_g\ln 2kr)}}{ikr}\int_{\ell=kR^\star_\odot}^\infty
\ell d\ell \, J_0(\ell\theta)e^{i\big(2\sigma_\ell+\frac{\ell^2}{2kr}+
2\xi_b(\vec b)\big)}
 \Big(1+{\cal O}\big(\theta,\frac{r_g}{r},r_g^2,J_2\big)\Big).
  \label{eq:beta*3=}
\end{eqnarray}

In the case of a pure gravitational monopole, the eikonal phase $\xi_b(\vec b)$ in (\ref{eq:alpha*3=})--(\ref{eq:beta*3=}) is absent. In that case, these integrals can be evaluated using the method of stationary phase, leading to the well-known result \cite{Nambu:2013a,Turyshev-Toth:2017,Turyshev-Toth:2019} with the PSF $\propto J^2_0$ of a monopole (i.e., a point mass or spherically-symmetric) lens.

However, in the presence of $\xi_b(\vec b)$, the method of stationary phase is not applicable as the expressions (\ref{eq:alpha*3=})--(\ref{eq:beta*3=}) now have angular dependence that is not captured by the integrals. Therefore, we need a method that can address this to by transforming these integrals into an appropriate form that captures such a dependence.

To evaluate these integrals, we developed what we call the {\em angular eikonal method}. This approach entails replacing the Bessel functions $J_0$ and $J_1$ with their integral representations\footnote{Note that we can use the same representations of these functions with positive sign in the phase. The result is identical as it
only replaces the integrand with its complex conjugate, but it leaves the real-valued result unaffected.
}:
{}
\begin{eqnarray}
J_0(\ell\theta)&=& \frac{1}{2\pi}\int_0^{2\pi} d\phi_\xi e^{-i\ell\theta\cos(\phi_\xi-\phi)},
\label{eq:J0}
\qquad
J_1(\ell\theta)= \frac{i}{2\pi}\int_0^{2\pi} d\phi_\xi \cos(\phi_\xi-\phi) e^{-i\ell\theta\cos(\phi_\xi-\phi)}.
\label{eq:J1}
\end{eqnarray}
These expressions recognize the fact that, to describe the geometry of the problem, we selected a heliocentric coordinate system whose $z$ axis is co-linear with the wavevector, $\vec k$, of the incident EM wave. The expressions in (\ref{eq:J0}) are integrals over the azimuthal angle $\phi_\xi$, representing the orientation of the unit vector of the impact parameter $\vec b$, as given by (\ref{eq:notes}). The expressions (\ref{eq:J0}) allow us to rewrite (\ref{eq:alpha*3=})--(\ref{eq:beta*3=}), to ${\cal O}\big(\theta,{r_g}/{r},r_g^2\big)$, in the following 2-dimensional form:
{}
\begin{eqnarray}
\alpha(r,\theta,\phi) &=& E_0\frac{ue^{ik(r+r_g\ln 2kr)}}{ik^2r^2}  \frac{1}{2\pi}\int_0^{2\pi} d\phi_\xi \cos(\phi_\xi-\phi)\int_{\ell=kR^\star_\odot}^\infty \ell^2 d\ell e^{i\big(2\sigma_\ell+\frac{\ell^2}{2kr}+2\xi_b(\vec b)-\ell\theta\cos(\phi_\xi-\phi)\big)},
  \label{eq:alpha*3}\\
\gamma(r,\theta,\phi) &=&
E_0\frac{ue^{ik(r+r_g\ln 2kr)}}{ikr} \frac{1}{2\pi}\int_0^{2\pi} d\phi_\xi \int_{\ell=kR^\star_\odot}^\infty
\ell d\ell e^{i\big(2\sigma_\ell+\frac{\ell^2}{2kr}+2\xi_b(\vec b)-\ell\theta\cos(\phi_\xi-\phi)\big)}.
  \label{eq:beta*3}
\end{eqnarray}

The step presented above correctly captures the functional form of the integrand, which is azimuthally perturbed by the eikonal phase shift, $2\xi_b(\vec b)$, whose presence breaks the spherical symmetry present in the case of a gravitational monopole. Technically, this step could have been done much earlier, in the Debye potential of the incident wave (\ref{eq:Pi_ie*+*=})  that still possesses the symmetries representative of Coulomb-scattering. However, doing it that early would obscure the presentation of the overall solution.  As it is known, solving the time-independent Schr\"odinger  equation in the presence of the Coulomb potential (representing a point source)  is a well-posed problem. As demonstrated by (\ref{eq:Pi*})--(\ref{eq:Pi_ie*+*=}), this problem reduces to solving the relevant wave equation by implementing separation of variables that results in a well-known solution \cite{Turyshev-Toth:2017}. In the case when the scattering potential is not spherically symmetric, separation of variables, in general, is not possible. Thus, other methods are needed.

For gravitational lensing in a weak gravitational field, characterized by a scattering potential with only small deviations from spherical symmetry, we may use the eikonal approximation to identify the eikonal phase shift that corresponds to that particular scattering potential (see details in Section~\ref{sec:eik-wfr}). This eikonal phase shift is effectively a representation of the well-known thin lens approximation \cite{Schneider-Ehlers-Falco:1992}.  However, our variables in (\ref{eq:alpha*3=})--(\ref{eq:beta*3=}) were still given in terms of the monopole case. This is where we recognized that the integral expressions (\ref{eq:J0}) may be used to solve the problem for small deviations from spherical symmetry, which was done in (\ref{eq:alpha*3})--(\ref{eq:beta*3}).  At this point, it is clear that the integrals over $d\phi_\xi$ in (\ref{eq:alpha*3})--(\ref{eq:beta*3}) act not only the monopole part of the phase, $-\ell\theta\cos(\phi_\xi-\phi)$ as in (\ref{eq:alpha*3=})--(\ref{eq:beta*3=}), but on the entire phase, which now includes the eikonal phase shift term $2\xi_b(\vec b)$ due to the gravitational multipoles. This outlines the logic behind the {\em angular eikonal method}.

Lastly, we mentioned that the resulting quantities (\ref{eq:alpha*3})--(\ref{eq:beta*3}) determine the EM field in the strong interference region of the SGL. Below, we find that these expressions can be evaluated using the method of stationary phase. Furthermore, as we know \cite{Turyshev-Toth:2017,Turyshev-Toth:2019}, in the interference region the factor $\alpha$ determining the radial components of the EM field is very small, behaving as $\alpha(r,\theta,\phi) \simeq \sqrt{{2r_g}/{r}}\,\gamma(r,\theta,\phi)$. Thus, this factor will yield a negligible contribution to the Poynting vector and it may be omitted. Therefore, in the discussion below we focus on the factor $\beta(r,\theta,\phi) $ only.

\subsection{Integral over the vector impact parameter}
\label{sec:in-imact-par}

As we mentioned above, the integral over $d\phi_\xi$ in (\ref{eq:beta*3}) properly acts not only on the monopole term of the phase, $-\ell\theta\cos(\phi_\xi-\phi)$, but on the entire phase  $2\sigma_\ell+\frac{\ell^2}{2k\tilde r}+2\xi_b-\ell\theta\cos(\phi_\xi-\phi)$, that now includes the contributions from the parts that perturb the spherically symmetric gravitational potential via the eikonal phase, $\xi_b$. In the case of an axisymmetric gravitational field, this perturbation is given by (\ref{eq:eik-ph-axi*}):
{}
\begin{eqnarray}
\xi_b(\vec b)=-kr_g\sum_{n=2}^\infty\frac{J_n}{n}\Big(\frac{R_\odot}{b}\Big)^n \sin^n\beta_s\cos[n(\phi_\xi-\phi_s)].
  \label{eq:xi_b}
\end{eqnarray}

For convenience, we introduce
{}
\begin{eqnarray}
\xi_b(\vec b)=-kr_g\psi(\vec b),\, \qquad
\psi(\vec b)=\sum_{n=2}^\infty\frac{J_n}{n}\Big(\frac{R_\odot}{b}\Big)^n \sin^n\beta_s\cos[n(\phi_\xi-\phi_s)].
  \label{eq:psi}
\end{eqnarray}

Furthermore, for $\ell\gg kr_g$, we evaluate $\sigma_\ell$ as \cite{Turyshev-Toth:2018-grav-shadow}:
{}
\begin{eqnarray}
\sigma_\ell&=& -kr_g\ln \ell.
\label{eq:sig-l*}
\end{eqnarray}
This form agrees with the other known forms of $\sigma_\ell$ \cite{Cody-Hillstrom:1970,Barata:2009ma} that are approximated for large $\ell$.

We rely on the semiclassical approximation that connects the partial momentum, $\ell$, to the impact parameter, $b$ for small angles $\theta$ (or large distances from the Sun, $R_\odot/r<b/r\ll 1$ -- see \cite{Turyshev-Toth:2017} for details). Using semi-classical form that connects the partial momentum and the impact parameter \cite{Schiff:1968,Landau-Lifshitz:1989,Messiah:1968,Turyshev-Toth:2017}
{}
\begin{equation}
\ell\simeq kb,
\label{eq:S-l-pri-p-g}
\end{equation}
we obtain
{}
\begin{eqnarray}
\varphi(\vec b)&=&2\sigma_\ell+\frac{\ell^2}{2k r}+2\xi_b(\vec b)-\ell\theta\cos(\phi_\xi-\phi)\big)= \nonumber\\
&=&
k\Big\{\frac{b^2}{2 r} -b\theta\cos(\phi_\xi-\phi)-2r_g\Big(\ln kb+\sum_{n=2}^\infty\frac{J_n}{n}\Big(\frac{R_\odot}{b}\Big)^n \sin^n\beta_s\cos[n(\phi_\xi-\phi_s)]\Big)\Big\},
  \label{eq:beta*3+4=}
\end{eqnarray}
or, compactly, using (\ref{eq:psi}):
{}
\begin{eqnarray}
\varphi(\vec b)&=&
k\Big\{\frac{b^2}{2 r} -b\theta\cos(\phi_\xi-\phi)-2r_g\big(\ln kb+\psi(\vec b)\big)\Big\}.
  \label{eq:beta*3+4}
\end{eqnarray}

We recognize that the vector of the impact parameter has the form ${\vec b}=b(\cos\phi_\xi,\sin\phi_\xi,0)$. Also, we define the vector ${\vec \theta}$ to a point on the image plane with coordinates $(r, \theta,\phi)$ that has the form ${\vec \theta}=\theta(\cos\phi,\sin\phi,0)$. With these definitions, we see that $b\theta\cos(\phi_\xi-\phi)=({\vec b}\cdot{\vec \theta})$, therefore
{}
\begin{eqnarray}
\varphi(\vec b)&=&
k\Big\{\frac{1}{2 r}\big(b^2 -2b r \theta\cos(\phi_\xi-\phi)\big)-2r_g\big(\ln kb+\psi(\vec b)\big)\Big\}=\nonumber\\
&\simeq&
k\Big\{\frac{1}{2 r}\big({\vec b} - r \vec \theta\big)^2-2r_g\big(\ln kb+\psi(\vec b)\big)\Big\}.
  \label{eq:beta*3+}
\end{eqnarray}
Thus, the phase $\varphi(\vec b)$ represents the Fermat potential that governs the gravitational lensing phenomena \cite{Liebes:1964,Refsdal:1964,Schneider-Ehlers-Falco:1992,Schneider-etal:2006}.

As a result, we can present (\ref{eq:beta*3}) as
{}
\begin{eqnarray}
\gamma(r,\theta,\phi) &=&
E_0ue^{ik(z+r_g\ln 2kr)} \frac{k}{ir}\frac{1}{2\pi}\int d^2\vec b \,\exp\Big[ik\Big(\frac{1}{2 r}({\vec b} - r \vec \theta)^2-2r_g\big(\ln kb+\psi(\vec b)\big)\Big)\Big].
  \label{eq:gamma0+}
\end{eqnarray}

The integral in (\ref{eq:gamma0+}) is known rather well. It was obtained using different methods and tools by several authors. For instance, a similar integral formula for the lensed wave amplitude was obtained using the scalar theory of light in \cite{Deguchi-Watson:1987,Nambu:2013,Nambu:2013a,Matsunaga-Yamamoto:2006}; by using the Fresnel--Kirchhoff diffraction formula \cite{Born-Wolf:1999}; and it was also obtained using the path integral formalism \cite{Feynman:1948,Feynman-Hibbs:1991} in \cite{Nakamura-Deguchi:1999,Yamamoto:2017}. However, all previous efforts discussed primarily a monopole case. Our expression (\ref{eq:gamma0+}) generalizes these previously obtained results via the presence of the eikonal phase shift term, $-2r_g\psi(\vec b)$, to the case of a lens with arbitrary multipole structure, which is explicitly captured by (\ref{eq:psi}).

We also note that all the previous results were obtained using the scalar theory, considering only the amplitude of the EM wave. A unique feature of our approach is that we are able  to reconstruct the entire vector structure of the EM field (e.g., using (\ref{eq:DB-sol00p*}) together with (\ref{eq:alpha*})--(\ref{eq:gamma*})).  This is an important capability when we consider the three-dimensional behavior of a vector theory, for instance, polarization of the EM wave as it propagates through a refractive medium without spherical symmetry. Thus, our approach supersedes and generalizes all previous results obtained for gravitation lensing in post-Newtonian gravity.

\begin{figure}
\includegraphics{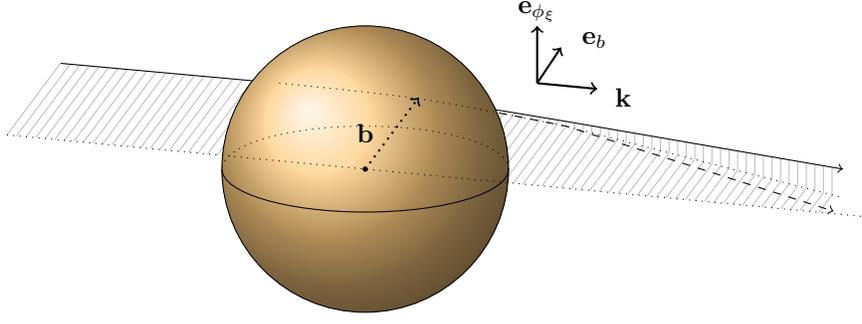}
\caption{\label{fig:basis}The basis vectors used to characterize the vector deflection angle. ${\vec e}_b$ and ${\vec k}$ are in the plane spanned by the incident light ray and the center of the lens; ${\vec e}_{\phi_\xi}$ is normal to this plane. The multipole moments change the amount by which the incident ray is deflected compared to the effect of the monopole (dashed arrow), but also lift the ray out of the plane spanned by the incident ray and the center of the lens.}
\end{figure}

To put the entire problem in the proper context related to the geometry of light propagation in the vicinity of a gravitating body, we consider the total gravitational delay, $d(\vec b)$, acquired by the EM wave as it travels in the gravity field of the extended lens. This delay contributes to the phase shift $\xi_b(\vec b)=kd(\vec b)$ and using   (\ref{eq:beta*3+4=}) is identified as
{}
\begin{eqnarray}
d(\vec b)&=&-2r_g\Big(\ln kb+\sum_{n=2}^\infty\frac{J_n}{n}\Big(\frac{R_\odot}{b}\Big)^n \sin^n\beta_s\cos[n(\phi_\xi-\phi_s)]\Big).
  \label{eq:delay*}
\end{eqnarray}
This is the generalization of the  classic Shapiro time delay to the case of an extended axisymmetric gravitational lens. This delay corresponds to the total gravitational deflection angle acquired by a light ray or, equivalently, rotation of the wavefront of the EM wave.
Using the expression (\ref{eq:b0}) for the radius vector of the EM wave, together with  $\vec b$ given by  (\ref{eq:notes-b}), we compute this angle as
{}
\begin{eqnarray}
\vec \theta_g =-\vec \nabla d(\vec b)=-\Big\{{\vec e}_b \frac{\partial  d(\vec b)}{\partial b}+{\vec e}_{\phi_\xi} \frac{\partial  d(\vec b)}{b\, \partial \phi_\xi}+{\vec k} \frac{\partial  d(\vec b)}{\partial \tau}\Big\},
  \label{eq:grad-b*}
\end{eqnarray}
where the basis vector ${\vec e}_b$ is the unit vector in the direction of the vector of the impact parameter  $\vec{b}$ and ${\vec e}_{\phi_\xi}$ is the unit vector in the azimuthal direction and is orthogonal to ${\vec b}$ and ${\vec k}$ (Fig.~\ref{fig:basis}).

Note that expression (\ref{eq:xi_b}) for the eikonal phase obtained in Appendix~\ref{sec:eik-phase-Jn}  was derived under the thin lens approximation where the distances travelled by the EM wave from the source to the lens, $\tau_0=(\vec k\cdot{\vec r}_0)$, and from the lens to the observer, $\tau=(\vec k\cdot{\vec r})$, are much larger than the impact parameter, namely $b/|\tau_0|\ll1$ and $b/|\tau|\ll1$. This is the reason why (\ref{eq:delay*}) does not depend on $\tau$, thus yielding a vanishing derivative with respect to $\tau$ in (\ref{eq:grad-b*}).

With these considerations in mind, we compute the vector of the total  angle of the gravitational deflection of light as the light ray passes in the vicinity of an extended axisymmetric lens:
{}
\begin{eqnarray}
\vec \theta_g &=&\frac{2r_g}{b}\Big\{{\vec e}_b-
\sum_{n=2}^\infty J_n\Big(\frac{R_\odot}{b}\Big)^n \sin^n\beta_s\Big({\vec e}_b\cos[n(\phi_\xi-\phi_s)]+{\vec e}_{\phi_\xi}\sin[n(\phi_\xi-\phi_s)]\Big)\Big\}.
  \label{eq:def-angl*}
\end{eqnarray}

The first term in (\ref{eq:def-angl*}) is the Einstein deflection angle in the gravity field of a spherically symmetric matter distribution (i.e., in the presence of a monopole or point mass).  The second term with $J_n$ describes the effect of the multipole moments as a sum of
\begin{inparaenum}[a)]
\item an additional deflection toward or away from the optical axis (the line parallel to the incoming ray of light that intersects the lens at the center), and
\item a deflection away from the plane defined by the incoming ray and the center of the lens.
\end{inparaenum}

To further appreciate the geometry that is represented by this $\vec{e}_{\phi_\xi}$ term, consider the $J_2$ case, when $n=2$. There are four principal directions (which depend on the orientation of the impact parameter given by the angle $\phi_\xi$) for which this second term is zero. These directions correspond to the cusps well-known astroid caustic of the quadrupole lens (see Sec.~\ref{sec:SGL-imaging} below). For all other angles, light is deflected away from the optical axis, lifted out of the plane spanned by the direction of the incident light ray and the center of the lens (Fig.~\ref{fig:basis}). These rays of light never intersect the optical axis; therefore, an observer at the optical axis sees only light from the principal directions. Thus we can instantly see how Eq.~(\ref{eq:def-angl*}) gives rise to the famous Einstein-cross that appears in images formed by gravitational lenses that do not possess spherical symmetry. This result demonstrates the utility and power of the {angular eikonal method}.

Result (\ref{eq:def-angl*}) is new. It correctly accounts for the vector nature of the impact parameter and its orientation with respect to the body's rotational axis. Its magnitude is consistent with that reported in \cite{LePoncinLafitte:2007tx} where a different approach was used. Eq.~(\ref{eq:def-angl*})  generalizes previous expressions that characterized the deflection of light by the gravitational field of a compact object. These earlier results mostly dealt with the monopole \cite{Herlt-Stephani:1976,Deguchi-Watson:1986,Nambu:2012}, (see \cite{Will_book93} and references therein), and  with the quadrupole contributions \cite{Klioner:1991SvA,Klioner-Kopeikin:1992,Zschocke-Klioner:2010}. Our expression describes the deflection of light in the presence of an axisymmetric gravitational field with an arbitrary set of zonal harmonics and for an arbitrary direction of the impact parameter vector.

\subsection{Reducing the double integral using the method of stationary phase}

We continue to work with the integral (\ref{eq:gamma0+}). We evaluate one of the two integrals using the method of stationary phase. We deal with the integral over the magnitude of the impact parameter, $b$. Introducing $\theta=\rho/r$, we see that (\ref{eq:gamma0+}) has the following explicit form that is useful for practical consideration:
{}
\begin{eqnarray}
\gamma(r,\theta,\phi) &=&
E_0ue^{ik(r+r_g\ln 2kr)} \frac{k}{ir}\frac{1}{2\pi}\int_0^{2\pi} d\phi_\xi \int_{b=R^\star_\odot}^\infty
b db \, e^{ik\big(\frac{1}{2r}\big(b^2 -2b\rho \cos(\phi_\xi-\phi)\big)-2r_g\big(\ln kb+\psi(\vec b)\big)\big)}.
  \label{eq:gamma0}
\end{eqnarray}

We recognize that (\ref{eq:gamma0}) is a double integral with respect to the impact parameter, $\vec b$: namely, $d{\vec b}^2=d\phi_\xi bdb$. One may be tempted to try to evaluate this integral using the 2-dimensional method of stationary phase. However,  the presence of higher multipoles leads to appearance of caustics so that such a stationary phase solution will not be accurate, especially at the cusps. Thus, only one of the two integrals in (\ref{eq:gamma0}) should be approximate using the method of stationary phase. We choose to approximate the integral over $db$, leaving the integral over $d\phi_\xi$ unchanged.

Considering (\ref{eq:gamma0}), we see that the points of stationary phase, where $d\varphi(\vec b)/db=0$, are given by the equation
{}
\begin{equation}
\frac{d\varphi}{db }=
k\Big\{\frac{1}{r}\big(b-\rho\cos(\phi_\xi-\phi)\big)-\frac{2r_g}{b}\Big(1-\sum_{n=2}^\infty J_n\Big(\frac{R_\odot}{b}\Big)^n \sin^n\beta_s\cos[n(\phi_\xi-\phi_s)]\big)\Big)\Big\}=0,
\label{eq:IFp-phase*1}
\end{equation}
which may be transformed as
{}
\begin{equation}
b^2-b\rho\cos(\phi_\xi-\phi)-2r_gr\Big(1-\sum_{n=2}^\infty J_n\Big(\frac{R_\odot}{b}\Big)^n \sin^n\beta_s\cos[n(\phi_\xi-\phi_s)]\big)\Big)=0.
\label{eq:IFp-d*1}
\end{equation}

We solve this equation iteratively, by presenting the impact parameter as $b=b^{[0]}+b^{[1]}$, where  $b^{[0]}$ is the solution involving only the monopole term, while $b^{[1]}$ is due to the eikonal phase. Substituting this trial solution in  (\ref{eq:IFp-d*1}) and equating same orders we get:
{}
\begin{eqnarray}
b^{[0]}{}^2-b^{[0]}\rho\cos(\phi_\xi-\phi)-2r_gr&=&0,
\label{eq:bb0}\\
b^{[1]}\Big(2b^{[0]}-\rho\cos(\phi_\xi-\phi)\Big)+2r_gr\sum_{n=2}^\infty J_n\Big(\frac{R_\odot}{b^{[0]}}\Big)^n \sin^n\beta_s\cos[n(\phi_\xi-\phi_s)]&=&0.
\label{eq:bb1}
\end{eqnarray}

From (\ref{eq:bb0}), we have
{}
\begin{equation}
b^{[0]}=\pm\sqrt{2r_gr+\big({\textstyle\frac{1}{2}}\rho\cos(\phi_\xi-\phi)\big)^2}+{\textstyle\frac{1}{2}}\rho\cos(\phi_\xi-\phi).
\label{eq:bb0*pm}
\end{equation}

In the region of strong interference, the relations $0\leq \theta\simeq \sqrt{2r_g/r}$ are satisfied, so that this solution may be given only to ${\cal O}(\rho^2)$. Also, as the magnitude of the impact parameter may only be positive, we choose the positive sign in (\ref{eq:bb0*pm}), which yields
{}
\begin{equation}
b^{[0]}=\sqrt{2r_gr}+{\textstyle\frac{1}{2}}\rho\cos(\phi_\xi-\phi) +{\cal O}(\rho^2).
\label{eq:bb0*=}
\end{equation}
Substituting this solution into (\ref{eq:bb1}), we get
{}
\begin{eqnarray}
b^{[1]}\sqrt{2r_gr}+r_gr\sum_{n=2}^\infty J_n \Big(\frac{R_\odot}{\sqrt{2r_gr}}\Big)^n \sin^n\beta_s\cos[n(\phi_\xi-\phi_s)]&=&0.
\label{eq:bb13=}
\end{eqnarray}
Thus, we have
{}
\begin{eqnarray}
b^{[1]}=-{\textstyle\frac{1}{2}}\sqrt{2r_gr} \sum_{n=2}^\infty J_n \Big(\frac{R_\odot }{\sqrt{2r_gr}}\Big)^n\sin^n\beta_s\cos[n(\phi_\xi-\phi_s)] +{\cal O}\big(\frac{\rho}{\sqrt{2r_gr}}J_n\big).
\label{eq:bb13}
\end{eqnarray}

Ultimately, we have the following solution for the impact parameter, $b=b^{[0]}+b^{[1]}+{\cal O}(r_g^2, \rho^2,{\rho J_n}/{\sqrt{2r_gr}})$:
{}
\begin{eqnarray}
b&=&\sqrt{2r_gr}+{\textstyle\frac{1}{2}}\rho\cos(\phi_\xi-\phi)-
{\textstyle\frac{1}{2}}\sqrt{2r_gr}\sum_{n=2}^\infty J_n \Big(\frac{R_\odot }{\sqrt{2r_gr}}\Big)^n\sin^n\beta_s\cos[n(\phi_\xi-\phi_s)].
\label{eq:bb0*}
\end{eqnarray}

We compute the stationary phase, $\varphi ({\vec b})$ from (\ref{eq:beta*3+4}) for the values of $b$ given by (\ref{eq:bb0*}):
{}
\begin{equation}
\varphi(\vec b)= k\Big\{r_g-2r_g\ln k\sqrt{2r_gr}-\sqrt{\frac{2r_g}{r}}\Big(\rho\cos(\phi_\xi-\phi)+
\sqrt{2r_gr}\sum_{n=2}^\infty \frac{J_n}{n} \Big(\frac{R_\odot }{\sqrt{2r_gr}}\Big)^n\sin^n\beta_s\cos[n(\phi_\xi-\phi_s)]\Big)\Big\}.
\label{eq:IFp-phase*5}
\end{equation}
Computing the second derivative of the phase $\varphi ({\vec b})$ from (\ref{eq:IFp-phase*1}) with respect to $b$, we have (as we do not account for the impact of the multipoles on the amplitude of the EM field, we need to know this quantity only to ${\cal O}(J_n)$):
{}
\begin{eqnarray}
\frac{d^2\varphi}{db^2}=k\Big(\frac{1}{r} +\frac{2r_g}{b^2}+{\cal O}(J_n)\Big).
\label{eq:S-l2+p}
\end{eqnarray}
Now, using $b$ from (\ref{eq:bb0*}), to ${\cal O}(r_g^2, \rho^2,J_n)$, we have
{}
\begin{eqnarray}
\varphi''(b_0)&=&\dfrac{2k}{r}\Big(1-{\textstyle\frac{1}{2}} \frac{\rho\cos(\phi_\xi-\phi)}{\sqrt{2r_gr}}\Big)
\qquad \Rightarrow \qquad
\sqrt{\frac{2\pi}{|\varphi''(b_0)|}}=\sqrt{\frac{\pi r}{k}}\Big(1+{\textstyle\frac{1}{4}} \frac{\rho\cos(\phi_\xi-\phi)}{\sqrt{2r_gr}}\Big).
\label{eq:S-l220-g}
\end{eqnarray}

We now may compute the amplitude of the integrand in (\ref{eq:gamma0}), which for $b$ from (\ref{eq:bb0*}) may be given as
{}
\begin{eqnarray}
A(b_0)\sqrt{\frac{2\pi}{|\varphi''(b_0)|}}&=&b_0\sqrt{\frac{2\pi}{|\varphi''(\ell_0)|}}=\sqrt{2r_gr}\Big(1+{\textstyle\frac{1}{2}} \frac{\rho\cos(\phi_\xi-\phi)}{\sqrt{2r_gr}}\Big)
\sqrt{\frac{\pi r}{k}}\Big(1+{\textstyle\frac{1}{4}} \frac{\rho\cos(\phi_\xi-\phi)}{\sqrt{2r_gr}}
\Big)=\nonumber\\
&=&\sqrt{2r_gr}
\sqrt{\frac{\pi r}{k}}\Big(1+
{\cal O}\Big(r_g^2,\frac{\rho}{\sqrt{2r_gr}},J_n\Big)\Big).~~~~~
\label{eq:IFp-phase*7}
\end{eqnarray}

As a result, the expression for the factor $\gamma(r,\theta,\phi) $ given by (\ref{eq:gamma0}) takes the form
{}
\begin{eqnarray}
\gamma(r,\theta,\phi) &=&
E_0\sqrt{2\pi kr_g}e^{i\sigma_0}\, e^{ikz} \times\nonumber\\
&&\hskip-40pt\, \times\,
\frac{1}{2\pi}\int_0^{2\pi} d\phi_\xi \exp\Big[-ik\sqrt{\frac{2r_g}{r}}\Big(\rho\cos(\phi_\xi-\phi)+
\sqrt{2r_gr}\sum_{n=2}^\infty \frac{J_n}{n} \Big(\frac{R_\odot }{\sqrt{2r_gr}}\Big)^n\sin^n\beta_s\cos[n(\phi_\xi-\phi_s)]\Big)\Big],
  \label{eq:beta*4d+}
\end{eqnarray}
valid to the order of ${\cal O}\big(r_g^2,{\rho}/{\sqrt{2r_gr}}, ({J_2}/{r^3})\Pi\big)$ and the constant $\sigma_0=-kr_g\ln(kr_g/e)-{\textstyle\frac{\pi}{4}}$ (see \cite{Turyshev-Toth:2017,Turyshev-Toth:2019} for details.)

We can present  expression (\ref{eq:beta*4d+}) in the following compact form:
{}
\begin{eqnarray}
\gamma(r,\theta,\phi) &=&
E_0\sqrt{2\pi kr_g}e^{i\sigma_0}\, e^{ikz}  B(\rho,\phi) +{\cal O}\Big(r_g^2,\frac{\rho}{\sqrt{2r_gr}},\frac{J_2}{r^3}\Pi\Big),
  \label{eq:beta*4d+f}
\end{eqnarray}
where $B(\rho,\phi)=B(\vec x)$, with $\vec x=\rho(\cos\phi,\sin\phi,0)$  being the coordinates on the image plane, is the complex amplitude of the EM field that has the form
{}
\begin{eqnarray}
B(\vec x) &=&
\frac{1}{2\pi}\int_0^{2\pi} d\phi_\xi \exp\Big[-ik\Big(\sqrt{\frac{2r_g}{r}}\rho\cos(\phi_\xi-\phi)+
2r_g\sum_{n=2}^\infty \frac{J_n}{n} \Big(\frac{R_\odot }{\sqrt{2r_gr}}\Big)^n\sin^n\beta_s\cos[n(\phi_\xi-\phi_s)]\Big)\Big].~~~~
  \label{eq:B2}
\end{eqnarray}

The quantity $B(\vec x)$  is the complex amplitude of the EM field after it scatters on the  gravitational field of an extended axisymmetric lens that is represented by a set of gravitational multipoles.  If the presence of the gravitational zonal harmonics be neglected, the result (\ref{eq:B2}) reduces to the familiar form for the monopole (see \cite{Turyshev-Toth:2017} and references therein), $J_0(k\sqrt{{2r_g}/{r}}\rho)$.  Eq.~(\ref{eq:B2}) is a new diffraction integral formula that extends the previous wave-theoretical description of gravitational lensing phenomena to the case of a lens with an arbitrary axisymmetric gravitational potential.  This result offers a new, powerful tool to study gravitational lensing in the limit of weak gravitational fields, at the first post-Newtonian approximation of the general theory of relativity.

\subsection{The EM field in the interference region}
\label{sec:amp_func-IF}

Now we are ready to present the components of the EM field in the interference region. The total field in accord with (\ref{eq:DB-sol00p*}) to ${\cal O}(r_g^2, r_g/r)$ has the form
{}
\begin{eqnarray}
  \left( \begin{aligned}
{  \hat D}_\theta& \\
{  \hat B}_\theta& \\
  \end{aligned} \right) &=&  \left( \begin{aligned}
{  \hat B}^{\tt }_\phi& \\
-{  \hat D}^{\tt }_\phi & \\
  \end{aligned} \right)=E_0
 \sqrt{2\pi kr_g}e^{i\sigma_0}B(\rho,\phi)
 e^{i(kz-\omega t)}\left( \begin{aligned}
\cos\phi \\
\sin\phi  \\
  \end{aligned} \right),
  \label{eq:DB-tot-th}
\end{eqnarray}
with radial components of the EM field behaving as $({\hat D}_r, {\hat B}_r)\simeq {\cal O}(r_g^2, r_g/r)$. The radial components of the EM field are negligibly small compared to the other two components, which is consistent with the fact that while passing through the gravity field of higher multipoles  the EM wave preserves its transverse structure.

Expression (\ref{eq:DB-tot-th}) describes the EM field in the interference region of the SGL  in the spherical coordinate system.  To study this field in the image plane, we need to transform this result  to a cylindrical coordinate system. To do that, we follow the approach demonstrated in  \cite{Turyshev-Toth:2017}, where instead of spherical coordinates $(r,\theta,\phi)$, we introduced a cylindrical coordinate system $(\rho,\phi,z)$, more convenient for these purposes.  In the region $r \gg r_g$, this was done by defining $R=ur = r+{r_g}/{2}+{\cal O}(r_g^2)$ and introducing the coordinate transformations $ \rho=R\sin\theta,$ $ z=R\cos\theta$, which, from (\ref{eq:metric-gen}), result in the following line element:
{}
\begin{eqnarray}
ds^2&=&u^{-2}c^2dt^2-u^2\big(dr^2+r^2(d\theta^2+\sin^2\theta d\phi^2)\big)=u^{-2}c^2dt^2-\big(d\rho^2+\rho^2d\phi^2+nu^2dz^2\big)+{\cal O}(r_g^2).
\label{eq:cyl_coord}
\end{eqnarray}

As a result, using  (\ref{eq:DB-tot-th}), for a high-frequency EM wave (i.e., neglecting terms $\propto(kr)^{-1}$) and for $r\gg r_g$, we derive the field near the optical axis, which up to terms of ${\cal O}(\rho^2/z^2)$, takes the form
{}
\begin{eqnarray}
    \left( \begin{aligned}
{E}_\rho& \\
{H}_\rho& \\
  \end{aligned} \right) =\left( \begin{aligned}
{H}_\phi& \\
-{E}_\phi& \\
  \end{aligned} \right) &=&
 E_0
 \sqrt{2\pi kr_g}e^{i\sigma_0}B(\rho,\phi)
  e^{i(kz -\omega t)}
 \left( \begin{aligned}
 \cos\phi& \\
 \sin\phi& \\
  \end{aligned} \right),
  \label{eq:DB-sol-rho}
\end{eqnarray}
with $({E}_z, {H}_z)= {\cal O}({\rho}/{z})$ and where $r=\sqrt{z^2+\rho^2}=z(1+{\rho^2}/{2z^2})=z+{\cal O}(\rho^2/z)$) and $\theta=\rho/z+{\cal O}(\rho^2/z^2)$. Note that these expressions were obtained  using the approximations (\ref{eq:tau-l=}) and are valid for forward scattering when $\theta\leq\sqrt{2r_g/r}$, or when $\rho\leq r_g$. For completeness, one may obtain a more general expression that will be valid for much larger deviations from the optical axis, say $\rho\sim R_\odot$. This work is ongoing and results will be reported.

Considering the image plane, we see that the quantity $B(\rho,\phi)$ in (\ref{eq:DB-sol-rho}) given by (\ref{eq:B2}) is a function of the coordinates on the image plane, $\vec x=\rho(\cos\phi,\sin\phi,0)$. Therefore, the entire amplitude of the EM wave, as a function of the coordinates on the image plane $B(\rho,\phi)\equiv B({\vec x}) $,  is given by a single integral (\ref{eq:B2}).

This is our main result. It determines the amplitude of the EM field in the image plane in the strong interference region of the SGL.
This function determines the structure of the point-spread function (PSF) of the SGL, which governs the optical properties of the SGL as far as imaging is concerned.  This expression describes light from a distant point source, projected onto the image plane by the SGL. Furthermore, it is presented in a form using units and parameters that relate directly to physically relevant quantities, making the result readily applicable to study gravitational lensing by real astrophysical objects, such as the Sun.

\subsection{Multipole contributions}

 \begin{figure}
\raisebox{3.2in}{a)}~\includegraphics{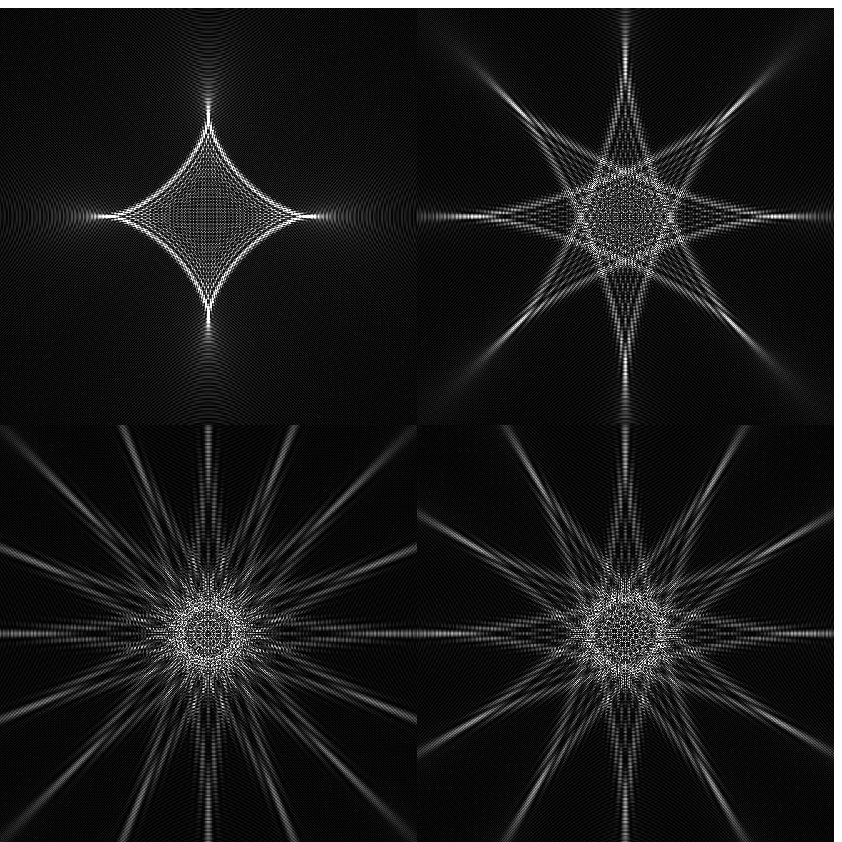}~\raisebox{3.2in}{b)}~\includegraphics{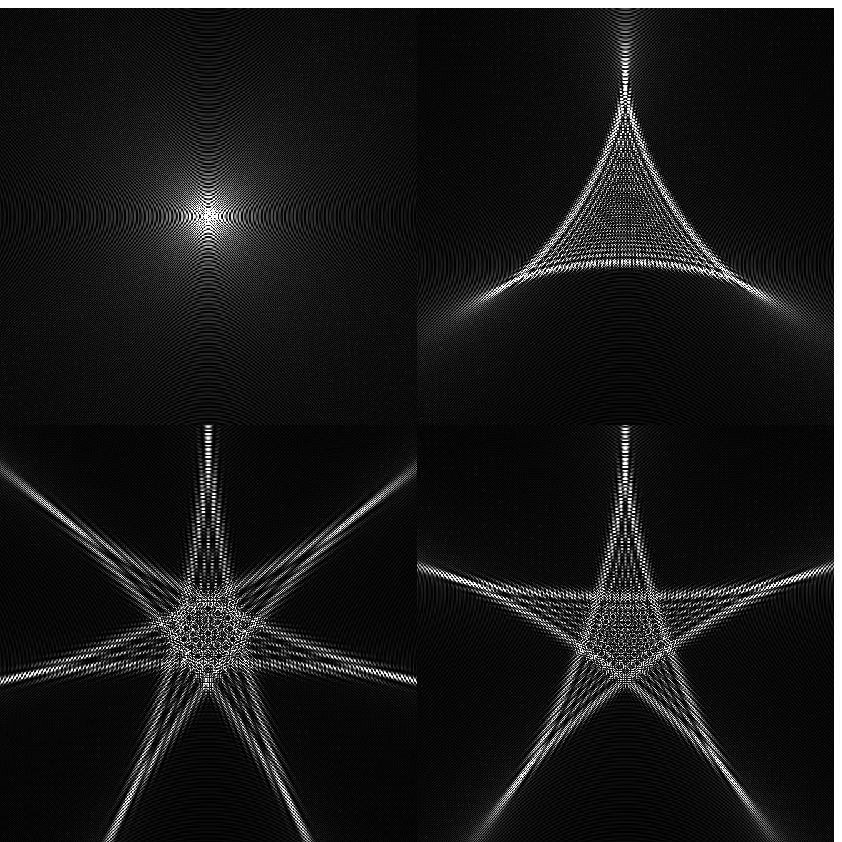}
\caption{\label{fig:caustics} Even and odd caustics representing individual contributions of the multipoles of a gravitational field to the PSF of the extended axisymmetric gravitational lens, obtained through numerical integration of ${\rm PSF}=|B({\vec x})|^2$ with $B({\vec x})$ from (\ref{eq:B20}). Images of a point source formed in the image plane of the lens. From top left, clockwise: a) $J_2$, $J_4$, $J_6$ and $J_8$; b) monopole, $J_3$, $J_5$ and $J_7$. For the odd-numbered caustics, a change in sign flips the image in the north-south direction.
}
\end{figure}

\begin{figure}
\includegraphics{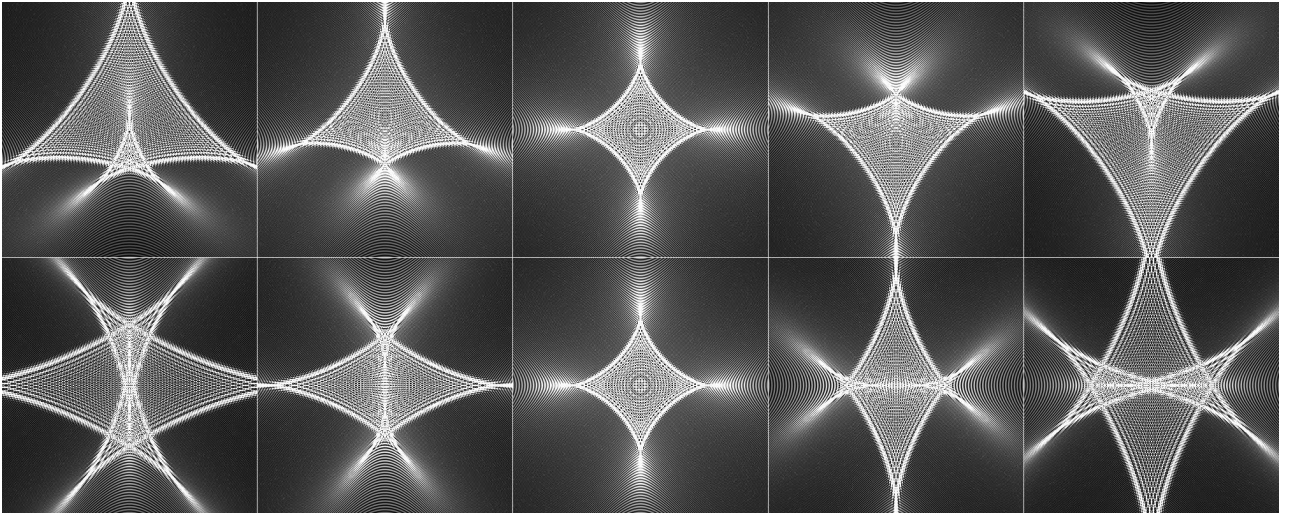}
\caption{\label{fig:J2J3J4}Interaction between caustics and the effects of sign, calculated by numerically integrating $|B({\vec x})|^2$ using (\ref{eq:B20}). Top row depicts the effect of $J_3$, distorting the $J_2$ astroid starting with a negative value similar in magnitude to $J_2$, going through 0 and reaching a positive value. Bottom row depicts the effect of $J_4$ on $J_2$ in a similar fashion. These images demonstrate that the sign of the $J_3$ caustic reverses its vertical, ``north--south'' orientation, whereas the sign of the $J_4$ caustic determines if it is the astroid's vertical or horizontal pair of cusps that are ``split'' as the astroid is stretched in the horizontal vs. vertical direction.}
\end{figure}

Using the result (\ref{eq:DB-sol-rho}), we may now compute the energy flux in the image region of the lens. With overline and brackets denoting time-averaging and ensemble averaging, the relevant components of the time-averaged Poynting vector for the EM field in the image volume may be given in the following form (see \cite{Turyshev-Toth:2017,Turyshev-Toth:2019,Turyshev-Toth:2020-extend} for details):
 {}
\begin{eqnarray}
S_z({\vec x})=\frac{c}{4\pi}\big<\overline{[{\rm Re}{\vec E}\times{\rm Re}{\vec H}]}_z\big>=\frac{c}{4\pi}E_0^2\,{2\pi kr_g}\,
\big<\overline{\big({\rm Re}\big[{ B}({\vec x})e^{i(kz-\omega t)}\big]\big)^2}\big>,
  \label{eq:S_z*6z}
\end{eqnarray}
with ${\bar S}_\rho= {\bar S}_\phi=0$ for all practical purposes. Defining the light amplification as usual \cite{Turyshev-Toth:2017,Turyshev-Toth:2019,Turyshev-Toth:2020-extend}, $\mu_z({\vec x})=S_z({\vec x})/|\vec S_0({\vec x})|$, where $\vec S_0({\vec x})$ being the Poynting vector carried by a plane wave in a vacuum in a flat space-time, we have the light amplification factor of the lens that, for short wavelengths (i.e., $kr_g\gg1$) is given as
 {}
\begin{eqnarray}
\mu_z({\vec x})={2\pi kr_g}  \, |B({\vec x})|^2,
  \label{eq:S_mu}
\end{eqnarray}
with $|B({\vec x})|^2=B({\vec x})B^*({\vec x})$, where $B^*(\vec x)$ is the complex conjugate of $B({\vec x})$, given by
(\ref{eq:B2}) that we repeat here for convenience:
{}
\begin{eqnarray}
B({\vec x}) &=&
\frac{1}{2\pi}\int_0^{2\pi} d\phi_\xi \exp\Big[-ik\Big(\sqrt{\frac{2r_g}{r}}\rho\cos(\phi_\xi-\phi)+
2r_g\sum_{n=2}^\infty \frac{J_n}{n} \Big(\frac{R_\odot }{\sqrt{2r_gr}}\Big)^n\sin^n\beta_s\cos[n(\phi_\xi-\phi_s)]\Big)\Big].~~~~
  \label{eq:B20}
\end{eqnarray}

As we see, the amplification of the lens is driven by the factor ${2\pi kr_g}$ in (\ref{eq:S_mu}). However, in the case of the monopole, the complex amplitude of the EM field (\ref{eq:B20}) was $\propto J_0(\sqrt{{2r_g}/{r}}\rho)$; thus it was reaching its maximum of 1 on the optical axis, $\rho=0$. In the case of an extended gravitating body, the complex amplitude is given by (\ref{eq:B20}), where $B({\vec x})$, in general, is a complex quantity whose magnitude is $|B({\vec x})|< 1$. As we see in Fig.~\ref{fig:caustics}, it reaches its maximum value not on the optical axis, but on the caustic that is formed in the image plane. For lenses dominated by the contribution of a single multipole moment, these caustics acquire the shapes of hypocycloids (e.g., the astroid, characterizing the $J_2$ quadrupole). However, when several multipole moments are present, their interaction results in more complex shapes; see Fig.~\ref{fig:J2J3J4} for some examples. In general, all the light from an extended source is still present in the image plane, but now it is scrambled. Image reconstruction will require deconvolution tools.

\begin{figure}
\includegraphics{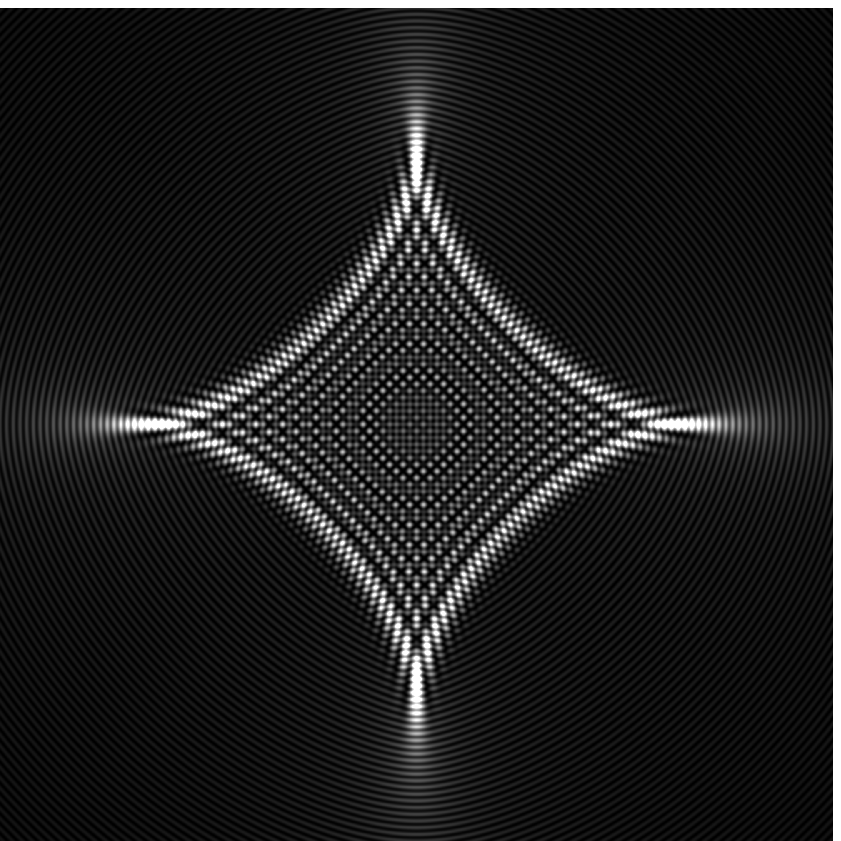}~\includegraphics{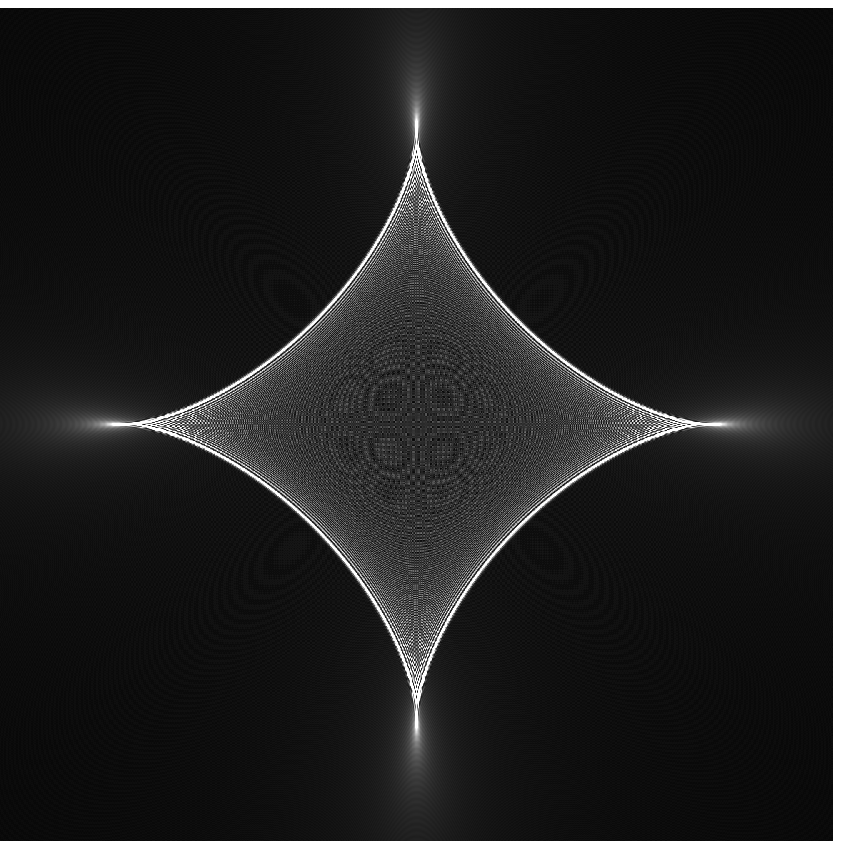}
\caption{\label{fig:example}The PSF of the SGL with multipole gravitational moments, obtained by integrating $|B({\vec x})|^2$ using (\ref{eq:B20}), using realistic solar parameters. These examples show light from a $\lambda=1~\mu$m point source, projected by the SGL to 650 AU from the Sun. Left: $\sin\beta_s=0.1$ (i.e., $\beta_s\sim 5.74^\circ$ from the solar axis of rotation) in an $8\times 8$ meter area; at a resolution of 4~mm, fine details due to diffraction are visible both inside and outside the caustic boundary. Right: $\sin\beta_s=0.387$ ($\beta_s\sim 22.78^\circ$) in a $120\times 120$~m area, at 6~cm resolution.
}
\end{figure}

The quantity $|B({\vec x})|^2$ is the point-spread function (PSF) that characterizes the optical properties of the gravitational lens and can be used to assess its imaging capabilities. The PSF of the lens is  extended from the $J^2_0(k\sqrt{{2r_g}/{r}}\rho)$ form for the monopole lens, discussed in  \cite{Turyshev-Toth:2017} and takes the form of $|B(\rho,\phi)|^2$ that now provides a complete description of the intensity distribution in the image plane and accounts for gravitational lensing by an arbitrary axisymmetric gravitational potential.

\subsection{Extended Sun contribution to image formation}
\label{sec:SGL-imaging}

When applying these results to the SGL, we need to recognize the fact that the Sun axisymmetric rotating body that also has north-south symmetry. As such, it will have only even multipole moments $J_{2n}$. The solar multipole moments are determined using available tracking data from interplanetary spacecraft: $J_2=(2.25\pm0.09)\times 10^{-7}$ \cite{Park-etal:2017},  and $J_4=-4.44\times 10^{-9}$, $J_6=-2.79\times 10^{-10}$, $J_8=1.48\times 10^{-11}$ \cite{Roxburgh:2001}. The deflection of light by these multipoles may lead to light rays missing the optical axis by many meters, resulting in large caustics on the image plane in the strong interference region of the SGL. With the contribution from $J_2$ being the dominant one to consider (see Fig.~\ref{fig:example}), depending on the  target's position with respect to the solar rotational axis (captured by the angle $\beta_s$), some of these multipoles may be needed for developing a comprehensive physical model needed for image deconvolution. The multipole moments of the Sun may also be varying temporally \cite{Rozelot:2019}, which requires further analysis. On the other hand, the magnitudes of light deflection due to $J_{10}$ and higher multipoles are very small at IR, optical or longer wavelengths. These fall within the diffraction pattern of the solar monopole, and thus may be omitted.

With these considerations in mind, the most comprehensive form of the complex amplitude of the EM field in the strong interference region of the SGL is given by (\ref{eq:B20}) were multipole summation is from $n=2$ to $n=8$, which is correct to the order of ${\cal O}(J_{10})$.

Clearly, there is no closed-form analytical solution for this integral. It can, however, be readily evaluated using numerical methods. It is also clear that, as $J_4, J_6, J_8$ are small, the resulting diffraction pattern will be dominated by the quadrupole with other multipoles contributing only small corrections (see Fig.~\ref{fig:example}).

\begin{figure}
\includegraphics{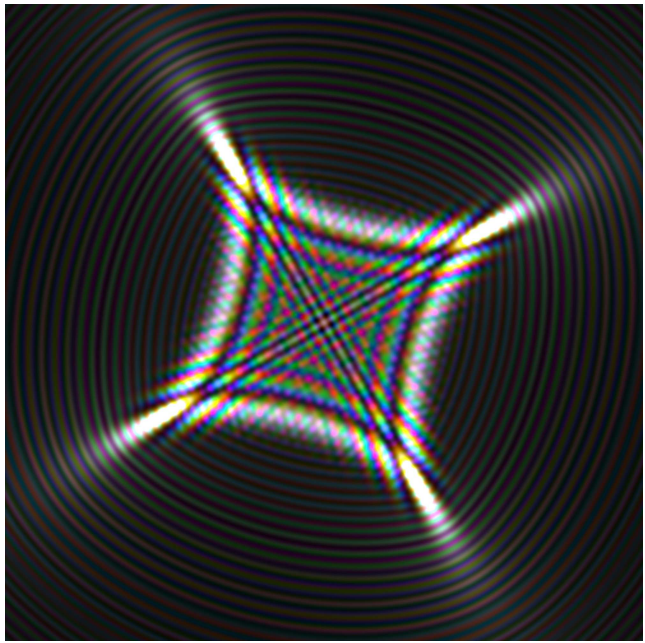}
\caption{\label{fig:rainbow}The PSF of the SGL in color, evaluated in multiple wavelengths between 400 and 675~nm in 25~nm increments, each assigned an approximate RGB color. The parameters used are $\sin\beta_s=0.05$ ($\beta_s\sim 2.87^\circ$) at 650~AU from the Sun, $\phi_s=30^\circ$; a $2\times 2$ meter area is depicted at 1~mm resolution. Whereas a rainbow pattern is visible in many parts of the image, the cusps are achromatic white, indicating that their position and appearance is not wavelength-dependent.
}
\end{figure}

The result (\ref{eq:B20}) depends on the wavelength of incident light. However, the geometric shape of the resulting caustic is wavelength-independent. This becomes evident when we evaluate (\ref{eq:B20}) in multiple wavelengths (Fig.~\ref{fig:rainbow}). Wavelength-dependent features (represented by approximate RGB color in this figure) are clearly evident both inside and outside the caustic boundary. However, the caustic boundary's location does not change, and the cusps, in particular, are achromatic white.

Finally, we mention that expression (\ref{eq:B20}) may be easily generalized to the case of extended sources at large but finite distances from the Sun.  Examining this integral, we see that it contains the expression $\rho\cos(\phi_\xi-\phi)$, which may be transformed as
{}
\begin{eqnarray}
\rho\cos(\phi_\xi-\phi)=({\vec n}_\xi\cdot{\vec x}),~~~{\rm where}~~~ {\vec x}=\rho(\cos\phi, \sin\phi, 0).
\label{eq:rn}
\end{eqnarray}

The form (\ref{eq:B20}) allows us to extend the new formulation to the case of sources at large but finite distances, $z_0$. A formal way to extend the result (\ref{eq:B20}) to the case of extended source is to rotate the coordinate system by a small angle $({\bar z}/{z_0})\vec x'$, as discussed in \cite{Turyshev-Toth:2019-fin-difract}, where ${\bar z}$ is the heliocentric distance to the image in the strong interference region of the SGL, $z_0$ is the heliocentric distance to the source plane and $\vec x'$ is a particular point on that source plane. As a  result, to deal with extended sources we start with (\ref{eq:B20}) and extend the argument as follows:
{}
\begin{eqnarray}
\vec x \qquad \Rightarrow \qquad \vec x+\frac{\bar z}{z_0}\vec x',~~~{\rm where}~~~ {\vec x}'=\rho'(\cos\phi', \sin\phi', 0),
\label{eq:rn0ex}
\end{eqnarray}
which, equivalently, may be expressed as $({\vec n}_\xi\cdot{\vec x}) \rightarrow ({\vec n}_\xi\cdot{\vec x})+({\bar z}/{z_0})({\vec n}_\xi\cdot{\vec x}')$, where $\vec n_\xi=(\cos\phi_\xi,\sin\xi_\xi,0)$. As a result, this rotation leads to a modification of the expression (\ref{eq:B20}) for the amplitude of the EM field, which now takes the form
{}
\begin{eqnarray}
B({\vec x},{\vec x}')&=&
\frac{1}{2\pi}\int_0^{2\pi} d\phi_\xi \exp\Big[-ik\Big\{\sqrt{\frac{2r_g}{\bar z}}\Big({\vec n}_\xi\cdot\big({\vec x}+\frac{\bar z}{z_0}{\vec x}'\big)\Big)+\nonumber\\
&&\hskip 100pt +\,
2r_g\sum_{n=2}^\infty \frac{J_n}{n} \Big(\frac{R_\odot }{\sqrt{2r_g\bar z}}\Big)^n\sin^n\beta_s\cos[n(\phi_\xi-\phi_s)]\Big\}\Big].~~~~
  \label{eq:B22+}
\end{eqnarray}

This expression allows us to consider imaging of extended bodies that are positioned at large, but finite distances from the SGL, with the SGL now treated as that produced by a gravitating body that is axisymmetric and rotating thus admitting characterization of its external gravitational field by zonal harmonics.

\section{Discussion and Conclusions}
\label{sec:end}

This paper represents a continuation of our efforts to provide a reliable, accurate, complete theoretical description of the image formation capabilities of gravitational lenses within the post-Newtonian approximation of the general theory of relativity. This work is especially relevant to our on-going work on the study of the optical properties of the SGL in the context of use for a resolved imaging of distant faint sources.

In previous papers \cite{Turyshev:2017,Turyshev-Toth:2017}, we offered a complete wave-theoretical description of the SGL under the simplifying assumption that the Sun's gravitational field is accurately represented as a gravitational monopole that was modeled as a point mass. Clearly, this is not exactly the case: the actual gravitational field of the Sun deviates from the monopole slightly. Though the effect is very small compared to the size of the solar system, it has considerable impact on the image formation capabilities of the SGL. Therefore, an accurate and complete description of the SGL must properly take into account these small deviations from spherical symmetry.

It was long understood that the tools of geometric optics are limited when it comes to caustics and the full wave-optical treatment is required \cite{Berry-Upstill:1982,Berry:1992}. It was in light of this limitation that we developed our new method to describe gravitational lensing within the weak-field and slow motion (i.e., a post-Newtonian) approximation of the general theory of relativity. The new method addresses light propagation in a weak gravitational field of arbitrary shape, not restricted by spherical symmetry. Our formalism allows us to describe the contribution of deviations from spherical symmetry on the optical properties of corresponding lens using the language of spherical harmonics. In particular, we can use zonal harmonics in the case of an axisymmetric body, such as the Sun.

Key to our approach is what we dubbed the {\em angular eikonal method}: a convolution of the eikonal phase (which is used to characterize deviations from spherical symmetry) and integral representations of Bessel-functions (that rely on the symmetries that exist in the case of a monopole lens). This allows us to correctly capture the functional dependence of the integrand and, in effect, to solve the wave equations within a slightly modified symmetry that is extended from spherical to an azimuthally perturbed one (that is due to the presence of the multipole moments.) The method is consistent with the thin lens or eikonal approximations used with in the scalar theory of diffraction \cite{Landau-Lifshitz:1988,Born-Wolf:1999,Sharma-etal:1988,Sharma-Sommerford:1990,Sharma-Sommerford-book:2006} that are frequently used to describe gravitational  lensing.

Our method preserves the structure and vector nature of the EM field and allows us to treat the diffracted EM field using regular tools of modern optics \cite{Born-Wolf:1999}. The entire diffraction behavior is captured in the form of a single integral (\ref{eq:B20}), which extends the set of analytical tools developed for gravitational lenses. The approach that we present and and the resulting expressions are applicable to a wide variety of astrophysical lenses.
Applying the method to the case of an axisymmetric body, represented using zonal harmonics, we arrive at our main result, Eq.~(\ref{eq:B22+}), which reduces the problem of the finding the EM field in the image plane placed in the string interference region of the SGL. This solution preserves the vector nature of the EM field, thus, going beyond the approaches relying on the scalar theories.

An important outcomes of the new solution is that it allows us to evaluate the behavior of the PSF of an extended gravitating lens. Applying this method to the SGL, we treat the solar gravitational field as that produced by an axisymmetric rotating body whose external gravity field is determined by the infinite set of zonal spherical harmonics. The PSF of the SGL is now determined by a single, well-behaved integral that can be readily evaluated using numerical methods, especially near the optical axis of the gravitational lens in what we call the region of strong interference.

Concerning the imaging of extended sources with the SGL of the extended Sun, we note that the total energy deposited in the image plane is still almost the same as it was in the case of the monopole SGL. However, the PSF of the extended SGL scrambles light on the image plane more than it did in the case of treating the Sun as the point mass. This will adversely affect the signal-to-noice ratio as far as as the realistic imaging capabilities of the extended SGL are concerned. The impact on the observing scenario and the integration time  are being investigated.

Concluding, we note that the new method can be used to investigate image formation processes for extended sources by the SGL, at a variety of wavelengths, using physically realistic observational scenarios. The approach that we presented may also be used in reverse: observing astrophysical lensing of distant objects may allow one to reconstruct the multipoles of the gravitational field of the lens and infer its mass distribution, possibly offering a new practical method in modern astrophysics. Our solution may also help in other areas, such as the modeling of particle collisions in high energy particle physics experiments on potentials with complex structure. Results of our studies in these and other directions, once available, will be published elsewhere.

\begin{acknowledgments}
This work in part was performed at the Jet Propulsion Laboratory, California Institute of Technology, under a contract with the National Aeronautics and Space Administration.
VTT acknowledges the generous support of Plamen Vasilev and other Patreon patrons.

\end{acknowledgments}


\begin{thebibliography}{84}
\expandafter\ifx\csname natexlab\endcsname\relax\def\natexlab#1{#1}\fi
\expandafter\ifx\csname bibnamefont\endcsname\relax
  \def\bibnamefont#1{#1}\fi
\expandafter\ifx\csname bibfnamefont\endcsname\relax
  \def\bibfnamefont#1{#1}\fi
\expandafter\ifx\csname citenamefont\endcsname\relax
  \def\citenamefont#1{#1}\fi
\expandafter\ifx\csname url\endcsname\relax
  \def\url#1{\texttt{#1}}\fi
\expandafter\ifx\csname urlprefix\endcsname\relax\def\urlprefix{URL }\fi
\providecommand{\bibinfo}[2]{#2}
\providecommand{\eprint}[2][]{\url{#2}}

\bibitem[{\citenamefont{{Einstein}}(1916)}]{Einstein-1916}
\bibinfo{author}{\bibfnamefont{A.}~\bibnamefont{{Einstein}}},
  \bibinfo{journal}{Annalen der Physik} \textbf{\bibinfo{volume}{49}},
  \bibinfo{pages}{146} (\bibinfo{year}{1916}).

\bibitem[{\citenamefont{Einstein}(1936)}]{Einstein:1936}
\bibinfo{author}{\bibfnamefont{A.}~\bibnamefont{Einstein}},
  \bibinfo{journal}{Science} \textbf{\bibinfo{volume}{84}},
  \bibinfo{pages}{506} (\bibinfo{year}{1936}).

\bibitem[{\citenamefont{Liebes}(1964)}]{Liebes:1964}
\bibinfo{author}{\bibfnamefont{S.}~\bibnamefont{Liebes}},
  \bibinfo{journal}{Phys. Rev.} \textbf{\bibinfo{volume}{133}},
  \bibinfo{pages}{B835} (\bibinfo{year}{1964}).

\bibitem[{\citenamefont{{Refsdal}}(1964)}]{Refsdal:1964}
\bibinfo{author}{\bibfnamefont{S.}~\bibnamefont{{Refsdal}}},
  \bibinfo{journal}{MNRAS} \textbf{\bibinfo{volume}{128}}, \bibinfo{pages}{307}
  (\bibinfo{year}{1964}).

\bibitem[{\citenamefont{{Schneider} et~al.}(1992)\citenamefont{{Schneider},
  {Ehlers}, and {Falco}}}]{Schneider-Ehlers-Falco:1992}
\bibinfo{author}{\bibfnamefont{P.~S.} \bibnamefont{{Schneider}}},
  \bibinfo{author}{\bibfnamefont{J.}~\bibnamefont{{Ehlers}}}, \bibnamefont{and}
  \bibinfo{author}{\bibfnamefont{E.}~\bibnamefont{{Falco}}},
  \emph{\bibinfo{title}{Gravitational Lenses}}
  (\bibinfo{publisher}{Springer-Verlag Berlin Heidelberg},
  \bibinfo{year}{1992}).

\bibitem[{\citenamefont{{Fock}}(1959)}]{Fock-book:1959}
\bibinfo{author}{\bibfnamefont{V.~A.} \bibnamefont{{Fock}}},
  \emph{\bibinfo{title}{The Theory of Space, Time and Gravitation}}
  (\bibinfo{publisher}{Fizmatgiz}, \bibinfo{address}{Moscow (in Russian)},
  \bibinfo{year}{1959}), \bibinfo{note}{{[English translation (1959), Pergamon,
  Oxford]}}.

\bibitem[{\citenamefont{{Landau} and
  {Lifshitz}}(1988{\natexlab{a}})}]{Landau-Lifshitz:1988}
\bibinfo{author}{\bibfnamefont{L.~D.} \bibnamefont{{Landau}}} \bibnamefont{and}
  \bibinfo{author}{\bibfnamefont{E.~M.} \bibnamefont{{Lifshitz}}},
  \emph{\bibinfo{title}{The Classical Theory of Fields.}}
  (\bibinfo{publisher}{7th edition. Nauka: Moscow (in Russian)},
  \bibinfo{year}{1988}{\natexlab{a}}).

\bibitem[{\citenamefont{{Turyshev}}(2017)}]{Turyshev:2017}
\bibinfo{author}{\bibfnamefont{S.~G.} \bibnamefont{{Turyshev}}},
  \bibinfo{journal}{Phys. Rev. D} \textbf{\bibinfo{volume}{95}},
  \bibinfo{pages}{084041} (\bibinfo{year}{2017}), \eprint{arXiv:1703.05783
  [gr-qc]}.

\bibitem[{\citenamefont{{Turyshev} and {Toth}}(2017)}]{Turyshev-Toth:2017}
\bibinfo{author}{\bibfnamefont{S.~G.} \bibnamefont{{Turyshev}}}
  \bibnamefont{and} \bibinfo{author}{\bibfnamefont{V.~T.}
  \bibnamefont{{Toth}}}, \bibinfo{journal}{Phys. Rev. D}
  \textbf{\bibinfo{volume}{96}}, \bibinfo{pages}{024008}
  (\bibinfo{year}{2017}), \eprint{arXiv:1704.06824 [gr-qc]}.

\bibitem[{\citenamefont{{Turyshev} and
  {Toth}}(2020{\natexlab{a}})}]{Turyshev-Toth:2020-extend}
\bibinfo{author}{\bibfnamefont{S.~G.} \bibnamefont{{Turyshev}}}
  \bibnamefont{and} \bibinfo{author}{\bibfnamefont{V.~T.}
  \bibnamefont{{Toth}}}, \bibinfo{journal}{Phys. Rev. D}
  \textbf{\bibinfo{volume}{102}}, \bibinfo{pages}{024038}
  (\bibinfo{year}{2020}{\natexlab{a}}), \bibinfo{note}{arXiv:2002.06492
  [astro-ph.IM]}.

\bibitem[{\citenamefont{Turyshev and Toth}(2019)}]{Turyshev-Toth:2019}
\bibinfo{author}{\bibfnamefont{S.~G.} \bibnamefont{Turyshev}} \bibnamefont{and}
  \bibinfo{author}{\bibfnamefont{V.~T.} \bibnamefont{Toth}},
  \bibinfo{journal}{Phys. Rev. D} \textbf{\bibinfo{volume}{99}},
  \bibinfo{pages}{024044} (\bibinfo{year}{2019}), \eprint{arXiv:1810.06627
  [gr-qc]}.

\bibitem[{\citenamefont{{Turyshev} and
  {Toth}}(2019{\natexlab{a}})}]{Turyshev-Toth:2018-plasma}
\bibinfo{author}{\bibfnamefont{S.~G.} \bibnamefont{{Turyshev}}}
  \bibnamefont{and} \bibinfo{author}{\bibfnamefont{V.~T.}
  \bibnamefont{{Toth}}}, \bibinfo{journal}{J. Optics}
  \textbf{\bibinfo{volume}{21}}, \bibinfo{pages}{045601}
  (\bibinfo{year}{2019}{\natexlab{a}}), \bibinfo{note}{arXiv:1805.00398
  [physics.optics]}.

\bibitem[{\citenamefont{{Turyshev} and
  {Toth}}(2019{\natexlab{b}})}]{Turyshev-Toth:2019-fin-difract}
\bibinfo{author}{\bibfnamefont{S.~G.} \bibnamefont{{Turyshev}}}
  \bibnamefont{and} \bibinfo{author}{\bibfnamefont{V.~T.}
  \bibnamefont{{Toth}}}, \bibinfo{journal}{Phys. Rev. D}
  \textbf{\bibinfo{volume}{100}}, \bibinfo{pages}{084018}
  (\bibinfo{year}{2019}{\natexlab{b}}), \bibinfo{note}{arXiv:1908.01948
  [gr-qc]}.

\bibitem[{\citenamefont{{Turyshev} and
  {Toth}}(2020{\natexlab{b}})}]{Turyshev-Toth:2020-photom}
\bibinfo{author}{\bibfnamefont{S.~G.} \bibnamefont{{Turyshev}}}
  \bibnamefont{and} \bibinfo{author}{\bibfnamefont{V.~T.}
  \bibnamefont{{Toth}}}, \bibinfo{journal}{Phys. Rev. D}
  \textbf{\bibinfo{volume}{101}}, \bibinfo{pages}{044025}
  (\bibinfo{year}{2020}{\natexlab{b}}), \bibinfo{note}{arXiv:1909.03116
  [gr-qc]}.

\bibitem[{\citenamefont{{Turyshev} and
  {Toth}}(2020{\natexlab{c}})}]{Turyshev-Toth:2020-image}
\bibinfo{author}{\bibfnamefont{S.~G.} \bibnamefont{{Turyshev}}}
  \bibnamefont{and} \bibinfo{author}{\bibfnamefont{V.~T.}
  \bibnamefont{{Toth}}}, \bibinfo{journal}{Phys. Rev. D}
  \textbf{\bibinfo{volume}{101}}, \bibinfo{pages}{044048}
  (\bibinfo{year}{2020}{\natexlab{c}}), \bibinfo{note}{arXiv:1911.03260
  [gr-qc]}.

\bibitem[{\citenamefont{{Toth} and {Turyshev}}(2020)}]{Toth-Turyshev:2020}
\bibinfo{author}{\bibfnamefont{V.~T.} \bibnamefont{{Toth}}} \bibnamefont{and}
  \bibinfo{author}{\bibfnamefont{S.~G.} \bibnamefont{{Turyshev}}},
  \bibinfo{journal}{Phys. Rev. D}  (\bibinfo{year}{2020}),
  \bibinfo{note}{arXiv:2012.05477 [gr-qc]}.

\bibitem[{\citenamefont{{Turyshev} et~al.}(2020)\citenamefont{{Turyshev},
  {Shao}, {Toth}, and et~al.}}]{Turyshev-etal:2020-PhaseII}
\bibinfo{author}{\bibfnamefont{S.~G.} \bibnamefont{{Turyshev}}},
  \bibinfo{author}{\bibfnamefont{M.}~\bibnamefont{{Shao}}},
  \bibinfo{author}{\bibfnamefont{V.~T.} \bibnamefont{{Toth}}},
  \bibnamefont{and} \bibinfo{author}{\bibnamefont{et~al.}},
  \emph{\bibinfo{title}{Direct multipixel imaging and spectroscopy of an
  exoplanet with a solar gravity lens mission}} (\bibinfo{year}{2020}),
  \bibinfo{note}{arXiv:1908.01948 [gr-qc]}.

\bibitem[{\citenamefont{{Turyshev} and {Toth}}(2015)}]{Turyshev-Toth:2013}
\bibinfo{author}{\bibfnamefont{S.~G.} \bibnamefont{{Turyshev}}}
  \bibnamefont{and} \bibinfo{author}{\bibfnamefont{V.~T.}
  \bibnamefont{{Toth}}}, \bibinfo{journal}{Int. J. Mod. Phys.}
  \textbf{\bibinfo{volume}{D24}}, \bibinfo{pages}{1550039}
  (\bibinfo{year}{2015}), \eprint{arXiv:1304.8122 [gr-qc]}.

\bibitem[{\citenamefont{Will}(1993)}]{Will_book93}
\bibinfo{author}{\bibfnamefont{C.~M.} \bibnamefont{Will}},
  \emph{\bibinfo{title}{Theory and Experiment in Gravitational Physics}}
  (\bibinfo{publisher}{Cambridge University Press},
  \bibinfo{address}{Cambridge, UK}, \bibinfo{year}{1993}).

\bibitem[{\citenamefont{{Asada} and {Kasai}}(2020)}]{Asada-Kasai:2020}
\bibinfo{author}{\bibfnamefont{H.}~\bibnamefont{{Asada}}} \bibnamefont{and}
  \bibinfo{author}{\bibfnamefont{M.}~\bibnamefont{{Kasai}}},
  \bibinfo{journal}{Prog. Theor. Phys.} \textbf{\bibinfo{volume}{104}},
  \bibinfo{pages}{95} (\bibinfo{year}{2020}).

\bibitem[{\citenamefont{Mie}(1908)}]{Mie:1908}
\bibinfo{author}{\bibfnamefont{G.}~\bibnamefont{Mie}},
  \bibinfo{journal}{Annalen der Physik} \textbf{\bibinfo{volume}{25}},
  \bibinfo{pages}{377} (\bibinfo{year}{1908}).

\bibitem[{\citenamefont{{Born} and {Wolf}}(October 13, 1999)}]{Born-Wolf:1999}
\bibinfo{author}{\bibfnamefont{M.}~\bibnamefont{{Born}}} \bibnamefont{and}
  \bibinfo{author}{\bibfnamefont{E.}~\bibnamefont{{Wolf}}},
  \emph{\bibinfo{title}{Principles of Optics: Electromagnetic Theory of
  Propagation, Interference and Diffraction of Light}}
  (\bibinfo{publisher}{Cambridge University Press; 7th edition},
  \bibinfo{year}{October 13, 1999}).

\bibitem[{\citenamefont{{Schiff}}(1968)}]{Schiff:1968}
\bibinfo{author}{\bibfnamefont{L.~I.} \bibnamefont{{Schiff}}},
  \emph{\bibinfo{title}{Quantum mechanics.}} (\bibinfo{publisher}{McGraw-Hill},
  \bibinfo{year}{1968}).

\bibitem[{\citenamefont{{Landau} and {Lifshitz}}(1989)}]{Landau-Lifshitz:1989}
\bibinfo{author}{\bibfnamefont{L.~D.} \bibnamefont{{Landau}}} \bibnamefont{and}
  \bibinfo{author}{\bibfnamefont{E.~M.} \bibnamefont{{Lifshitz}}},
  \emph{\bibinfo{title}{Quantum mechanics. Non-Relativistic Theory.}}
  (\bibinfo{publisher}{4th edition. Nauka: Moscow (in Russian)},
  \bibinfo{year}{1989}).

\bibitem[{\citenamefont{Messiah}(1968)}]{Messiah:1968}
\bibinfo{author}{\bibfnamefont{A.}~\bibnamefont{Messiah}},
  \emph{\bibinfo{title}{Quantum Mechanics, Vol 1}} (\bibinfo{publisher}{John
  Wiley \& Sons}, \bibinfo{year}{1968}).

\bibitem[{\citenamefont{{Friedrich}}(2006)}]{Friedrich-book-2006}
\bibinfo{author}{\bibfnamefont{H.}~\bibnamefont{{Friedrich}}},
  \emph{\bibinfo{title}{Theoretical Atomic Physics, 3-ed}}
  (\bibinfo{publisher}{Springer-Verlag}, \bibinfo{address}{Berlin, Heidelberg},
  \bibinfo{year}{2006}).

\bibitem[{\citenamefont{{Friedrich}}(2013)}]{Friedrich-book-2013}
\bibinfo{author}{\bibfnamefont{H.}~\bibnamefont{{Friedrich}}},
  \emph{\bibinfo{title}{Scattering Theory}}
  (\bibinfo{publisher}{Springer-Verlag}, \bibinfo{address}{Berlin, Heidelberg},
  \bibinfo{year}{2013}).

\bibitem[{\citenamefont{{Burke}}(2011{\natexlab{a}})}]{Burke-book-2011}
\bibinfo{author}{\bibfnamefont{P.~G.} \bibnamefont{{Burke}}},
  \emph{\bibinfo{title}{R-Matrix Theory of Atomic Collisions. Application to
  Atomic, Molecular and Optical Processes}}
  (\bibinfo{publisher}{Springer-Verlag}, \bibinfo{address}{Berlin, Heidelberg},
  \bibinfo{year}{2011}{\natexlab{a}}).

\bibitem[{\citenamefont{{Sharma} et~al.}(1988)\citenamefont{{Sharma}, {Roy},
  and {Sommerford}}}]{Sharma-etal:1988}
\bibinfo{author}{\bibfnamefont{S.~K.} \bibnamefont{{Sharma}}},
  \bibinfo{author}{\bibfnamefont{T.~K.} \bibnamefont{{Roy}}}, \bibnamefont{and}
  \bibinfo{author}{\bibfnamefont{D.~J.} \bibnamefont{{Sommerford}}},
  \bibinfo{journal}{Journal of Physics D: Applied Physics}
  \textbf{\bibinfo{volume}{21}}, \bibinfo{pages}{1685} (\bibinfo{year}{1988}).

\bibitem[{\citenamefont{{Sharma} and
  {Sommerford}}(1990)}]{Sharma-Sommerford:1990}
\bibinfo{author}{\bibfnamefont{S.~K.} \bibnamefont{{Sharma}}} \bibnamefont{and}
  \bibinfo{author}{\bibfnamefont{D.~J.} \bibnamefont{{Sommerford}}},
  \bibinfo{journal}{Il Nuovo Cimento D} \textbf{\bibinfo{volume}{12}},
  \bibinfo{pages}{719} (\bibinfo{year}{1990}).

\bibitem[{\citenamefont{{Sharma} and
  {Sommerford}}(2006)}]{Sharma-Sommerford-book:2006}
\bibinfo{author}{\bibfnamefont{S.~K.} \bibnamefont{{Sharma}}} \bibnamefont{and}
  \bibinfo{author}{\bibfnamefont{D.~J.} \bibnamefont{{Sommerford}}},
  \emph{\bibinfo{title}{Light Scattering by Optically Soft Particles: Theory
  and Applications}} (\bibinfo{publisher}{Springer-Verlag},
  \bibinfo{address}{Berlin, Heidelberg, New York}, \bibinfo{year}{2006}).

\bibitem[{\citenamefont{{Semon} and {Taylor}}(1977)}]{Semon-Taylor:1977}
\bibinfo{author}{\bibfnamefont{M.~D.} \bibnamefont{{Semon}}} \bibnamefont{and}
  \bibinfo{author}{\bibfnamefont{J.~R.} \bibnamefont{{Taylor}}},
  \bibinfo{journal}{Phys. Rev. A} \textbf{\bibinfo{volume}{16}},
  \bibinfo{pages}{33} (\bibinfo{year}{1977}).

\bibitem[{\citenamefont{{Grandy Jr.}}(2000)}]{Grandy-book-2000}
\bibinfo{author}{\bibfnamefont{W.~T.} \bibnamefont{{Grandy Jr.}}},
  \emph{\bibinfo{title}{Scattering of Waves from Large Spheres}}
  (\bibinfo{publisher}{Cambridge University Press}, \bibinfo{year}{2000}).

\bibitem[{\citenamefont{{Kopeikin}}(1997)}]{Kopeikin:1997}
\bibinfo{author}{\bibfnamefont{S.~M.} \bibnamefont{{Kopeikin}}},
  \bibinfo{journal}{J. Math. Phys.} \textbf{\bibinfo{volume}{38}},
  \bibinfo{pages}{2587} (\bibinfo{year}{1997}).

\bibitem[{\citenamefont{{Kopeikin} et~al.}(2011)\citenamefont{{Kopeikin},
  {Efroimsky}, and {Kaplan}}}]{Kopeikin-book-2011}
\bibinfo{author}{\bibfnamefont{S.~M.} \bibnamefont{{Kopeikin}}},
  \bibinfo{author}{\bibfnamefont{M.}~\bibnamefont{{Efroimsky}}},
  \bibnamefont{and} \bibinfo{author}{\bibfnamefont{G.}~\bibnamefont{{Kaplan}}},
  \emph{\bibinfo{title}{{Relativistic Celestial Mechanics of the Solar
  System}}} (\bibinfo{publisher}{Wiley-VCH, Berlin}, \bibinfo{year}{2011}).

\bibitem[{\citenamefont{{Feynman}}(1948)}]{Feynman:1948}
\bibinfo{author}{\bibfnamefont{R.~P.} \bibnamefont{{Feynman}}},
  \bibinfo{journal}{Rev. of Mod. Phys.} \textbf{\bibinfo{volume}{20}},
  \bibinfo{pages}{367} (\bibinfo{year}{1948}).

\bibitem[{\citenamefont{{Feynman} and {Hibbs}}(1965)}]{Feynman-Hibbs:1991}
\bibinfo{author}{\bibfnamefont{R.}~\bibnamefont{{Feynman}}} \bibnamefont{and}
  \bibinfo{author}{\bibfnamefont{A.}~\bibnamefont{{Hibbs}}},
  \emph{\bibinfo{title}{Quantum Mechanics and Path Integrals}}
  (\bibinfo{publisher}{McGraw-Hill}, \bibinfo{address}{New York},
  \bibinfo{year}{1965}).

\bibitem[{\citenamefont{{Nakamura} and
  {Deguchi}}(1999)}]{Nakamura-Deguchi:1999}
\bibinfo{author}{\bibfnamefont{T.~T.} \bibnamefont{{Nakamura}}}
  \bibnamefont{and}
  \bibinfo{author}{\bibfnamefont{S.}~\bibnamefont{{Deguchi}}},
  \bibinfo{journal}{Prog. Theor. Phys. Supp.} \textbf{\bibinfo{volume}{133}},
  \bibinfo{pages}{137} (\bibinfo{year}{1999}).

\bibitem[{\citenamefont{{Yamamoto}}(2017)}]{Yamamoto:2017}
\bibinfo{author}{\bibfnamefont{K.}~\bibnamefont{{Yamamoto}}},
  \bibinfo{journal}{Int. J. Astron. Astrophys.} \textbf{\bibinfo{volume}{7}},
  \bibinfo{pages}{221} (\bibinfo{year}{2017}).

\bibitem[{\citenamefont{{Roxburgh}}(2001)}]{Roxburgh:2001}
\bibinfo{author}{\bibfnamefont{I.~W.} \bibnamefont{{Roxburgh}}},
  \bibinfo{journal}{Astron. Astrophys.} \textbf{\bibinfo{volume}{377}},
  \bibinfo{pages}{688} (\bibinfo{year}{2001}).

\bibitem[{\citenamefont{Le~Poncin-Lafitte and
  Teyssandier}(2008)}]{LePoncinLafitte:2007tx}
\bibinfo{author}{\bibfnamefont{C.}~\bibnamefont{Le~Poncin-Lafitte}}
  \bibnamefont{and}
  \bibinfo{author}{\bibfnamefont{P.}~\bibnamefont{Teyssandier}},
  \bibinfo{journal}{Phys. Rev. D} \textbf{\bibinfo{volume}{77}},
  \bibinfo{pages}{044029} (\bibinfo{year}{2008}).

\bibitem[{\citenamefont{{Hull Jr.} and {Breit}}(1959)}]{Hull-Breit:1959}
\bibinfo{author}{\bibfnamefont{M.~H.} \bibnamefont{{Hull Jr.}}}
  \bibnamefont{and} \bibinfo{author}{\bibfnamefont{D.~G.}
  \bibnamefont{{Breit}}}, in \emph{\bibinfo{booktitle}{Handbuch der Physik,
  vol. 41/1}}, edited by
  \bibinfo{editor}{\bibfnamefont{S.}~\bibnamefont{{Flugge}}}
  (\bibinfo{publisher}{Springer}, \bibinfo{address}{Berlin, Gottingen,
  Heidelberg}, \bibinfo{year}{1959}), p. \bibinfo{pages}{408}.

\bibitem[{\citenamefont{{Thompson} and
  {Nunes}}(2009)}]{Thomson-Nunes-book:2009}
\bibinfo{author}{\bibfnamefont{I.~J.} \bibnamefont{{Thompson}}}
  \bibnamefont{and} \bibinfo{author}{\bibfnamefont{F.~M.}
  \bibnamefont{{Nunes}}}, \emph{\bibinfo{title}{Nuclear Reactions for
  Astrophysics: Principles, Calculation and Applications of Low-Energy
  Reactions}} (\bibinfo{publisher}{Cambridge University Press},
  \bibinfo{year}{2009}), \bibinfo{edition}{1st} ed.

\bibitem[{\citenamefont{{Turyshev} and
  {Toth}}(2018{\natexlab{a}})}]{Turyshev-Toth:2018-grav-shadow}
\bibinfo{author}{\bibfnamefont{S.~G.} \bibnamefont{{Turyshev}}}
  \bibnamefont{and} \bibinfo{author}{\bibfnamefont{V.~T.}
  \bibnamefont{{Toth}}}, \bibinfo{journal}{Phys. Rev. D}
  \textbf{\bibinfo{volume}{98}}, \bibinfo{pages}{104015}
  (\bibinfo{year}{2018}{\natexlab{a}}), \bibinfo{note}{arXiv:1805.10581
  [gr-qc]}.

\bibitem[{\citenamefont{Herlt and Stephani}(1976)}]{Herlt-Stephani:1976}
\bibinfo{author}{\bibfnamefont{E.}~\bibnamefont{Herlt}} \bibnamefont{and}
  \bibinfo{author}{\bibfnamefont{H.}~\bibnamefont{Stephani}},
  \bibinfo{journal}{Int. J. Theor. Phys.} \textbf{\bibinfo{volume}{15}},
  \bibinfo{pages}{45} (\bibinfo{year}{1976}).

\bibitem[{\citenamefont{{Grandy Jr.}}(2005)}]{Grandy-book-2005}
\bibinfo{author}{\bibfnamefont{W.~T.} \bibnamefont{{Grandy Jr.}}},
  \emph{\bibinfo{title}{Scattering of Waves from Large Spheres}}
  (\bibinfo{publisher}{Cambridge University Press},
  \bibinfo{address}{Cambridge, UK}, \bibinfo{year}{2005}).

\bibitem[{\citenamefont{{van de Hulst}}(1981)}]{vandeHulst-book-1981}
\bibinfo{author}{\bibfnamefont{H.~C.} \bibnamefont{{van de Hulst}}},
  \emph{\bibinfo{title}{Light Scattering by Small Particles}}
  (\bibinfo{publisher}{Dover Publications}, \bibinfo{address}{New York},
  \bibinfo{year}{1981}).

\bibitem[{\citenamefont{{Kerker}}(1969)}]{Kerker-book:1969}
\bibinfo{author}{\bibfnamefont{M.}~\bibnamefont{{Kerker}}},
  \emph{\bibinfo{title}{The scattering of light, and other electromagnetic
  radiation}} (\bibinfo{publisher}{Academic Press, New York},
  \bibinfo{year}{1969}).

\bibitem[{\citenamefont{Morse and Feshbach}(1953)}]{Morse-Feshbach:1953}
\bibinfo{author}{\bibfnamefont{P.~M.} \bibnamefont{Morse}} \bibnamefont{and}
  \bibinfo{author}{\bibfnamefont{H.}~\bibnamefont{Feshbach}},
  \emph{\bibinfo{title}{Methods of Theoretical Physics, Part I}}
  (\bibinfo{publisher}{McGraw-Hill Science, New York}, \bibinfo{year}{1953}).

\bibitem[{\citenamefont{{Turyshev} and
  {Toth}}(2018{\natexlab{b}})}]{Turyshev-Toth:2018}
\bibinfo{author}{\bibfnamefont{S.~G.} \bibnamefont{{Turyshev}}}
  \bibnamefont{and} \bibinfo{author}{\bibfnamefont{V.~T.}
  \bibnamefont{{Toth}}}, \bibinfo{journal}{Phys. Rev. A}
  \textbf{\bibinfo{volume}{97}}, \bibinfo{pages}{033810}
  (\bibinfo{year}{2018}{\natexlab{b}}), \eprint{arXiv:1801.06253
  [physics.optics]}.

\bibitem[{\citenamefont{{Burke}}(2011{\natexlab{b}})}]{Burke:2011}
\bibinfo{author}{\bibfnamefont{P.~G.} \bibnamefont{{Burke}}},
  \emph{\bibinfo{title}{Potential Scattering in Atomic Physics.}}
  (\bibinfo{publisher}{Springer}, \bibinfo{year}{2011}{\natexlab{b}}).

\bibitem[{\citenamefont{{Newton}}(2013)}]{Newton-book-2013}
\bibinfo{author}{\bibfnamefont{R.~G.} \bibnamefont{{Newton}}},
  \emph{\bibinfo{title}{Scattering Theory of Waves and Particles}}
  (\bibinfo{publisher}{Dover Books on Physics. 2-nd Edition},
  \bibinfo{year}{2013}).

\bibitem[{\citenamefont{Bateman et~al.}(1953)\citenamefont{Bateman,
  Erd{\'e}lyi, and {Bateman Manuscript Project}}}]{Bateman-Erdelyi:1953}
\bibinfo{author}{\bibfnamefont{H.}~\bibnamefont{Bateman}},
  \bibinfo{author}{\bibfnamefont{A.}~\bibnamefont{Erd{\'e}lyi}},
  \bibnamefont{and} \bibinfo{author}{\bibnamefont{{Bateman Manuscript
  Project}}}, \emph{\bibinfo{title}{Higher transcendental functions}},
  vol.~\bibinfo{volume}{1} of \emph{\bibinfo{series}{Higher Transcendental
  Functions}} (\bibinfo{publisher}{McGraw-Hill, New York},
  \bibinfo{year}{1953}).

\bibitem[{\citenamefont{Korn and Korn}(1968)}]{Korn-Korn:1968}
\bibinfo{author}{\bibfnamefont{G.~A.} \bibnamefont{Korn}} \bibnamefont{and}
  \bibinfo{author}{\bibfnamefont{T.~M.} \bibnamefont{Korn}},
  \emph{\bibinfo{title}{Mathematical Handbook for Scientists and Engineers:
  Definitions, Theorems, and Formulas for Reference and Review}}
  (\bibinfo{publisher}{McGraw-Hill Book Co., New York}, \bibinfo{year}{1968}).

\bibitem[{\citenamefont{{Abramowitz} and
  {Stegun}}(1965)}]{Abramovitz-Stegun:1965}
\bibinfo{author}{\bibfnamefont{M.}~\bibnamefont{{Abramowitz}}}
  \bibnamefont{and} \bibinfo{author}{\bibfnamefont{I.~A.}
  \bibnamefont{{Stegun}}}, \emph{\bibinfo{title}{Handbook of Mathematical
  Functions: with Formulas, Graphs, and Mathematical Tables.}}
  (\bibinfo{publisher}{Dover Publications, New York; revised edition},
  \bibinfo{year}{1965}).

\bibitem[{\citenamefont{{Bakaleinikov} and
  {Silbergleit}}(2020)}]{Bakaleinikov:2020}
\bibinfo{author}{\bibfnamefont{L.}~\bibnamefont{{Bakaleinikov}}}
  \bibnamefont{and}
  \bibinfo{author}{\bibfnamefont{A.}~\bibnamefont{{Silbergleit}}},
  \bibinfo{journal}{J. Math. Phys.} \textbf{\bibinfo{volume}{61}},
  \bibinfo{pages}{083503} (\bibinfo{year}{2020}).

\bibitem[{\citenamefont{{Nambu}}(2013{\natexlab{a}})}]{Nambu:2013a}
\bibinfo{author}{\bibfnamefont{Y.}~\bibnamefont{{Nambu}}}, in
  \emph{\bibinfo{booktitle}{J. Phys. Conf. Ser.}}
  (\bibinfo{year}{2013}{\natexlab{a}}), vol. \bibinfo{volume}{410} of
  \emph{\bibinfo{series}{J. Phys. Conf. Ser}}, p. \bibinfo{pages}{012036}.

\bibitem[{\citenamefont{{Cody} and {Hillstrom}}(1970)}]{Cody-Hillstrom:1970}
\bibinfo{author}{\bibfnamefont{W.~J.} \bibnamefont{{Cody}}} \bibnamefont{and}
  \bibinfo{author}{\bibfnamefont{K.~E.} \bibnamefont{{Hillstrom}}},
  \bibinfo{journal}{Mathematics of Computation} \textbf{\bibinfo{volume}{24}},
  \bibinfo{pages}{671} (\bibinfo{year}{1970}).

\bibitem[{\citenamefont{Barata et~al.}(2011)\citenamefont{Barata, Canto, and
  Hussein}}]{Barata:2009ma}
\bibinfo{author}{\bibfnamefont{J.~C.~A.} \bibnamefont{Barata}},
  \bibinfo{author}{\bibfnamefont{L.~F.} \bibnamefont{Canto}}, \bibnamefont{and}
  \bibinfo{author}{\bibfnamefont{M.~S.} \bibnamefont{Hussein}},
  \bibinfo{journal}{Braz. J. Phys.} \textbf{\bibinfo{volume}{41}},
  \bibinfo{pages}{50} (\bibinfo{year}{2011}).

\bibitem[{\citenamefont{{Schneider} et~al.}(2006)\citenamefont{{Schneider},
  {Kochanek}, and {Wambsganss}}}]{Schneider-etal:2006}
\bibinfo{author}{\bibfnamefont{P.}~\bibnamefont{{Schneider}}},
  \bibinfo{author}{\bibfnamefont{C.}~\bibnamefont{{Kochanek}}},
  \bibnamefont{and}
  \bibinfo{author}{\bibfnamefont{J.}~\bibnamefont{{Wambsganss}}},
  \emph{\bibinfo{title}{Gravitational Lensing: Strong, Weak and Micro: Saas-Fee
  Advanced Course 33}} (\bibinfo{publisher}{Springer},
  \bibinfo{address}{Berlin}, \bibinfo{year}{2006}).

\bibitem[{\citenamefont{{Deguchi} and {Watson}}(1987)}]{Deguchi-Watson:1987}
\bibinfo{author}{\bibfnamefont{S.}~\bibnamefont{{Deguchi}}} \bibnamefont{and}
  \bibinfo{author}{\bibfnamefont{W.~D.} \bibnamefont{{Watson}}},
  \bibinfo{journal}{\apj} \textbf{\bibinfo{volume}{315}}, \bibinfo{pages}{440}
  (\bibinfo{year}{1987}).

\bibitem[{\citenamefont{{Nambu}}(2013{\natexlab{b}})}]{Nambu:2013}
\bibinfo{author}{\bibfnamefont{Y.}~\bibnamefont{{Nambu}}},
  \bibinfo{journal}{Int. J. Astron. Astrophys.} \textbf{\bibinfo{volume}{3}},
  \bibinfo{pages}{1} (\bibinfo{year}{2013}{\natexlab{b}}).

\bibitem[{\citenamefont{{Matsunaga} and
  {Yamamoto}}(2006)}]{Matsunaga-Yamamoto:2006}
\bibinfo{author}{\bibfnamefont{N.}~\bibnamefont{{Matsunaga}}} \bibnamefont{and}
  \bibinfo{author}{\bibfnamefont{K.}~\bibnamefont{{Yamamoto}}},
  \bibinfo{journal}{JCAP} \textbf{\bibinfo{volume}{2006}}, \bibinfo{eid}{023}
  (\bibinfo{year}{2006}).

\bibitem[{\citenamefont{{Deguchi} and {Watson}}(1986)}]{Deguchi-Watson:1986}
\bibinfo{author}{\bibfnamefont{S.}~\bibnamefont{{Deguchi}}} \bibnamefont{and}
  \bibinfo{author}{\bibfnamefont{W.~D.} \bibnamefont{{Watson}}},
  \bibinfo{journal}{\apj} \textbf{\bibinfo{volume}{307}}, \bibinfo{pages}{30}
  (\bibinfo{year}{1986}).

\bibitem[{\citenamefont{{Nambu}}(2013{\natexlab{c}})}]{Nambu:2012}
\bibinfo{author}{\bibfnamefont{Y.}~\bibnamefont{{Nambu}}}, \bibinfo{journal}{J.
  Phys. Conf. Ser.} \textbf{\bibinfo{volume}{410}}, \bibinfo{pages}{012036}
  (\bibinfo{year}{2013}{\natexlab{c}}).

\bibitem[{\citenamefont{{Klioner}}(1991)}]{Klioner:1991SvA}
\bibinfo{author}{\bibfnamefont{S.~A.} \bibnamefont{{Klioner}}},
  \bibinfo{journal}{Sov. Astron.} \textbf{\bibinfo{volume}{35}},
  \bibinfo{pages}{523} (\bibinfo{year}{1991}).

\bibitem[{\citenamefont{{Klioner} and
  {Kopeikin}}(1992)}]{Klioner-Kopeikin:1992}
\bibinfo{author}{\bibfnamefont{S.~A.} \bibnamefont{{Klioner}}}
  \bibnamefont{and} \bibinfo{author}{\bibfnamefont{S.~M.}
  \bibnamefont{{Kopeikin}}}, \bibinfo{journal}{Astron. J.}
  \textbf{\bibinfo{volume}{104}}, \bibinfo{pages}{897} (\bibinfo{year}{1992}).

\bibitem[{\citenamefont{{Zschocke} and
  {Klioner}}(2010)}]{Zschocke-Klioner:2010}
\bibinfo{author}{\bibfnamefont{S.}~\bibnamefont{{Zschocke}}} \bibnamefont{and}
  \bibinfo{author}{\bibfnamefont{S.~A.} \bibnamefont{{Klioner}}},
  \bibinfo{journal}{CQG} \textbf{\bibinfo{volume}{28}}, \bibinfo{pages}{015009}
  (\bibinfo{year}{2010}).

\bibitem[{\citenamefont{{Park} et~al.}(2017)\citenamefont{{Park}, {Folkner},
  {Konopliv}, {Williams}, {Smith}, and {Zuber}}}]{Park-etal:2017}
\bibinfo{author}{\bibfnamefont{R.~S.} \bibnamefont{{Park}}},
  \bibinfo{author}{\bibfnamefont{W.~M.} \bibnamefont{{Folkner}}},
  \bibinfo{author}{\bibfnamefont{A.~S.} \bibnamefont{{Konopliv}}},
  \bibinfo{author}{\bibfnamefont{J.~G.} \bibnamefont{{Williams}}},
  \bibinfo{author}{\bibfnamefont{D.~E.} \bibnamefont{{Smith}}},
  \bibnamefont{and} \bibinfo{author}{\bibfnamefont{M.~T.}
  \bibnamefont{{Zuber}}}, \bibinfo{journal}{Astron. J.}
  \textbf{\bibinfo{volume}{153}}, \bibinfo{pages}{121} (\bibinfo{year}{2017}).

\bibitem[{\citenamefont{Rozelot and Eren}(2020)}]{Rozelot:2019}
\bibinfo{author}{\bibfnamefont{J.}~\bibnamefont{Rozelot}} \bibnamefont{and}
  \bibinfo{author}{\bibfnamefont{S.}~\bibnamefont{Eren}},
  \bibinfo{journal}{Adv. Space Res.} \textbf{\bibinfo{volume}{65}},
  \bibinfo{pages}{2821 } (\bibinfo{year}{2020}).

\bibitem[{\citenamefont{{Berry} and {Upstill}}(1982)}]{Berry-Upstill:1982}
\bibinfo{author}{\bibfnamefont{M.~V.} \bibnamefont{{Berry}}} \bibnamefont{and}
  \bibinfo{author}{\bibfnamefont{C.}~\bibnamefont{{Upstill}}},
  \bibinfo{journal}{Optics Laser Technology} \textbf{\bibinfo{volume}{14}},
  \bibinfo{pages}{257} (\bibinfo{year}{1982}).

\bibitem[{\citenamefont{{Berry}}(1992)}]{Berry:1992}
\bibinfo{author}{\bibfnamefont{M.~V.} \bibnamefont{{Berry}}}, in
  \emph{\bibinfo{booktitle}{Huygens’ Principle 1690-1990: Theory and
  Applications}}, edited by \bibinfo{editor}{\bibnamefont{{H. Block}}},
  \bibinfo{editor}{\bibnamefont{{H.A. Ferwerda}}}, \bibnamefont{and}
  \bibinfo{editor}{\bibnamefont{{H.K. Kuiken}}} (\bibinfo{publisher}{Elsevier
  Science Publishers B.V.}, \bibinfo{year}{1992}), pp.
  \bibinfo{pages}{97--111}.

\bibitem[{\citenamefont{{Blanchet} and {Damour}}(1986)}]{Blanchet-Damour:1986}
\bibinfo{author}{\bibfnamefont{L.}~\bibnamefont{{Blanchet}}} \bibnamefont{and}
  \bibinfo{author}{\bibfnamefont{T.}~\bibnamefont{{Damour}}},
  \bibinfo{journal}{Philos. Trans. R. Soc. London Ser. A}
  \textbf{\bibinfo{volume}{320}}, \bibinfo{pages}{379} (\bibinfo{year}{1986}).

\bibitem[{\citenamefont{{Blanchet} and {Damour}}(1989)}]{Blanchet-Damour:1989}
\bibinfo{author}{\bibfnamefont{L.}~\bibnamefont{{Blanchet}}} \bibnamefont{and}
  \bibinfo{author}{\bibfnamefont{T.}~\bibnamefont{{Damour}}},
  \bibinfo{journal}{Ann. Inst. Henri Poincar\'e} \textbf{\bibinfo{volume}{50}},
  \bibinfo{pages}{377} (\bibinfo{year}{1989}).

\bibitem[{\citenamefont{Linet and Teyssandier}(2002)}]{Linet_2002}
\bibinfo{author}{\bibfnamefont{B.}~\bibnamefont{Linet}} \bibnamefont{and}
  \bibinfo{author}{\bibfnamefont{P.}~\bibnamefont{Teyssandier}},
  \bibinfo{journal}{Phys. Rev. D} \textbf{\bibinfo{volume}{66}}
  (\bibinfo{year}{2002}).

\bibitem[{\citenamefont{{Mathis} and {Le
  Poncin-Lafitte}}(2007)}]{Mathis-LePoncinLafitte:2007}
\bibinfo{author}{\bibfnamefont{S.}~\bibnamefont{{Mathis}}} \bibnamefont{and}
  \bibinfo{author}{\bibfnamefont{C.}~\bibnamefont{{Le Poncin-Lafitte}}},
  \bibinfo{journal}{Astron. \& Astrophys.} \textbf{\bibinfo{volume}{497}},
  \bibinfo{pages}{889} (\bibinfo{year}{2007}).

\bibitem[{\citenamefont{Thorne}(1980)}]{Thorne:1980}
\bibinfo{author}{\bibfnamefont{K.~S.} \bibnamefont{Thorne}},
  \bibinfo{journal}{Rev. Mod. Phys.} \textbf{\bibinfo{volume}{52}},
  \bibinfo{pages}{299} (\bibinfo{year}{1980}).

\bibitem[{\citenamefont{{Turyshev} et~al.}(2013)\citenamefont{{Turyshev},
  {Toth}, and {Sazhin}}}]{Turyshev:2012nw}
\bibinfo{author}{\bibfnamefont{S.~G.} \bibnamefont{{Turyshev}}},
  \bibinfo{author}{\bibfnamefont{V.~T.} \bibnamefont{{Toth}}},
  \bibnamefont{and} \bibinfo{author}{\bibfnamefont{M.~V.}
  \bibnamefont{{Sazhin}}}, \bibinfo{journal}{Phys. Rev. D}
  \textbf{\bibinfo{volume}{87}}, \bibinfo{pages}{024020}
  (\bibinfo{year}{2013}), \eprint{arXiv:1212.0232 [gr-qc]}.

\bibitem[{\citenamefont{{Turyshev} et~al.}(2014)\citenamefont{{Turyshev},
  {Sazhin}, and {Toth}}}]{Turyshev-GRACE-FO:2014}
\bibinfo{author}{\bibfnamefont{S.~G.} \bibnamefont{{Turyshev}}},
  \bibinfo{author}{\bibfnamefont{M.~V.} \bibnamefont{{Sazhin}}},
  \bibnamefont{and} \bibinfo{author}{\bibfnamefont{V.~T.}
  \bibnamefont{{Toth}}}, \bibinfo{journal}{Phys. Rev. D}
  \textbf{\bibinfo{volume}{89}}, \bibinfo{pages}{105029}
  (\bibinfo{year}{2014}), \eprint{arXiv:1402.7111 [gr-qc]}.

\bibitem[{\citenamefont{Turyshev et~al.}(2010)\citenamefont{Turyshev, Farr,
  Folkner, Girerd, Hemmati, Murphy, Williams, and Degnan}}]{Turyshev-PLR:2010}
\bibinfo{author}{\bibfnamefont{S.~G.} \bibnamefont{Turyshev}},
  \bibinfo{author}{\bibfnamefont{W.}~\bibnamefont{Farr}},
  \bibinfo{author}{\bibfnamefont{W.~M.} \bibnamefont{Folkner}},
  \bibinfo{author}{\bibfnamefont{A.~R.} \bibnamefont{Girerd}},
  \bibinfo{author}{\bibfnamefont{H.}~\bibnamefont{Hemmati}},
  \bibinfo{author}{\bibfnamefont{T.~W.~J.} \bibnamefont{Murphy}},
  \bibinfo{author}{\bibfnamefont{J.~G.} \bibnamefont{Williams}},
  \bibnamefont{and} \bibinfo{author}{\bibfnamefont{J.~J.}
  \bibnamefont{Degnan}}, \bibinfo{journal}{Exper. Astron.}
  \textbf{\bibinfo{volume}{28}}, \bibinfo{pages}{209} (\bibinfo{year}{2010}),
  \eprint{arXiv:1003.4961 [gr-qc]}.

\bibitem[{\citenamefont{{Mott}}(1928)}]{Mott:1928}
\bibinfo{author}{\bibfnamefont{N.~F.} \bibnamefont{{Mott}}},
  \bibinfo{journal}{Proc. Royal Soc. London Ser. A}
  \textbf{\bibinfo{volume}{118}}, \bibinfo{pages}{542} (\bibinfo{year}{1928}).

\bibitem[{\citenamefont{{Gordon}}(1928)}]{Gordon:1928}
\bibinfo{author}{\bibfnamefont{W.}~\bibnamefont{{Gordon}}},
  \bibinfo{journal}{Zeitschrift f\"ur Physik} \textbf{\bibinfo{volume}{48}},
  \bibinfo{pages}{180} (\bibinfo{year}{1928}).

\bibitem[{\citenamefont{{Landau} and
  {Lifshitz}}(1988{\natexlab{b}})}]{Landau-Lifshitz:1988m}
\bibinfo{author}{\bibfnamefont{L.~D.} \bibnamefont{{Landau}}} \bibnamefont{and}
  \bibinfo{author}{\bibfnamefont{E.~M.} \bibnamefont{{Lifshitz}}},
  \emph{\bibinfo{title}{Mechanics.}} (\bibinfo{publisher}{4th edition. Nauka:
  Moscow (in Russian)}, \bibinfo{year}{1988}{\natexlab{b}}).

\bibitem[{\citenamefont{{Chaichian} and
  {Demichev}}(2001)}]{Chaichian-Demichev:2001}
\bibinfo{author}{\bibfnamefont{M.}~\bibnamefont{{Chaichian}}} \bibnamefont{and}
  \bibinfo{author}{\bibfnamefont{A.}~\bibnamefont{{Demichev}}},
  \emph{\bibinfo{title}{Path Integrals in Physics. Volume I: Stochastic
  Processes and Quantum Mechanics}} (\bibinfo{publisher}{IOP Publishing},
  \bibinfo{address}{Bristol and Philadelphia}, \bibinfo{year}{2001}).

\end{thebibliography}

\appendix

\section{Representation of the field in terms of Debye potentials}
\label{app:Debye}

To represent the EM field equations in terms of Debye potentials, we start with (\ref{eq:rotE_fl})--(\ref{eq:rotH_fl}), where we treat gravity to be static, thus $\dot u=0$. Assuming, as usual (we follow closely the discussion presented in \cite{Born-Wolf:1999,Turyshev-Toth:2017,Turyshev-Toth:2019}, adapted for the gravitational lens), the time dependence of the field in the form $\exp(-i\omega t)$ where $k= \omega/c$, the time-independent parts of the electric and magnetic vectors satisfy Maxwell's equations:
\begin{eqnarray}
{\rm curl}\,{\vec D}&=&ik\,u^2\,{\vec B}+{\cal O}(G^2),
\label{eq:rotH_fl**=}\\[3pt]
{\rm curl}\,{\vec B}&=&-ik\,u^2\,{\vec D}+{\cal O}(G^2).
\label{eq:rotE_fl**=}
\end{eqnarray}

As shown in  \cite{Turyshev-Toth:2017}, in spherical  coordinates, the field equations (\ref{eq:rotH_fl**=})--(\ref{eq:rotE_fl**=}) to ${\cal O}(G^2)$ become
{}
\begin{eqnarray}
-ik\,u^2{\hat D}_r&=&\frac{1}{r^2\sin\theta}\Big(
\frac{\partial}{\partial \theta}(r\sin\theta \hat B_\phi)-\frac{\partial}{\partial \phi}(r\hat B_\theta)\Big),
\label{eq:Dr}\\[3pt]
-ik\,u^2{\hat D}_\theta&=&\frac{1}{r\sin\theta}
\Big(
\frac{\partial \hat B_r}{\partial \phi}-\frac{\partial}{\partial r}(r\sin\theta \hat B_\phi)\Big),
\label{eq:Dt}\\[3pt]
-ik\,u^2{\hat D}_\phi&=&\frac{1}{r}
\Big(
\frac{\partial}{\partial r}(r \hat B_\theta)-\frac{\partial \hat B_r}{\partial \theta}\Big),
\label{eq:Dp}\\[3pt]
ik\,u^2{\hat B}_r&=&\frac{1}{r^2\sin\theta}\Big(\frac{\partial}{\partial \theta}(r\sin\theta \hat D_\phi)-\frac{\partial}{\partial \phi}(r\hat D_\theta)\Big),
\label{eq:Br}\\[3pt]
ik\,u^2{\hat B}_\theta&=&
\frac{1}{r\sin\theta}
\Big(\frac{\partial \hat D_r}{\partial \phi}-\frac{\partial}{\partial r}(r\sin\theta \hat D_\phi)\Big),
\label{eq:Bt}\\[3pt]
ik\,u^2{\hat B}_\phi&=&\frac{1}{r}
\Big(\frac{\partial}{\partial r}(r \hat D_\theta)-\frac{\partial \hat D_r}{\partial \theta}\Big),
\label{eq:Bp}
\end{eqnarray}
while the remaining two equations from Eq.~(\ref{eq:rotE_fl})--(\ref{eq:rotH_fl}) take the form
{}
\begin{eqnarray}
\frac{\partial}{\partial r}\Big(u^2r^2\sin\theta B_r\Big)+\frac{\partial}{\partial \theta}\Big(u^2r\sin\theta B_\theta\Big)+\frac{\partial}{\partial \phi}\Big(u^2r B_\phi\Big)&=&0,
\label{eq:divB*3}\\
\frac{\partial}{\partial r}\Big(u^2r^2\sin\theta D_r\Big)+\frac{\partial}{\partial \theta}\Big(u^2r\sin\theta D_\theta\Big)+\frac{\partial}{\partial \phi}\Big(u^2r D_\phi\Big)&=&0.
\label{eq:divD*3}
\end{eqnarray}

Our goal is to find a general solution to these equations in the form of a superposition of two linearly independent solutions $\big({}^e{\vec D}, {}^e{\vec B}\big)$ and $\big({}^m{\vec D}, {}^m{\vec B}\big)$ that satisfy the following relationships:
{}
\begin{eqnarray}
{}^e{\hskip -2pt}{\hat D}_r &=& {\hat D}_r, \qquad
{}^e{\hskip -2pt}{\hat B}_r=0, \\
\label{eq:electr}
\hskip 18pt
{}^m{\hskip -2pt}{\hat D}_r&=&0, \qquad ~\, {}^m{\hskip -2pt}{\hat B}_r={\hat B}_r.
\label{eq:magnet}
\end{eqnarray}
With $\hat B_r={}^e{\hskip -2pt}\hat B_r=0$, (\ref{eq:Dt}) and (\ref{eq:Dp}) become
{}
\begin{eqnarray}
ik\, u^2\,{}^e{\hskip -2pt}{\hat D}_\theta&=&\frac{1}{r}
\frac{\partial}{\partial r}
\big(r \,{}^e{\hskip -2pt}\hat B_\phi\big),
\label{eq:Dt*}\\
ik\, u^2\,{}^e{\hskip -2pt}{\hat D}_\phi&=&-\frac{1}{r}
\frac{\partial}{\partial r}\big(r \,{}^e{\hskip -2pt}\hat B_\theta\big).
\label{eq:Dp*}
\end{eqnarray}

Substituting these relationships into (\ref{eq:Bt}) and (\ref{eq:Bp}) we obtain
{}
\begin{eqnarray}
\frac{\partial}{\partial r}\Big[\frac{1}{u^2}\frac{\partial}{\partial r}\big(r\,{}^e{\hskip -2pt}{\hat B}_\theta\big)\Big]+
k^2u^2(r\,{}^e{\hskip -2pt}{\hat B}_\theta)&=&-
\frac{ik}{\sin\theta}\frac{\partial \,{}^e{\hskip -2pt}\hat D_r}{\partial \phi},
\label{eq:Bp+}\\
\frac{\partial}{\partial r}\Big[\frac{1}{u^2}\frac{\partial}{\partial r}\big(r\,{}^e{\hskip -2pt}{\hat B}_\phi\big)\Big]+
k^2u^2(r\,{}^e{\hskip -2pt}{\hat B}_\phi) &=&
ik \frac{\partial \,{}^e{\hskip -2pt}\hat D_r}{\partial \theta}.
\label{eq:Bt+}
\end{eqnarray}

From ${\rm div} (u^2{}^e{\vec B})=0 $ given by Eq.~(\ref{eq:rotH_fl}) (which in the expanded form is given by Eq.~(\ref{eq:divB*3})) and using our assumption that $\,{}^e{\hskip -2pt}\hat B_r=0$, we have
{}
\begin{eqnarray}
\frac{\partial}{\partial \theta}\big(u^2 \sin\theta \,{}^e{\hskip -2pt}\hat B_\theta\big)+
\frac{\partial }{\partial \phi}\big(u^2\,{}^e{\hskip -2pt}\hat B_\phi\big)&=&0,
\label{eq:divB_fl+}
\end{eqnarray}
which ensures that (\ref{eq:Br}) is also satisfied at the needed level of accuracy. As we know, this equation is valid for a spherically symmetric gravitational field. Terms that characterize deviations from the monopole in the generic form of the Newtonian potential, $U$, lack spherical symmetry. For these terms, the condition (\ref{eq:divB_fl+}) may be satisfied only approximately. Indeed, after substitution from (\ref{eq:Dt*}), (\ref{eq:Dp*}),  in (\ref{eq:Br}), we have
{}
\begin{eqnarray}
\frac{1}{r^2\sin\theta}\Big(\frac{\partial}{\partial \theta}\big(r\sin\theta \,{}^e{\hskip -2pt}\hat D_\phi\big)-\frac{\partial}{\partial \phi}\big(r\,{}^e{\hskip -2pt}\hat D_\theta\big)\Big)&=&\nonumber\\
&&\hskip-163pt =\,-\,
\frac{1}{ik\,u^2r^2\sin\theta}\Big\{ \frac{\partial}{\partial r}\Big[\frac{r}{u^2}\Big(\frac{\partial}{\partial \theta}\big(u^2\sin\theta \,{}^e{\hskip -2pt}\hat B_\theta\big)+\frac{\partial }{\partial \phi}\big(u^2{}^e{\hskip -2pt}\hat B_\phi\big)\Big)\Big]+\nonumber\\
&&\hskip-100pt +\,
2\frac{\partial}{\partial r}\Big[r\Big(\sin\theta  \,{}^e{\hskip -2pt}\hat B_\theta \frac{\partial \ln u^2}{\partial \theta} +{}^e{\hskip -2pt}\hat B_\phi \frac{\partial \ln u^2}{\partial \phi}\Big)\Big]-r\Big(\sin\theta \,{}^e{\hskip -2pt}\hat B_\theta \frac{\partial^2 \ln u^2}{\partial r \partial\theta} +\,{}^e{\hskip -2pt}\hat B_\phi \frac{\partial^2 \ln u^2}{\partial r \partial\phi} \Big)\Big\}=
\nonumber\\
&&\hskip-163pt =
\,\frac{1}{ik\,u^2r^2\sin\theta}\Big\{
2\frac{\partial}{\partial r}\Big[r\Big(\sin\theta  \,{}^e{\hskip -2pt}\hat B_\theta \frac{\partial \ln u^2}{\partial \theta} +{}^e{\hskip -2pt}\hat B_\phi \frac{\partial \ln u^2}{\partial \phi}\Big)\Big]-r\Big(\sin\theta \,{}^e{\hskip -2pt}\hat B_\theta \frac{\partial^2 \ln u^2}{\partial r \partial\theta} +\,{}^e{\hskip -2pt}\hat B_\phi \frac{\partial^2 \ln u^2}{\partial r \partial\phi} \Big)\Big\}.
\label{eq:Br+}
\end{eqnarray}

The first term in this expression vanishes because of  (\ref{eq:divB_fl+}) (as it was in the case of a monopole, see \cite{Turyshev-Toth:2017}). Considering the remaining terms, and taking into account the form of  $u$ from (\ref{eq:w-PN}) with the Newtonian potential, $U$, given either (\ref{eq:pot_stf}) or (\ref{eq:pot_st3}),  we see that the following relation is true:
{}
\begin{eqnarray}
\frac{\partial}{\partial \theta}\big(r\sin\theta \,{}^e{\hskip -2pt}\hat D_\phi\big)-\frac{\partial}{\partial \phi}\big(r\,{}^e{\hskip -2pt}\hat D_\theta\big)&=&\nonumber\\
&&\hskip-180pt =
\,\frac{1}{ik\,u^2}\Big\{
2\frac{\partial}{\partial r}\Big[r\Big(\sin\theta  \,{}^e{\hskip -2pt}\hat B_\theta \frac{\partial \ln u^2}{\partial \theta} +{}^e{\hskip -2pt}\hat B_\phi \frac{\partial \ln u^2}{\partial \phi}\Big)\Big]-r\Big(\sin\theta \,{}^e{\hskip -2pt}\hat B_\theta \frac{\partial^2 \ln u^2}{\partial r \partial\theta} +\,{}^e{\hskip -2pt}\hat B_\phi \frac{\partial^2 \ln u^2}{\partial r \partial\phi} \Big)\Big\}\simeq \frac{1}{ik\,u^2}{\cal O}\Big(\frac{J_2 R^2_\odot}{r^3}\Big),
\label{eq:Br+2}
\end{eqnarray}
where $J_2$ is the gravitational quadrupole moment of the mass distribution inside the lens (which typically is the largest term after the monopole). Clearly, as $r$ gets larger, this expression vanishes, justifying the validity of  (\ref{eq:Br}), namely
{}
\begin{eqnarray}
\lim_{r\rightarrow\infty} \frac{1}{r^2\sin\theta}\Big(\frac{\partial}{\partial \theta}\big(r\sin\theta \,{}^e{\hskip -2pt}\hat D_\phi\big)-\frac{\partial}{\partial \phi}\big(r\,{}^e{\hskip -2pt}\hat D_\theta\big)\Big)\simeq \frac{1}{ik\,u^2} \frac{1}{r^2\sin\theta}{\cal O}\Big(\frac{J_2R^2_\odot}{r^3}\Big)\rightarrow 0.
\label{eq:Br+3}
\end{eqnarray}
In all practical scenarios, the limit (\ref{eq:Br+3}) is  satisfied for $r \gtrsim R_\odot$. Thus, for scenarios relevant for the SGL, (\ref{eq:Br+}) is equal to 0.  As a result, when describing the SGL and  considering light propagation in a weak gravitational field, we may neglect the effect of the gravitational field on the amplitude of the EM wave. In this case, our primary interest is the phase of the wave, thus this expression constitutes the condition consistent with the eikonal approximation. The complementary case with ${}^m{\hskip -2pt}\hat D_r=0$ is treated identically, in accordance with (\ref{eq:magnet}).

When the radial magnetic field vanishes, the solution is called {\it the electric wave} (or transverse magnetic wave); correspondingly, when the radial electric field vanishes, the solution is called {\it the magnetic wave} (or transverse electric wave). These can both be derived from the corresponding Debye scalar potentials ${}^e{\hskip -1pt}\Pi$ and ${}^m{\hskip -1pt}\Pi$.

Given ${}^e{\hskip -2pt}\hat B_r=0$, ${}^e{\hskip -2pt}\hat D_\phi$ and ${}^e{\hskip -2pt}\hat D_\theta$ in (\ref{eq:Br}) can be represented as a scalar field's gradient:
{}
\begin{eqnarray}
{}^e{\hskip -2pt}\hat D_\phi=\frac{1}{r\sin\theta}
\frac{\partial U}{\partial \phi}+\frac{1}{ik\,u^2} \frac{1}{r\sin\theta}{\cal O}\Big(\frac{J_2R^2_\odot}{r^3}\Big),\qquad
{}^e{\hskip -2pt}\hat D_\theta=\frac{1}{r}
\frac{\partial U}{\partial \theta}+\frac{1}{ik\,u^2} \frac{1}{r}{\cal O}\Big(\frac{J_2R^2_\odot}{r^3}\Big),
\label{eq:Dp-Dt}
\end{eqnarray}
where $U$ is some function. Introducing the electric Debye potential
${}^e{\hskip -1pt}\Pi$ that relates to $U$ as
{}
\begin{eqnarray}
U=\frac{1}{u^2}\frac{\partial }{\partial r}\big(r\,{}^e{\hskip -1pt}\Pi\big).
\label{eq:Pi}
\end{eqnarray}
We use this expression in (\ref{eq:Dp-Dt}) and obtain
{}
\begin{eqnarray}
{}^e{\hskip -2pt}\hat D_\theta=\frac{1}{u^2r}\Big\{
\frac{\partial^2 \big(r\,{}^e{\hskip -1pt}\Pi\big)}{\partial r\partial \theta}-\frac{\partial \ln u^2}{\partial \theta}\frac{\partial}{\partial r }\big(r{}^e\Pi\big)\Big\},
\qquad
{}^e{\hskip -2pt}\hat D_\phi=\frac{1}{u^2r\sin\theta}\Big\{
\frac{\partial^2 \big(r\,{}^e{\hskip -1pt}\Pi\big)}{\partial r\partial \phi}-\frac{\partial \ln u^2}{\partial \phi}\frac{\partial}{\partial r }\big(r{}^e\Pi\big)\Big\},
\label{eq:Dp-Dt+}
\end{eqnarray}
which satisfy Eq.~(\ref{eq:Br}) with ${}^e{\hskip -2pt}\hat B_r=0$ and thus also (\ref{eq:Br+2}).

It can be seen that (\ref{eq:Dt*}) and (\ref{eq:Dp*}) are satisfied by
{}
\begin{eqnarray}
{}^e{\hskip -2pt}{\hat B}_\phi&=&\frac{ik}{r}\frac{\partial \big(r \,{}^e{\hskip -1pt}\Pi\big)}{\partial \theta}+{\cal O}\Big(\frac{J_2}{r^3}{}^e\Pi\Big),
\qquad
{}^e{\hskip -2pt}{\hat B}_\theta=-\frac{ik}{r\sin\theta}
\frac{\partial \big(r \,{}^e{\hskip -1pt}\Pi\big)}{\partial \phi}+{\cal O}\Big(\frac{J_2}{r^3}{}^e\Pi\Big).
\label{eq:Bt*}
\end{eqnarray}
If we substitute both of (\ref{eq:Bt*}) into (\ref{eq:Dr}) we obtain
\begin{eqnarray}
\,{}^e{\hskip -2pt}{\hat D}_r&=&-\frac{1}{u^2r^2\sin\theta}\Big[\frac{\partial}{\partial \theta}\Big(\sin\theta \frac{\partial (r\,{}^e{\hskip -1pt}\Pi)}{\partial \theta}\Big)+\frac{1}{\sin\theta}\frac{\partial^2 (r\,{}^e{\hskip -1pt}\Pi)}{\partial \phi^2}\Big]+{\cal O}\Big(\frac{J_2}{r^3}{}^e\Pi\Big).
\label{eq:Dr*+}
\end{eqnarray}

Substituting expressions (\ref{eq:Bt*}) into (\ref{eq:Bp+})--(\ref{eq:Bt+}) yields
{}
\begin{eqnarray}
\dfrac{-ik}{\sin\theta} \dfrac{\partial}{\partial\phi}
\Big\{\dfrac{\partial}{\partial r}\Big[\dfrac{1}{u^2}\dfrac{\partial}{\partial r}
\big(r \,{}^e{\hskip -1pt}\Pi\big)\Big] + k^2u^2(r \,{}^e{\hskip -1pt}\Pi)
-{}^e{\hskip -2pt}\hat D_r\Big\}&=&{\cal O}\Big(\frac{J_2}{r^3}{}^e\Pi\Big),\\
ik\dfrac{\partial}{\partial\theta} \Big\{
\dfrac{\partial}{\partial r}\Big[\dfrac{1}{u^2}\dfrac{\partial}{\partial r}
\big(r\,{}^e{\hskip -1pt}\Pi\big) \Big]+ k^2u^2(r {}^e{\hskip -1pt}\Pi) -
{}^e{\hskip -2pt}\hat D_r \Big\} &=&{\cal O}\Big(\frac{J_2}{r^3}{}^e\Pi\Big),
\end{eqnarray}
i.e., the derivative of the same expression with respect to both $\phi$ and $\theta$ vanishes. This is clearly satisfied if we set the expression itself to ${\cal O}\big(({J_2}/{r^3}){}^e\Pi\big)$. Dividing by $u^2$ and using (\ref{eq:Dr*+}) leads to
{}
\begin{eqnarray}
\frac{1}{u^2}\frac{\partial}{\partial r}\Big[\frac{1}{u^2} \frac{\partial (r\,{}^e{\hskip -1pt}\Pi)}{\partial r}\Big]+\frac{1}{u^4r^2\sin\theta}\Big\{\frac{\partial}{\partial \theta}\Big(\sin\theta \frac{\partial (r\,{}^e{\hskip -1pt}\Pi)}{\partial \theta}\Big)+\frac{1}{\sin\theta}
\frac{\partial^2 (r\,{}^e{\hskip -1pt}\Pi)}{\partial \phi^2}\Big\}+k^2(r\,{}^e{\hskip -1pt}\Pi)={\cal O}\Big(\frac{J_2}{r^3}{}^e\Pi\Big).
\label{eq:Pi-eq}
\end{eqnarray}
Defining $u'=\partial u/\partial r$ and $u''=\partial^2 u/\partial r^2$, this equation may be rewritten as
\begin{eqnarray}
\frac{1}{r^2}\frac{\partial }{\partial r}\Big(r^2\frac{\partial}{\partial r} \Big[\frac{\,{}^e{\hskip -1pt}\Pi}{u}\Big]\Big)+
\frac{1}{r^2\sin\theta}\Big\{\frac{\partial}{\partial \theta}\Big(\sin\theta \frac{\partial}{\partial \theta} \Big[\frac{\,{}^e{\hskip -1pt}\Pi}{u}\Big]\Big)+
\frac{1}{\sin\theta}
\frac{\partial^2 }{\partial \phi^2}\Big[\frac{\,{}^e{\hskip -1pt}\Pi}{u}\Big]\Big\}+\Big(k^2u^4-u\big(\frac{1}{u}\big)''
\Big)\Big[\frac{\,{}^e{\hskip -1pt}\Pi}{u}\Big]={\cal O}\Big(\frac{J_2}{r^3}\frac{\,{}^e{\hskip -1pt}\Pi}{u}\Big),~~~
\label{eq:Pi-eq+weq}
\end{eqnarray}
which is the wave equation for the quantity ${\,{}^e{\hskip -1pt}\Pi}/{u}$:
\begin{eqnarray}
\Big(\Delta+k^2u^4-u\big(\frac{1}{u}\big)''\Big)\Big[\frac{\,{}^e{\hskip -1pt}\Pi}{u}\Big]={\cal O}\Big(r_g^2,\frac{J_2}{r^3}\frac{\,{}^e{\hskip -1pt}\Pi}{u}\Big).
\label{eq:Pi-eq+wew1}
\end{eqnarray}

We are concerned only with the field produced by the extended gravitational field. Thus, the quantity $u$ has the from $u({\vec r})=1+U/{c^2}+
{\cal O}(c^{-4}),$ as given by (\ref{eq:w-PN}). With this, we can rewrite (\ref{eq:Pi-eq+wew1}) as
{}
\begin{eqnarray}
\Big(\Delta +k^2\big(1+\frac{4U}{c^2}\big)-u\big(\frac{1}{u}\big)''
\Big)\Big[\frac{\,{}^e{\hskip -1pt}\Pi}{u}\Big]={\cal O}\Big(r_g^2,\frac{J_2}{r^3}\frac{\,{}^e{\hskip -1pt}\Pi}{u}\Big).
\label{eq:Pi-eq*=]*}
\end{eqnarray}

Equation~(\ref{eq:Pi-eq*=]*}) is similar to the Schr\"odinger equation of quantum mechanics, used to describe scattering on the Coulomb potential. However, this equation has an extra potential of $-u(u^{-1})''\simeq r_g/r^3$. It is known \cite{Messiah:1968} that the presence of potentials of $\propto 1/r^3$ in (\ref{eq:Pi-eq*=]*}) does not alter the asymptotic behavior of the solutions. Neglecting the $u(u^{-1})''\simeq r^{-3}$ term in (\ref{eq:Pi-eq*=]*}) reduces this equation to the form of a time-independent Schr\"odinger equation that describes scattering in a Newtonian potential:
{}
\begin{eqnarray}
\Big(\Delta +k^2\big(1+\frac{4U}{c^2}\big)\Big)\Big[\frac{\,{}^e{\hskip -1pt}\Pi}{u}\Big]={\cal O}\Big(r_g^2,\frac{J_2}{r^3}\frac{\,{}^e{\hskip -1pt}\Pi}{u}\Big).
\label{eq:Pi-eq*=+}
\end{eqnarray}
In the case of the SGL, we will always be at distances that are much larger than the Sun's Schwarzschild radius. Therefore, we may neglect the term $u(u^{-1})''\simeq {r_g}/{r^3}$ in (\ref{eq:Pi-eq*=]*}). We use (\ref{eq:Pi-eq*=+}) for the purposes of establishing the properties of the EM wave diffracted by the solar gravitational lens.  An identical equation may be obtained for  ${}^m{\hskip -1pt}\Pi$. This solution is consistent with the eikonal approximation, the use of which to describe the scattering of high-energy particles or processes related to the diffraction of light is well-justified.

By means of  (\ref{eq:Pi-eq}), Eq.~(\ref{eq:Dr*+}) may be written as
\begin{eqnarray}
\,{}^e{\hskip -2pt}{\hat D}_r&=&
\frac{\partial}{\partial r}\Big[\frac{1}{u^2} \frac{\partial (r\,{}^e{\hskip -1pt}\Pi)}{\partial r}\Big]+k^2u^2(r\,{}^e{\hskip -1pt}\Pi)+{\cal O}\Big(\frac{J_2}{r^3}\frac{\,{}^e{\hskip -1pt}\Pi}{u}\Big).
\label{eq:Dr**}
\end{eqnarray}
It can be verified by substituting (\ref{eq:Dp-Dt+})--(\ref{eq:Pi-eq}) and (\ref{eq:Dr**}) into (\ref{eq:Dr})--(\ref{eq:Bp}) that we have obtained a solution of our set of equations. In a similar way, we may consider the magnetic wave. We find that this wave can be derived from a potential ${}^m{\hskip -1pt}\Pi$ which satisfies the same differential equation (\ref{eq:Pi-eq}) as ${}^e{\hskip -1pt}\Pi$.

The complete solution of the EM field equations is obtained by adding the two fields (as discussed in \cite{Mie:1908,Born-Wolf:1999,Kerker-book:1969}), namely ${\vec D}={}^e{\hskip -1pt}{\vec D}+{}^m{\hskip -1pt}{\vec D}$; and ${\vec B}={}^e{\hskip -1pt}{\vec B}+{}^m{\hskip -1pt}{\vec B}$. This yields, to ${\cal O}\big(r_g^2,({J_2}/{r^3})({}^e\Pi/u)\big)$,
{}
\begin{eqnarray}
{\hat D}_r&=&\frac{1}{u}\Big\{\frac{\partial^2 }{\partial r^2}
\Big[\frac{r\,{}^e{\hskip -1pt}\Pi}{u}\Big]+\Big(k^2u^4-u\big(\frac{1}{u}\big)''\Big)\Big[\frac{r\,{}^e{\hskip -1pt}\Pi}{u}\Big]\Big\},
\label{eq:Dr-em}\\[3pt]
{\hat D}_\theta&=&\frac{1}{u^2r}
\frac{\partial^2 \big(r\,{}^e{\hskip -1pt}\Pi\big)}{\partial r\partial \theta}+\frac{ik}{r\sin\theta}
\frac{\partial\big(r\,{}^m{\hskip -1pt}\Pi\big)}{\partial \phi},
\label{eq:Dt-em}\\[3pt]
{\hat D}_\phi&=&\frac{1}{u^2r\sin\theta}
\frac{\partial^2 \big(r\,{}^e{\hskip -1pt}\Pi\big)}{\partial r\partial \phi}-\frac{ik}{r}
\frac{\partial\big(r\,{}^m{\hskip -1pt}\Pi\big)}{\partial \theta},
\label{eq:Dp-em}\\[3pt]
{\hat B}_r&=&\frac{1}{u}\Big\{\frac{\partial^2}{\partial r^2}\Big[\frac{r\,{}^m{\hskip -1pt}\Pi}{u}\Big]+\Big(k^2u^4-u\big(\frac{1}{u}\big)''\Big)\Big[\frac{r\,{}^m{\hskip -1pt}\Pi}{u}\Big]\Big\},
\label{eq:Br-em}\\[3pt]
{\hat B}_\theta&=&-\frac{ik}{r\sin\theta}
\frac{\partial\big(r\,{}^e{\hskip -1pt}\Pi\big)}{\partial \phi}+\frac{1}{u^2r}
\frac{\partial^2 \big(r\,{}^m{\hskip -1pt}\Pi\big)}{\partial r\partial \theta},
\label{eq:Bt-em}\\[3pt]
{\hat B}_\phi&=&\frac{ik}{r}
\frac{\partial\big(r\,{}^e{\hskip -1pt}\Pi\big)}{\partial \theta}+\frac{1}{u^2r\sin\theta}
\frac{\partial^2 \big(r\,{}^m{\hskip -1pt}\Pi\big)}{\partial r\partial \phi}.
\label{eq:Bp-em}
\end{eqnarray}

Both potentials $\Pi=\big({}^e{\hskip -1pt}\Pi, \,{}^m{\hskip -1pt}\Pi\big)$ are solutions of the differential equation (\ref{eq:Pi-eq+wew1}), which, in the case of the weak gravity characteristic for the SGL, is given by (\ref{eq:Pi-eq*=+}) and in terms of potential $\Pi$  takes the form:
{}
\begin{eqnarray}
\Big(\Delta +k^2\big(1+\frac{4U}{c^2}\big)\Big)\Big[\frac{\Pi}{u}\Big]
={\cal O}\Big(r_g^2,\frac{J_2}{r^3}\frac{\Pi}{u}\Big).
\label{eq:Pi-d+}
\end{eqnarray}
This completes decomposition of the Maxwell equations (\ref{eq:rotE_fl})--(\ref{eq:rotH_fl}) on the curved background in the weak gravitational field of the solar system.  Eqs.~(\ref{eq:Dr-em})--(\ref{eq:Bp-em}) together with (\ref{eq:Pi-d+}) is our primary result that will use throughout this paper. Again, note that in (\ref{eq:Pi-d+}), we discarded the term $u(u^{-1})'' \sim1/r^3$, representing the tail of the gravitational potential, as insignificant (see discussions in Appendix~F of Ref.~\cite{Turyshev-Toth:2017} and in  Appendix~C of \cite{Turyshev-Toth:2019}).

Finally, for the components ${\hat D}_\theta, {\hat D}_\phi$ and ${\hat B}_\theta, {\hat B}_\phi$ to be continuous over a spherical surface at some large distance from the origin, $r=R_\star$, it is evidently sufficient that the four quantities \cite{Turyshev-Toth:2017}
\begin{eqnarray}
\epsilon (r\,{}^e{\hskip -1pt}\Pi), \qquad \mu (r\,{}^m{\hskip -1pt}\Pi),
\qquad \frac{\partial (r\,{}^e{\hskip -1pt}\Pi)}{\partial r},
\qquad \frac{\partial (r\,{}^m{\hskip -1pt}\Pi)}{\partial r},
\label{eq:bound_cond}
\end{eqnarray}
shall also be continuous over this surface. Thus, our boundary conditions also split into independent conditions on $\,{}^e{\hskip -1pt}\Pi$ and $\,{}^m{\hskip -1pt}\Pi$. Our problem is thus reduced to the problem of finding two mutually independent solutions of the equations (\ref{eq:Pi-eq}) with prescribed boundary conditions.

\section{Computing the eikonal phase}
\label{sec:eik-phase}
\subsection{Different forms of the gravitational potential}
\label{sec:potU}

Before we proceed with solving (\ref{eq:Pi-d+}), we recognize that the gravitational potential $U$ from (\ref{eq:w-PN}) in spherical coordinates $(r\equiv|{\vec x}|,\phi,\theta)$ may be given in the most general case in the form of spherical harmonics:
{}
\begin{eqnarray}
U&=&G\int \frac{\rho({\vec x'})d^3x'}{|{\vec x}-{\vec x'}|}+{\cal O}(c^{-4})=\frac{GM}{r}\Big(1+\sum_{\ell=2}^\infty\sum_{k=0}^{+\ell}\Big(\frac{R}{r}\Big)^\ell P_{\ell k}(\cos\theta)(C_{\ell k}\cos k\phi+S_{\ell k}\sin k\phi)\Big)+
{\cal O}(c^{-4}),
\label{eq:pot_w_0}
\end{eqnarray}
where $\rho({\vec x})$ is the mass density inside the body,  $M$ is its mass, $R$ is its radius, $P_{\ell k}$ are the Legendre polynomials, while $C_{\ell k}$ and $S_{\ell k}$ are relativistic normalized spherical harmonic coefficients that characterize the body.

In the case of an axisymmetric body (i.e., the Sun), its external gravitational potential is reduced to the $k=0$ zonal harmonics and may be expressed \cite{Roxburgh:2001,LePoncinLafitte:2007tx} in terms of the usual dimensionless multipole moments $J_\ell$:
{}
\begin{eqnarray}
U&=&\frac{GM}{r}\Big\{1-\sum_{\ell=2}^\infty J_\ell \Big(\frac{R}{r}\Big)^\ell P_\ell\Big(\frac{{\vec k}_3\cdot{\vec x}}{r}\Big)\Big\}+
{\cal O}(c^{-4}),
\label{eq:pot_stis}
\end{eqnarray}
where ${\vec k}_3$ denotes the unit vector along the $x^3$-axis, $P_\ell$ are the Legendre polynomials. Furthermore, in the case of an axisymmetric and rotating body with ``north-south symmetry'', such as the Sun, the expression (\ref{eq:pot_stis}) contains only the $\ell=2,4,6,8...$ even moments \cite{Roxburgh:2001}.

Following \cite{Linet_2002}, we take into account the identity
{}
\begin{eqnarray}
\frac{\partial^\ell}{\partial z^\ell}\Big(\frac{1}{r}\Big)&=&\frac{(-1)^\ell \ell!}{r^{1+\ell}} P_\ell \Big(\frac{{\vec k}_3\cdot{\vec x}}{r}\Big), \qquad z=x^3,
\label{eq:iden}
\end{eqnarray}
and present  $U$ as the following expansion in a series of derivatives of $1/r$
{}
\begin{eqnarray}
U&=&GM\Big\{\frac{1}{r}-\sum_{\ell=2}^\infty \frac{(-1)^\ell}{ \ell!} J_\ell  R^\ell\frac{\partial^\ell}{\partial z^\ell}\Big(\frac{1}{r}\Big)\Big\}+
{\cal O}(c^{-4}).
\label{eq:pot_st3}
\end{eqnarray}
As we shall see below, this form is much more convenient  for the computation of integrals involving  $U$.

Considering the generic case, it was shown \cite{Mathis-LePoncinLafitte:2007} that the scalar gravitational potential (\ref{eq:pot_w_0}) may equivalently be given in the following form:
{}
\begin{eqnarray}
U&=&GM\Big\{\frac{1}{r}+\sum_{\ell=2}^\infty \frac{(-1)^\ell}{\ell!} {\cal T}^{<a_1...a_\ell>}\frac{\partial^\ell}{\partial x^{<a_1...}\partial x^{a_\ell>}}\Big(\frac{1}{r}\Big)\Big\}+
{\cal O}(c^{-4}),
\label{eq:pot_stf}
\end{eqnarray}
where $r=|{\vec x}|$, $M$ is the post-Newtonian mass of the body, and ${\cal T}^{<a_1...a_\ell>}$ are the symmetric trace-free (STF) mass multipole moments of the body \cite{Thorne:1980,Blanchet-Damour:1986,Blanchet-Damour:1989,Kopeikin:1997} defined as
{}
\begin{eqnarray}
M&=&\int_{\tt V} d^3{\vec x}\, \rho({\vec x}),\qquad
{\cal T}^{<a_1...a_\ell>}=\frac{1}{M}\int_{\tt V} d^3{\vec x}\, \rho({\vec x})\, x^{<a_1...a_\ell>},
\label{eq:mom}
\end{eqnarray}
where $x^{<a_1...a_\ell>}=x^{<a_1}x^{a_2...}x^{a_\ell>}$, the angle  brackets $<...>$ denote the STF operator, and ${\tt V}$ means the total volume of the isolated gravitating body under consideration. The dipole moment ${\cal T}^a$ is absent in the expansion (\ref{eq:pot_stf}) since we took the origin of the coordinates to be at the center of mass of the body.

\subsection{Computing the eikonal phase}

Based on the form of the post-Newtonian potentials (\ref{eq:pot_w_0}), (\ref{eq:pot_stf}), and (\ref{eq:pot_st3}), it is convenient to separate the monopole term from the rest of the multipoles. As we know \cite{Turyshev-Toth:2017}, the action of the monopole term is similar to that of the Coulomb potential which is a long-range potential that is felt as far as the source.  The remaining multipole terms form the short-range potential, $V_{\tt sr}c^2$, yielding the decomposition $U/c^2= {r_g}/{2r}+ V_{\tt sr}$ for the Newtonian potential, which allows to present the potential term in (\ref{eq:Pi-d+}) in the following form:
{}
\begin{eqnarray}
\frac{4U}{c^2}&=& \frac{2r_g}{r}+ 4V_{\tt sr}.
\label{eq:shrtpot}
\end{eqnarray}
The short-range potential forms the eikonal phase given by (\ref{eq:eik7h}) that has the form
{}
\begin{equation}
\xi_b(\tau) =\frac{k}{2}\int^{\tau}_{\tau_0}  2V_{\tt sr}({\vec b},\tau') d\tau' +{\cal O}(r_g^2).
\label{eq:eik_pA}
\end{equation}

\subsubsection{Computing the eikonal phase for an axisymmetric body}
\label{sec:eik-phase-Jn}

Here we develop an expression for the eikonal phase in the case of an axisymmetric body, with its potential given by (\ref{eq:pot_st3}).
In this case, the decomposition of the post-Newtonian potential takes the from
{}
\begin{eqnarray}
\frac{4U}{c^2}&=&
\frac{2r_g}{r}-2r_g\sum_{\ell=2}^\infty \frac{(-1)^\ell}{ \ell!} J_\ell  R^\ell\frac{\partial^\ell}{\partial s^\ell}\Big(\frac{1}{r}\Big).
\label{eq:U-c2]=}
\end{eqnarray}
In the case the short-range potential, $V_{\tt sr}$ from (\ref{eq:pot_stf}) is given as
{}
\begin{eqnarray}
V_{\tt sr}&=&-
\frac{r_g}{2} \sum_{\ell=2}^\infty \frac{(-1)^\ell}{ \ell!} J_\ell  R^\ell\frac{\partial^\ell}{\partial s^\ell}\Big(\frac{1}{r}\Big).
\label{eq:V-sr-m2A}
\end{eqnarray}

We now compute the leading term of this expansion. For that, we define the vector $\vec s$ to be a unit vector in the direction of the axis of rotation. Remembering that $r=\sqrt{b^2+\tau^2}+{\cal O}(r_g)$ from (\ref{eq:b0}), we evaluate directional derivatives ${\partial}/{\partial s}$ along  $\vec s\equiv \vec k_3$, which have the form
{}
\begin{eqnarray}
\frac{\partial}{\partial { s}}=({\vec s}\cdot{\vec \nabla})=\Big({\vec s}\cdot\frac{\partial}{\partial {\vec r}}\Big).
\label{eq:dir-s}
\end{eqnarray}
This relation allows us to compute the relevant partial derivatives for the leading tersm in (\ref{eq:pot_st3}):
{}
\begin{eqnarray}
\frac{\partial}{\partial { s}} \, \frac{1}{r}&=&-\frac{({\vec s}\cdot{\vec r})}{r^3}, \qquad \frac{\partial^2}{\partial { s}^2} \, \frac{1}{r}=\frac{3({\vec s}\cdot{\vec r})^2}{r^5}-\frac{1}{r^3},
\qquad \frac{\partial^3}{\partial { s}^3} \, \frac{1}{r}=-3\Big(\frac{5({\vec s}\cdot{\vec r})^3}{r^7}-\frac{3({\vec s}\cdot{\vec r})}{r^5}\Big),
\label{eq:dir-s1}\\
\frac{\partial^4}{\partial { s^4}} \, \frac{1}{r}&=&3\Big(\frac{35({\vec s}\cdot{\vec r})^4}{r^9}-\frac{30({\vec s}\cdot{\vec r})^2}{r^7}+\frac{3}{r^5}\Big),
\qquad \frac{\partial^5}{\partial { s}^5} \, \frac{1}{r}=-15\Big(\frac{63({\vec s}\cdot{\vec r})^5}{r^{11}}-\frac{70({\vec s}\cdot{\vec r})^3}{r^9}+\frac{15({\vec s}\cdot{\vec r})}{r^7}\Big),
\label{eq:dir-s2}\\
\frac{\partial^6}{\partial { s}^6} \, \frac{1}{r}&=&45\Big(\frac{231({\vec s}\cdot{\vec r})^6}{r^{13}}-\frac{315({\vec s}\cdot{\vec r})^4}{r^{11}}+\frac{105({\vec s}\cdot{\vec r})^2}{r^9}-\frac{5}{r^7}\Big),
\\
\frac{\partial^7}{\partial { s}^7} \, \frac{1}{r}&=&-315\Big(\frac{429({\vec s}\cdot{\vec r})^7}{r^{15}}-\frac{693({\vec s}\cdot{\vec r})^5}{r^{13}}+\frac{315({\vec s}\cdot{\vec r})^3}{r^{11}}-\frac{35({\vec s}\cdot{\vec r})}{r^9}\Big),
\\
\frac{\partial^8}{\partial { s}^8} \, \frac{1}{r}&=&315\Big(\frac{6435({\vec s}\cdot{\vec r})^8}{r^{17}}-\frac{12012({\vec s}\cdot{\vec r})^6}{r^{15}}+\frac{6930({\vec s}\cdot{\vec r})^4}{r^{13}}-\frac{1260({\vec s}\cdot{\vec r})^2}{r^{11}}+\frac{35}{r^9}\Big).
\end{eqnarray}

Using these expressions in (\ref{eq:V-sr-m2A}), we have
{}
\begin{eqnarray}
2V_{\tt sr}({\vec b},\tau)&=&-
r_g \Big\{J_2R^2\frac{1}{2}\Big(\frac{3({\vec s}\cdot{\vec r})^2}{r^5}-\frac{1}{r^3}\Big)+
J_3R^3\frac{1}{2}\Big(\frac{5({\vec s}\cdot{\vec r})^3}{r^7}-\frac{3({\vec s}\cdot{\vec r})}{r^5}\Big)+\nonumber\\
&+&
J_4R^4\frac{1}{8}\Big(\frac{35({\vec s}\cdot{\vec r})^4}{r^9}-\frac{30({\vec s}\cdot{\vec r})^2}{r^7}+\frac{3}{r^5}\Big)+
J_5R^5\frac{1}{8}\Big(\frac{63({\vec s}\cdot{\vec r})^5}{r^{11}}-\frac{70({\vec s}\cdot{\vec r})^3}{r^9}+\frac{15({\vec s}\cdot{\vec r})}{r^7}\Big)+\nonumber\\
&+&
J_6R^6\frac{1}{16}\Big(\frac{231({\vec s}\cdot{\vec r})^6}{r^{13}}-\frac{315({\vec s}\cdot{\vec r})^4}{r^{11}}+\frac{105({\vec s}\cdot{\vec r})^2}{r^9}-\frac{5}{r^7}\Big)+\nonumber\\
&+&
J_7R^7\frac{1}{16}\Big(\frac{429({\vec s}\cdot{\vec r})^7}{r^{15}}-\frac{693({\vec s}\cdot{\vec r})^5}{r^{13}}+\frac{315({\vec s}\cdot{\vec r})^3}{r^{11}}-\frac{35({\vec s}\cdot{\vec r})}{r^9}\Big)+\nonumber\\
&+&
J_8R^8\frac{1}{128}\Big(\frac{6435({\vec s}\cdot{\vec r})^8}{r^{17}}-\frac{12012({\vec s}\cdot{\vec r})^6}{r^{15}}+\frac{6930({\vec s}\cdot{\vec r})^4}{r^{13}}-\frac{1260({\vec s}\cdot{\vec r})^2}{r^{11}}+\frac{35}{r^9}\Big)+\nonumber\\
&+&
\sum_{\ell=9}^\infty \frac{(-1)^\ell}{ \ell!} J_\ell  R^\ell\frac{\partial^\ell}{\partial s^\ell}\Big(\frac{1}{r}\Big)\Big\}.
\label{eq:V-sr-m2A3}
\end{eqnarray}
We can now substitute result (\ref{eq:V-sr-m2A3}) into expression (\ref{eq:eik_pA}) and integrate it. We observe that, technically, it is more straightforward to compute the eikonal phase shift  integration along the entire path from $\tau_0$ to $\tau$. Note that this way one computes the double shift, $2\xi_b(\tau)$. This integration results in many terms that  depend on the distance to the source, $r_0=\sqrt{b^2+\tau_0^2}$, and that to the image plane, $r=\sqrt{b^2+\tau^2}$. The resulting expression is greatly simplified in the case when both the source and the observer on the image plane are located at very large distances from the lens and the following inequalities are satisfied: $b/\sqrt{b^2+\tau^2}\simeq b/\tau\ll1$ and $b/\sqrt{b^2+\tau_0^2}\simeq b/\tau_0\ll1$. This step, essentially, constitutes the  thin lens approximation.\footnote{If needed, one can use all those terms to evaluate the eikonal phase, $\xi_b(\tau) $, for shorter distances, when $\tau\sim\tau_0\simeq b$. For problems related to gravitational lensing this is unnecessary, but may be needed for some solar system spacecraft tracking applications \cite{Turyshev:2012nw,Turyshev-GRACE-FO:2014,Turyshev-PLR:2010}.} This allows us to greatly simplify the result of the integration, yielding
{}
\begin{eqnarray}
\xi_b(\tau) &=&-
kr_g \Big\{J_2\Big(\frac{R}{b}\Big)^2 \frac{1}{2}\Big[2({\vec s}\cdot{\vec b})^2\frac{1}{b^2}+({\vec s}\cdot{\vec k})^2-1\Big]+
J_3\Big(\frac{R}{b}\Big)^3\frac{1}{3}\Big[\frac{({\vec s}\cdot{\vec b})}{b}\Big(4({\vec s}\cdot{\vec b})^2\frac{1}{b^2}+3({\vec s}\cdot{\vec k})^2-3\Big)\Big]+\nonumber\\
&+&
J_4\Big(\frac{R}{b}\Big)^4\frac{1}{4}\Big[\big(({\vec s}\cdot{\vec b})^2\frac{1}{b^2}+({\vec s}\cdot{\vec k})^2-1\big)({\vec s}\cdot{\vec b})^2\frac{8}{b^2}+\big(({\vec s}\cdot{\vec k})^2-1\big)^2\Big]+
\nonumber\\
&+&
J_5\Big(\frac{R}{b}\Big)^5\frac{1}{5}\Big[\frac{({\vec s}\cdot{\vec b})}{b}
\Big(({\vec s}\cdot{\vec b})^4\frac{16}{b^4}+\big(({\vec s}\cdot{\vec k})^2-1\big)({\vec s}\cdot{\vec b})^2\frac{20}{b^2}+5\big(({\vec s}\cdot{\vec k})^2-1\big)^2\Big)\Big]+\nonumber\\
&+&
J_6\Big(\frac{R}{b}\Big)^6\frac{1}{6}
\Big[({\vec s}\cdot{\vec b})^6\frac{32}{b^6}+\big(({\vec s}\cdot{\vec k})^2-1\big)({\vec s}\cdot{\vec b})^4\frac{48}{b^4}+\big(({\vec s}\cdot{\vec k})^2-1\big)^2({\vec s}\cdot{\vec b})^2\frac{18}{b^2}+\big(({\vec s}\cdot{\vec k})^2-1\big)^3\Big]+\nonumber\\
&+&
J_7\Big(\frac{R}{b}\Big)^7\frac{1}{7}\Big[\frac{({\vec s}\cdot{\vec b})}{b}
\Big(({\vec s}\cdot{\vec b})^6\frac{64}{b^6}+\big(({\vec s}\cdot{\vec k})^2-1\big)({\vec s}\cdot{\vec b})^4\frac{112}{b^4}+\big(({\vec s}\cdot{\vec k})^2-1\big)^2({\vec s}\cdot{\vec b})^2\frac{56}{b^2}+7\big(({\vec s}\cdot{\vec k})^2-1\big)^3\Big)\Big]+\nonumber\\
&+&
J_8\Big(\frac{R}{b}\Big)^8\frac{1}{8}
\Big[({\vec s}\cdot{\vec b})^8\frac{128}{b^8}+\big(({\vec s}\cdot{\vec k})^2-1\big)({\vec s}\cdot{\vec b})^6\frac{256}{b^6}+\big(({\vec s}\cdot{\vec k})^2-1\big)^2({\vec s}\cdot{\vec b})^4\frac{160}{b^4}+\nonumber\\
&&\hskip 100pt +\,\big(({\vec s}\cdot{\vec k})^2-1\big)^3({\vec s}\cdot{\vec b})^2\frac{32}{b^2}+\big(({\vec s}\cdot{\vec k})^2-1\big)^4\Big]+\nonumber\\
&+&
\sum_{\ell=9}^\infty \frac{(-1)^\ell}{ 2\ell!} J_\ell  R^\ell \int^{\tau}_{\tau_0} \frac{\partial^\ell}{\partial s^\ell}\Big(\frac{1}{r}\Big) d\tau' \Big\}+{\cal O}(r_g^2).
\label{eq:V-xi}
\end{eqnarray}

Note that a similar result for the quadrupole $J_2$ term was obtained in \cite{Klioner:1991SvA,Klioner-Kopeikin:1992,Zschocke-Klioner:2010}. Expression (\ref{eq:V-xi}) extends all the previous computations to the higher order terms including $J_8$.

We use the heliocentric spherical coordinate system and define the vectors of impact parameter, $\vec b$, the wavevector, $\vec k$, and the unit vector long the solar rotational axis, $\vec s$, as follows:
{}
\begin{eqnarray}
{\vec b}&=& b(\cos\phi_\xi,\sin\phi_\xi,0),\qquad {\vec k}=(0,0,1),
\qquad {\vec s}=(\sin\beta_s\cos\phi_s,\sin\beta_s\sin\phi_s,\cos\beta_s).
\label{eq:vec3}
\end{eqnarray}
With these definitions,  the eikonal phase (\ref{eq:delta-D*-av0WKB+1*}) for the case of axisymmetric body whose gravitational potential is given by (\ref{eq:pot_st3}) as\footnote{To derive the results in a compact form we used multiple angle formulae: {\tt https://mathworld.wolfram.com/Multiple-AngleFormulas.html}  and also {\tt https://www.anirdesh.com/math/trig/cosine-identities.php} }
{}
\begin{eqnarray}
\xi_b(r)
&=& -kr_g\Big\{J_2\frac{1}{2}\Big(\frac{R_\odot}{b}\Big)^2 \sin^2\beta_s\cos[2(\phi_\xi-\phi_s)]+
J_3\frac{1}{3}\Big(\frac{R_\odot}{b}\Big)^3 \sin^3\beta_s\cos[3(\phi_\xi-\phi_s)]+\nonumber\\
&&\hskip 20pt +\,
J_4\frac{1}{4}\Big(\frac{R_\odot}{b}\Big)^4\sin^4\beta_s\cos[4(\phi_\xi-\phi_s)]+
J_5\frac{1}{5}\Big(\frac{R_\odot}{b}\Big)^5 \sin^5\beta_s\cos[5(\phi_\xi-\phi_s)]+
\nonumber\\
&&\hskip 20pt +\,
J_6\frac{1}{6}\Big(\frac{R_\odot}{b}\Big)^6\sin^6\beta_s\cos[6(\phi_\xi-\phi_s)]+
J_7\frac{1}{7}\Big(\frac{R_\odot}{b}\Big)^7\sin^7\beta_s\cos[7(\phi_\xi-\phi_s)]+\nonumber\\
&&\hskip 20pt +\,J_8\frac{1}{8}\Big(\frac{R_\odot}{b}\Big)^8\sin^8\beta_s\cos[8(\phi_\xi-\phi_s)]+
\sum_{n=9}^\infty\frac{1}{n}J_n\Big(\frac{R_\odot}{b}\Big)^n \sin^n\beta_s\cos[n(\phi_\xi-\phi_s)]\Big\}
+{\cal O}(r_g^2).~~~
\label{eq:eik-ph-axi*=}
\end{eqnarray}
Assuming that the pattern evident in these expressions continues for higher multipoles, we obtain the following compact expression for the eikonal phase:
{}
\begin{eqnarray}
\xi_b(\vec b,\vec s)
&=& -kr_g\sum_{n=2}^\infty\frac{J_n}{n}\Big(\frac{R_\odot}{b}\Big)^n \sin^n\beta_s\cos[n(\phi_\xi-\phi_s)]
+{\cal O}(r_g^2).
\label{eq:eik-ph-axi*}
\end{eqnarray}
Note that the sum in (\ref{eq:eik-ph-axi*}) contains contributions from all multipole moments, $n=2,3,4,5...$ and is valid for any axisymmetric body with respect to the $z=x^3$ axis represented by ${\vec s}$. If in addition to being axisymmetric, that body also has ``north-south'' symmetry (symmetry under a reflection with respect to the plane of rotation), that sum contains only even terms, $n=2,4,6,8...,$ \cite{Roxburgh:2001}.

\subsubsection{Computing the eikonal phase using STF tensors}

Using the representations (\ref{eq:pot_w_0}), (\ref{eq:pot_stf}) or (\ref{eq:pot_st3}), it is convenient to separate the monopole term in the potential $U$.  In fact, to determine the solution to (\ref{eq:Pi-d+}), similarly to \cite{Turyshev-Toth:2018-plasma,Turyshev-Toth:2019}, we first separate the monopole contribution and present  the $U$-dependent term in (\ref{eq:Pi-d+}) as
{}
\begin{eqnarray}
\frac{4U}{c^2}&=&
\frac{2r_g}{r}+ 2r_g\sum_{\ell=2}^\infty \frac{(-1)^\ell}{\ell!} {\cal T}^{<a_1...a_\ell>}\frac{\partial^\ell}{\partial x^{<a_1...}\partial x^{a_\ell>}}\Big(\frac{1}{r}\Big),
\label{eq:U-c2]}
\end{eqnarray}
where $r_g=2GM/c^2$ is the Schwarzschild radius of the body and the short-range potential $V_{\tt sr}$ from (\ref{eq:pot_stf}) is given by
{}
\begin{eqnarray}
V_{\tt sr}&=&\frac{r_g}{2}\sum_{\ell=2}^\infty \frac{(-1)^\ell}{\ell!} {\cal T}^{<a_1...a_\ell>}\frac{\partial^\ell}{\partial x^{<a_1...}\partial x^{a_\ell>}}\Big(\frac{1}{r}\Big).
\label{eq:V-sr-m2}
\end{eqnarray}
As such, this form is valid for any deviation from spherical symmetry in the gravitational field.

Given $V_{\tt sr}({\vec r})$ from (\ref{eq:V-sr-m2}), we reduced the problem to evaluating a single integral to determine the Debye potential $\Pi({\vec r})$ from (\ref{eq:Pi-eq*0+*+1}), which is a great simplification. Given the fact that ${\vec b}$ is constant and by taking the short-range potential $V_{\tt sr}({\vec r})$ from (\ref{eq:V-sr-m2}), we evaluate (\ref{eq:eik7h}) as
{}
\begin{eqnarray}
\xi_b(r)
&=& kr_g\sum_{\ell=2}^\infty
\frac{(-1)^\ell}{2\ell!} {\cal T}^{<a_1...a_\ell>} \int^{\tau}_{\tau_0} \frac{\partial^\ell}{\partial x^{<a_1...}\partial x^{a_\ell>}}\Big(\frac{1}{r}\Big) d\tau'.
\label{eq:delta-D*-av0WKB+1*}
\end{eqnarray}

 In fact, we may generalize expression ${\vec \nabla}={\nabla}_b+{\vec k}\,d/d\tau +{\cal O}(r_g)$ and write
{}
\begin{eqnarray}
\frac{\partial^\ell}{\partial x^{<a_1...}\partial x^{a_\ell>}}&\equiv&
{\vec \nabla}^{<a_1....}{\vec \nabla}^{a_\ell>}=\sum_{p=0}^\ell \frac{\ell!}{p!(\ell-p)!}k_{<a_1}...k_{a_p} \partial_{a_{p+1}}... \partial_{a_\ell>} \frac{\partial^p}{\partial \tau^p}+{\cal O}(r_g),
\label{eq:derivA}
\end{eqnarray}
where a new shorthand notation $\partial_a\equiv \partial/\partial {\vec b}^a$ has been used and $\tau$ is defined by (\ref{eq:x-Newt*=0}).

 Using representation (\ref{eq:derivA}), we can compute the relevant integral (with $r=\sqrt{b^2+\tau^2}$ and $r_0=\sqrt{b^2+\tau_0^2}$, as discussed in Sec.~\ref{sec:eik-wfr})
 {}
\begin{eqnarray}
 \int^{\tau}_{\tau_0} \frac{\partial^\ell}{\partial x^{<a_1...}\partial x^{a_\ell>}}\Big(\frac{1}{r}\Big) d\tau'&=&\nonumber\\
 &&\hskip -100pt\,=
  \sum_{p=0}^\ell \frac{\ell!}{p!(\ell-p)!}k_{<a_1}...k_{a_p} \partial_{a_{p+1}}... \partial_{a_\ell>} \Big\{\frac{\partial^{p}}{\partial \tau^{p}}\ln \Big(\frac{\sqrt{b^2+\tau^2}+\tau}{b}\Big) +
\frac{\partial^{p}}{\partial \tau_0^{p}}\ln \Big(\frac{\sqrt{b^2+\tau_0^2}+|\tau_0|}{b}\Big)\Big\} =\nonumber\\
 &&\hskip -100pt\,=
 \partial_{<a_1}... \partial_{a_\ell>}\Big\{ \ln  \Big(\frac{\sqrt{b^2+\tau^2}+\tau}{b}\Big)+\ln \Big(\frac{\sqrt{b^2+\tau^2}+|\tau_0|}{b}\Big)\Big\}+\nonumber\\
 &&\hskip -80pt\,+
  \sum_{p=1}^\ell \frac{\ell!}{p!(\ell-p)!}k_{<a_1}...k_{a_p} \partial_{a_{p+1}}... \partial_{a_\ell>} 
  \Big\{\frac{\partial^{p-1}}{\partial \tau^{p-1}}  \frac{1}{\sqrt{b^2+\tau^2}}+\frac{\partial^{p-1}}{\partial \tau_0^{p-1}}  \frac{1}{\sqrt{b^2+\tau_0^2}}\Big\},
\label{eq:int+1*}
\end{eqnarray}
where we accounted for the fact that $\tau$ changes sign at $\tau=0$.

As a result, the eikonal phase (\ref{eq:delta-D*-av0WKB+1*}) takes the form:
{}
\begin{eqnarray}
2\xi_b(r)
&=& kr_g\sum_{\ell=2}^\infty
\frac{(-1)^\ell}{\ell!} {\cal T}^{<a_1...a_\ell>}   
\Big\{
 \partial_{<a_1}... \partial_{a_\ell>}\Big\{ \ln \Big(\frac{\sqrt{b^2+\tau^2}+\tau}{b}\Big)+\ln  \Big(\frac{\sqrt{b^2+\tau_0^2}+|\tau_0|}{b}\Big)\Big\}+\nonumber\\
 &&\hskip 20pt\,+
  \sum_{p=1}^\ell \frac{\ell!}{p!(\ell-p)!}k_{<a_1}...k_{a_p} \partial_{a_{p+1}}... \partial_{a_\ell>} 
  \Big\{\frac{\partial^{p-1}}{\partial \tau^{p-1}}  \frac{1}{\sqrt{b^2+\tau^2}}+\frac{\partial^{p-1}}{\partial \tau_0^{p-1}}  \frac{1}{\sqrt{b^2+\tau_0^2}}\Big\}\Big\}. 
\label{eq:eik-phA}
\end{eqnarray}

Applying the same approximations used to derive (\ref{eq:V-xi}), namely $b/\sqrt{b^2+\tau^2}\simeq b/\tau\ll1$ and $b/\sqrt{b^2+\tau_0^2}\simeq b/\tau_0\ll1$,  (\ref{eq:eik-phA}) takes a much simplified form 
{}
\begin{eqnarray}
\xi_b(r)
&=& -kr_g\sum_{\ell=2}^\infty
\frac{(-1)^\ell}{\ell!} {\cal T}^{<a_1...a_\ell>}   
 \partial_{<a_1}... \partial_{a_\ell>} \ln kb.
\label{eq:eik-ph}
\end{eqnarray}

We observe that in the case of an axisymmetric matter distribution within the lens result (\ref{eq:eik-ph}) reduces to  (\ref{eq:eik-ph-axi*}).  Note that expression (\ref{eq:eik-ph}) is derived for an arbitrary matter distribution within a gravitating body. One may use this result to derive the eikonal phase for any of the terms in the Newtonian potential (\ref{eq:pot_w_0}). Such a generic potential may be suitable for analysis of the gravitational lensing by a lens with an arbitrary intrinsic matter distribution.

\section{Using the path integral formalism}
\label{app:pathintegral}

As we mentioned before,  Eq.~(\ref{eq:Pi-eq+wew1*+}) is nearly identical to the time-independent Schr\"odinger equation that in nuclear physics describes the scattering problem on a potential $U$ \cite{Messiah:1968,Mott:1928,Gordon:1928}. Here we further explore this analogy. For that, we note that in the absence of the scattering potential, $U$, solution (\ref{eq:Pi-eq+wew1*+}) may be given in the from of a plane EM wave given as $\psi_0(\vec r)=E_0e^{i \vec k\cdot\vec r}$. Next, we introduce a cylindrical coordinate system $(\rho, \phi, z)$, whose $z$-axes is directed along the wavevector $\vec k$. Then, by defining the amplification factor due to scattering on the gravitational potential $U$ of the lens as $\mu(\vec r)=[\Pi(\vec r)/u]/\psi_0(\vec r)$,  we rewrite (\ref{eq:Pi-eq+wew1*+})  as
{}
\begin{eqnarray}
\Big\{\Delta \psi_0({\vec r})+k^2\psi_0({\vec r})\Big\}\mu({\vec r})+\psi_0({\vec r})\Delta \mu({\vec r})+
2\big({\vec \nabla}\psi_0({\vec r})\cdot{\vec \nabla} \mu({\vec r})\big)
+k^2\frac{4U({\vec r})}{c^2}\mu(\vec r)\psi_0({\vec r})&=&0.
\label{eq:eikpp}
\end{eqnarray}
As $\psi_0({\vec r})$ is the solution of the homogeneous wave equation in flat, vacuum spacetime, the first term in (\ref{eq:eikpp}) is zero. Then, we can divide the remaining terms of (\ref{eq:eikpp}) by $\psi_0({\vec r})$, which yields:
{}
\begin{eqnarray}
\Delta \mu({\vec r})+
2\Big({\vec \nabla}\ln \psi_0({\vec r})\cdot{\vec \nabla} \mu({\vec r})\Big)
+k^2\frac{4U({\vec r})}{c^2}\mu(\vec r)&=&0.
\label{eq:eikpp2}
\end{eqnarray}
Clearly, ${\vec \nabla}\ln \psi_0({\vec r})=ik{\vec k}$, where $k=\omega/c$ is the wavenumber and $\vec k$ is the unit vector in the direction of the wavevector. From discussion in Section~\ref{sec:eik-wfr}, we know that $({\vec k} \cdot {\vec \nabla})=d/d\tau$. We remember the form of the Laplacian in the cylindrical coordinate system $({\vec \rho}, z)$ that in our case is given as
{}
\begin{eqnarray}
\Delta \mu({\vec r})= \Delta_{\vec \rho} \mu({\vec r})+\frac{\partial^2 \mu({\vec r})}{\partial z^2},
\qquad {\rm where}\qquad
\Delta_{\vec \rho} \mu({\vec r})=\frac{1}{\rho}\frac{\partial}{\partial\rho}\Big(\rho \frac{\partial \mu({\vec r})}{\partial\rho} \Big)+\frac{1}{\rho^2}\frac{\partial^2 \mu({\vec r})}{\partial\phi^2}.
\label{eq:eikpp3}
\end{eqnarray}
Substituting these results into (\ref{eq:eikpp2}), we have
{}
\begin{eqnarray}
\frac{\partial^2 \mu({\vec r})}{\partial z^2}+\Delta_{\vec \rho} \mu({\vec r})+
2ik \frac{ d\mu({\vec r})}{d\tau}
+k^2\frac{4U({\vec r})}{c^2}\mu(\vec r)&=&0.
\label{eq:eikpp4}
\end{eqnarray}
We assume that $k/|\partial \ln \mu/\partial z|\sim$ (scale at which $\mu$ varies)/(wavelength) $\gg1$, we neglect the first term  compared with the second term, which constitutes the eikonal approximation.
Then Eq.~(\ref{eq:eikpp4}) takes a familiar form:
{}
\begin{eqnarray}
i\frac{d \mu(\tau,{\vec \rho})}{d\tau}&=&\Big(
-\frac{1}{2k}\Delta_{\vec \rho} -\frac{2kU(\tau,{\vec \rho})}{c^2} \Big)\mu(\tau,{\vec \rho}),
\label{eq:eikpp5}
\end{eqnarray}
which is the Schr\"odinger equation with the ``time'' coordinate $\tau$, the ``particle mass'' $k$, and the ``time-dependent potential'' $-2kc^{-2}U(\tau, \vec \rho)$. The corresponding Lagrangian that yields the classical equation of motion is given as
{}
\begin{eqnarray}
L(\tau,\vec \rho,\dot{\vec\rho})=k\Big({\textstyle\frac{1}{2}}\dot{\vec\rho}^2+2c^{-2}U(\tau,{\vec \rho})\Big),
\label{eq:eikpp6}
\end{eqnarray}
where $\dot{\vec\rho}=d{\vec\rho}/d\tau$. This Lagrangian describes the motion of a pendulum in a gravity field with potential $-2kc^{-2}U(\tau, \vec \rho)$. This is a mechanical analogy for forming the caustics on the image plane of the SGL. This is similar to a motion of a connected pendulum where each of the multipoles characterized by a unique natural spacial frequency, affects the motion of the entire pendulum in a carefully prescribed fashion \cite{Landau-Lifshitz:1988m}.

In the path integral formulation \cite{Feynman-Hibbs:1991}, the solution to (\ref{eq:eikpp5}) may formally be written as
{}
\begin{eqnarray}
\mu(\vec r)&=&\int {\cal D}\vec \rho(\tau)\exp\Big[i\int_{\tau_0}^\tau L(\tau,\vec \rho,\dot{\vec\rho})d\tau \Big].
\label{eq:eikpp7}
\end{eqnarray}
Following the established rules of evaluating path integrals \cite{Feynman-Hibbs:1991,Chaichian-Demichev:2001}, we have
{}
\begin{eqnarray}
\int_{\tau_0}^\tau k{\textstyle\frac{1}{2}}\dot{\vec\rho}^2d\tau=\frac{k}{2 \tau}\Big(\vec \rho(\tau) - \vec \rho(0)\Big)^2\simeq \frac{k}{2 r}({\vec b} - r \vec \theta)^2,
\label{eq:eikpp7a}
\end{eqnarray}
with $\vec \rho(\tau)=r \vec \theta$, $\vec \rho(0)={\vec b}$ and
where we realize that for very small angles $\theta=\rho/r$, $\tau=(\vec k\cdot\vec x)\simeq r$ is a valid approximation. Integrating (\ref{eq:eikpp7a}), we also use a thin lens approximation while assuming that the effect of the lens on light is instantaneous and affects light only after it has passed through the lens. Initially the light continues on a straight line, so that $\vec \rho(0) - \vec \rho(\tau_0)=0$, then, there is a sudden path change after which the light continues on a different straight line until it reaches the observer on the image plane. Integrating the potential terms, we use representation (\ref{eq:shart-pot}), that is, ${4U}/{c^2}= {2r_g}/{r}+ 4V_{\tt sr}$ and the approach presented in Appendix~\ref{sec:eik-phase-Jn}, which yeilds
{}
\begin{eqnarray}
\int_{\tau_0}^\tau k 2c^{-2}U(\tau, \vec \rho) d\tau=
kr_g\ln 2kr+kr_g\ln 2kr_0-2kr_g\big(\ln kb+\psi(\vec b)\big),
\label{eq:eikpp7b}
\end{eqnarray}
where $\psi(\vec b)$ is given by (\ref{eq:psi}).
As a result, after applying the appropriate normalization factor $A=\sqrt{k/2\pi ir}$ to each of the two dimensions involved \cite{Feynman-Hibbs:1991,Chaichian-Demichev:2001}, the expression (\ref{eq:eikpp7}) results in
{}
\begin{eqnarray}
\mu(\vec r) &=&
E_0e^{ikr_g\ln 2kr} \frac{k}{ir}\frac{1}{2\pi}\int d^2\vec b \,\exp\Big[ik\Big(\frac{1}{2 r}({\vec b} - r \vec \theta)^2-2r_g\big(\ln kb+\psi(\vec b)\big)\Big)\Big].
  \label{eq:gamma077}
\end{eqnarray}
Combining this expression with $\psi_0(\vec r)=E_0e^{i \vec k\cdot\vec r}$, we get for the Debye potential $\Pi(\vec r)$ an expression that is equivalent to (\ref{eq:gamma0+}) for the factor $\gamma(r,\theta,\phi) $, providing the connection between the two different methods used to derive this result.

The derivation presented here shows a deep connection between various methods of modern theoretical physics used to provide a wave-optical description of diffraction of light, namely the Kirchhoff--Fresnel diffraction formula \cite{Born-Wolf:1999,Landau-Lifshitz:1988,Nambu:2013}, the path integrals \cite{Feynman:1948,Feynman-Hibbs:1991,Chaichian-Demichev:2001,Nakamura-Deguchi:1999,Yamamoto:2017}, the Mie theory \cite{Mie:1908,Born-Wolf:1999,Turyshev-Toth:2017} relying on the Debye potentials and the eikonal approximation \cite{Sharma-etal:1988,Sharma-Sommerford:1990,Sharma-Sommerford-book:2006}. The approach that we presented in this paper has the advantage as it can be used to evaluate the vector nature of the EM field diffracted by the gravity field of an extended lens. This connection will be investigated further.

\end{document}